\documentclass[aps,prx,twocolumn,notitlepage,showkeys,showpacs,superscriptaddress,longbibliography,preprintnumbers]{revtex4-2}
\usepackage{epsf}
\usepackage{bm}
\usepackage{times}
\usepackage{amsfonts}
\usepackage{amssymb}
\usepackage{amsmath,mathtools}
\usepackage{array}
\usepackage{enumerate,dsfont,here}
\usepackage{dcolumn,multirow,bbold}
\usepackage{tikz-cd}
\usepackage{bbding}
\usepackage[colorlinks=true,citecolor=blue,linkcolor=blue]{hyperref} 
\usepackage[normalem]{ulem}
\usepackage{longtable}
\usepackage{orcidlink}

\newcommand{\creation}{\hat{c}^{\dagger}}

\newcommand{\bx}{\bm{x}}
\newcommand{\bk}{\bm{k}}
\newcommand{\bt}{\bm{t}}
\newcommand{\br}{\bm{r}}
\newcommand{\bR}{\bm{R}}

\newcommand{\bb}{\bm{b}}
\newcommand{\ba}{\bm{a}}

\newcommand{\mZ}{\mathbb{Z}}
\newcommand{\calT}{\mathcal{T}}
\newcommand{\calC}{\mathcal{C}}
\newcommand{\calP}{\mathcal{P}}
\newcommand{\calJ}{\mathcal{J}}
\newcommand{\calG}{\mathcal{G}}
\newcommand{\calM}{\mathcal{M}}

\newcommand{\calA}{\mathcal{A}}

\newcommand{\im}{\text{Im }}
\newcommand{\calGint}{\calG_{\text{int}}}

\begin{document}
	
	\title{Classification of time-reversal symmetric topological superconducting phases for conventional pairing symmetries}
	\author{Seishiro Ono\,\orcidlink{0000-0002-4908-2612}}
	\email{s-ono@g.ecc.u-tokyo.ac.jp}
	\affiliation{Department of Applied Physics, University of Tokyo, Tokyo 113-8656, Japan}
	
	\author{Ken Shiozaki}
	\email{ken.shiozaki@yukawa.kyoto-u.ac.jp}
	\affiliation{Yukawa Institute for Theoretical Physics, Kyoto University, Kyoto 606-8502, Japan}
	
	\author{Haruki Watanabe\,\orcidlink{0000-0002-8112-021X}}
	\email{hwatanabe@g.ecc.u-tokyo.ac.jp}
	\affiliation{Department of Applied Physics, University of Tokyo, Tokyo 113-8656, Japan}
	
	\preprint{YITP-22-62}

	\begin{abstract}
		Based on a recently developed framework, we conduct classifications of time-reversal symmetric topological superconductors with conventional pairing symmetries. Our real-space approach clarifies the nature of boundary modes in nontrivial phases. 
		The key difference from the calculations for topological crystalline insulators originates from the appearance of vortex zero modes on the interface of several two-dimensional topological superconductors. 
		 We find that our classification is complete in the $K$-theory sense for all rod groups, all layer groups, and 159 out of 230 space groups. 
		Our results shed new light on superconductors with conventional pairing as candidates for topological superconductors.
	\end{abstract}
	
\maketitle
\section{Introduction}

In the past decades, topological superconductors (TSCs) have attracted much attention because of their exotic surface states that realize Majorana fermions. 
Majorana fermions could be leveraged for fault-tolerant quantum computation~\cite{RevModPhys.87.137}.
Topological superconducting phases protected by internal symmetries, particle-hole symmetry (PHS) and time-reversal symmetries (TRS), are discovered in the early stages of studies of TSCs~\cite{PhysRevB.78.195125,Kitaev_bott,Ryu_2010}. 
Symmetry properties of Cooper pairs, often called pairing symmetries, are closely related to the topological nature of superconductivity. 

Crystalline symmetry plays a pivotal role in superconductors.
Pairing symmetries are classified into irreducible representations of point groups~\cite{Sigrist-Ueda}. 
Recent intensive studies reveal that unconventional superconductors, whose pairing symmetries are distinct from conventional Bardeen-Cooper-Schrieffer (BCS) superconductivity, have a great chance of being TSCs ~\cite{PhysRevB.89.020509,PhysRevB.90.100509,PhysRevLett.107.217001,PhysRevB.94.174502,PhysRevB.95.224514,PhysRevX.8.041026,PhysRevLett.122.227001,PhysRevLett.123.217001,PhysRevLett.124.207001,PhysRevLett.125.097001,PhysRevResearch.3.L032071,Tang-Ono-Wan-Watanabe2021}. 
For instance, odd-parity superconductors are promising candidates for TSCs~\cite{PhysRevB.81.220504,PhysRevLett.105.097001}.
Furthermore, thanks to the bloom in understanding of TSCs protected by crystalline symmetries~\cite{PhysRevB.79.214526,PhysRevLett.111.087002,PhysRevB.90.165114,Ando-Fu,SatoFujimoto,Sato_2017,Shiozaki-Sato-Gomi2017,PhysRevLett.119.246401,PhysRevB.97.205136,PhysRevB.97.205135,Ono-Watanabe2018,PhysRevX.9.011012,Cornfeld-Chapman,Ono-Yanase-Watanabe2019,Skurativska2019,Ono-Po-Watanabe2020,Ahn2019,SI_Luka,PhysRevResearch.2.043300,Ono-Po-Shiozaki2020,huang2020faithful,PhysRevResearch.3.013052,Ono-Shiozaki2022,PhysRevB.105.094518, 2111.07252,2108.04534,2109.02664,PhysRevB.104.094529,PhysRevB.105.104501,PhysRevB.105.104515}, it is known that topological superconducting phases can also exist for even-parity but unconventional pairing symmetries~\cite{CC-Shiozaki,Ono-Po-Watanabe2020,Ono-Po-Shiozaki2020,SI_Luka,Surface_Fang}. 
Unfortunately, however, unconventional superconductivity is quite rare, and most superconductors exhibit conventional pairing symmetries.

Then, it is crucial to understand the topological nature of superconductors with conventional pairing symmetries. Several studies actually show the existence of topological phases for conventional pairing symmetries. 
For example, for noncentrosymmetric superconductors in which even-parity and odd-parity components are allowed to be mixed, TSCs protected by TRS and PHS can be realized when the odd-parity component is comparable to or larger than the even-parity one~\cite{PhysRevB.79.094504}. 
More recently, Ref.~\cite{Fe-based_TSC} has reported a topological superconducting phase for conventional pairing symmetry in centrosymmetric space group $P4/nmm$ (No.~129). 
It should be noted that sign-changing pair potentials, dubbed extended $s$- or $s_{\pm}$-wave pairings, are required to realize this phase~\cite{PhysRevB.81.134508, PhysRevB.83.224511,Fe-based_TSC}.
However, despite intense research efforts, the list of topological phases for conventional pairing symmetries is still elusive in most space groups.

In principle, TSCs could be classified by $K$-theory~\cite{Karoubi,Freed2013,Shiozaki-Sato-Gomi2017,K-AHSS}, but the actual calculation is often challenging.
Recently, an alternative approach has been introduced based on a real-space perspective~\cite{TC_PRX,TC_PRB,Xiong_2018,R-AHSS_Song,R-AHSS_Shiozaki, TC_AII, defect_network, PhysRevB.101.100501,PhysRevB.101.085137,PhysRevB.101.165129}, which is called Atiyah-Hirzebruch spectral sequence (AHSS)~\cite{R-AHSS_Shiozaki} in real space or topological crystals~\cite{TC_AII, TC_MSG}.
The idea of this method is that any topological crystalline phase can be constructed by symmetrically placing lower-dimensional topological phases. 
It has actually succeeded in comprehensively classifying topological insulators~\cite{TC_AII, TC_MSG} and bosonic systems~\cite{TC_PRB,R-AHSS_Song,PhysRevB.101.085137,PhysRevB.101.165129}. 

In this work, we generalize the method to superconductors. 
There are two complications in the real-space classification for superconductors, compared with the case for insulators.
For one, we find one-dimensional building block TSCs that bring out some technical differences in the classification procedures.
In addition, we discover a new obstruction to constructing gapped phases. 
The existence of superconducting vortices with odd-integer windings can be enforced by crystalline symmetries. Consequentially, Majorana zero modes emerge at the vortex cores.

We overcome these difficulties and provide real-space classifications of time-reversal symmetric TSCs with conventional pairing symmetries in all rod groups, layer groups, and space groups. Our results for all rod, all layer, and 159 space groups are complete. 
Importantly, we find that there exist topological superconducting phases in 199 out of 230 space groups (61 centrosymmetric and 138 noncentrosymmetric space groups). 
Since pairing symmetries of most realistic materials are conventional as mentioned above, our study will stimulate further studies of topological superconductivity in realistic materials that have already been verified or will be discovered in the future.

\section{Overview of Real-Space Classification}
%\textit{Overview of Real-Space Classification.---}
Our classification of topological superconductors in this work is based on the framework developed in previous studies~\cite{TC_PRX,TC_PRB,R-AHSS_Song,R-AHSS_Shiozaki, defect_network}.
Consider a symmetry group $G=\calG_{\text{int}} \times \calG$, where $\calG_{\text{int}}$ and $\calG$ are an internal symmetry group and a space group.
The classification of topological crystalline phase with symmetry $G$ is based on a conjecture:~any topological crystalline phase can be adiabatically deformed into a $\calG$-symmetric patchwork of lower-dimensional invertible topological phases protected by internal symmetries of the subspace~\cite{TC_PRX,TC_PRB}. 
Under this assumption, the classification of gapped topological crystalline phases is equivalent to the classification of patchworks that do not contain any gapless mode in the bulk.
For superconductors with conventional pairing symmetries, this can be performed in the following steps (A)--(E).

\underline{(A): Decomposing unit cell.}
Three-dimensional (3D) space is decomposed into $p$-dimensional regions for $p= 0, 1, 2,$ and $3$, which are called $p$-cells.
Each $p$-cell should be chosen small enough so that any two points in a $p$-cell are not symmetry-related to each other. 
To define $3$-cells, we choose a closed and simply connected region that covers the entire space by symmetry operations exactly once.
Such a region is known as asymmetric unit (AU)~\cite{ITC} and can be chosen as a polyhedron. % for the case of space groups.
The AU is divided into $M$ subregions ($M\geq1$). 
The interiors of these subregions and their symmetric copies are $3$-cells.
In general, $M$ can be set $1$ but sometimes other choices are convenient.
The $2$-cells are polygons on faces of $3$-cells.
Similarly, $1$-cells are line segments along the edges of $2$-cells. 
Endpoints of $1$-cells are $0$-cells.
This decomposition is referred to as \textit{cell decomposition}. 
Although the cell decomposition for a given symmetry setting is not unique, the classification outcome does not depend on the choice.

\underline{(B): Identifying building blocks.} 
For each $p$-cell, we determine its symmetry group, which is composed of $\calG_{\text{int}}$ and the subgroup of $\calG$ that behaves as an internal symmetry group on the $p$-cell. For time-reversal symmetric superconductors with significant spin-orbit coupling (SOC), $\calG_{\text{int}}$ is generated by the time-reversal symmetry with $\mathcal{T}^2=-1$ and the particle-hole symmetry $\mathcal{C}^2=+1$. 
On each $p$-cell, we list all $p$-dimensional TSCs protected by the symmetry group of the $p$-cell.  According to Refs.~\cite{Kitaev_bott,Ryu_2010}, for conventional pairing symmetries, there are four building blocks:
3D $\mZ$-TSC on $3$-cells, 
2D $\mZ_2$-TSC on $2$-cells whose symmetry is $\calG_{\text{int}}$ only (i.e., without an additional mirror), 
1D $\mZ_2$-TSC or 1D $\mZ$-TSC on $1$-cells depending on the symmetry of the cell.

\underline{(C): Constructing boundary-gapped patchwork.}
We construct patchworks by symmetrically arranging these building blocks.   
A TSC on a $p$-cell exhibits $(p-1)$-dimensional gapless boundary modes. 
We demand that boundary modes in a patchwork be gapped among them.  
We refer to patchworks obtained this way as \textit{boundary-gapped patchworks}. We identify boundary-gapped patchworks that can be deformed into each other.
When the point group of $\calG$ contains an element whose determinant is $-1$, nontrivial TSCs on $3$-cells result in irremovable gapless surface states on $2$-cells and hence do not produce boundary-gapped patchworks.

\underline{(D): Checking the absence of vortex zero modes.}
We check if a boundary-gapped patchwork is  \textit{fully gapped}. 
Patchworks constructed by 2D TSCs may still contain vortex zero modes and we require the absence of such zero modes.
This step is unique to superconductors and was not necessary in previous studies for insulators and bosonic systems~\cite{R-AHSS_Song,TC_AII, TC_MSG}.
A fully gapped patchwork constructed from nontrivial TSCs on $p$-cell gives an element of an abelian group $E_{p,-p}^{\infty}=(\mZ_2)^{n_p}\times\mZ^{m_p}$ with nonnegative integers $n_p$ and $m_p$ [see Appendix~\ref{app:tabSG}-\ref{app:tabRG}]. 
When the determinant is $+1$ for all the elements of the point group of $\calG$, we find $E_{3,-3}^{\infty}=\mathbb{Z}$. Otherwise, $E_{3,-3}^{\infty}=0$.

\underline{(E): Determining $K$-theoretic classifications.}
As the last step we want to determine the abelian group of these topological phases, which is given by $K$ group.
The abelian group $E_{p,-p}^{\infty}$ was computed separately for each $p=1,2,3$ and, in general, $\oplus_{p=1}^3E_{p,-p}^{\infty}$ does not reproduce the $K$ group result, because $E_{p,-p}^{\infty}$ $(p=1,2,3)$ are not mutually independent.
For example, a patchwork obtained by stacking several copies of nontrivial patchworks in $E_{2,-2}^{\infty}$ is sometimes equivalent to a nontrivial patchwork in $E_{1,-1}^{\infty}$. To identify the group structure among $E_{p,-p}^{\infty}$, we have to solve the group extension problem~\cite{R-AHSS_Shiozaki,PhysRevB.99.085127}, as we will discuss later.

\begin{figure}[t]
	\begin{center}
		\includegraphics[width=0.99\columnwidth]{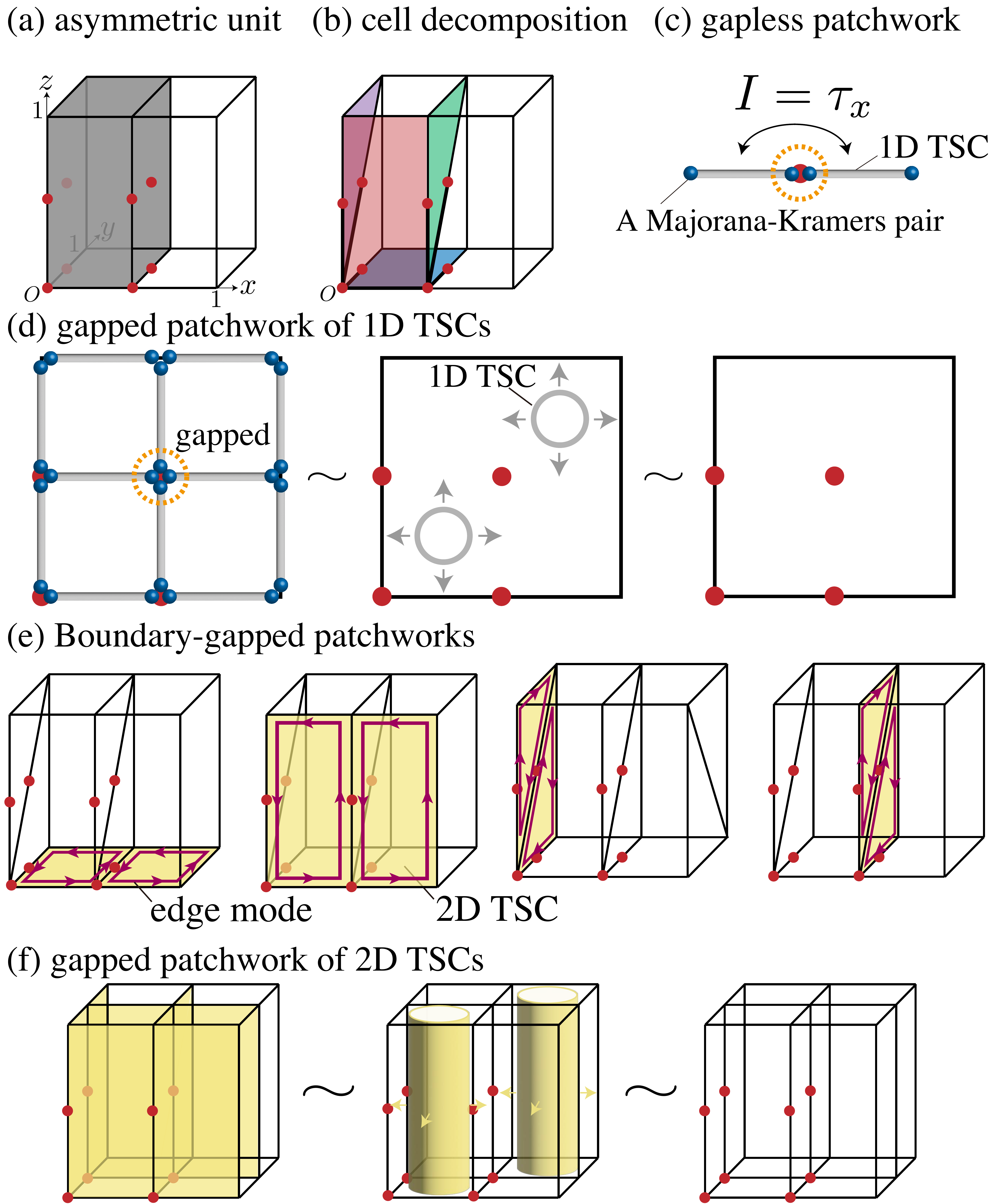}
		\caption{\label{fig1}\textbf{Illustration of 3D patchworks in space group $P\bar{1}$.} 
			(a) A choice of asymmetric unit. 
			(b) A cell decomposition. Symmetry-inequivalent $0$-cells, $1$-cells, and $2$-cells are represented by red solid circles, black bold lines segments, and the colored faces, respectively. 
			(c) An example of gapless patchworks. Here, gray lines and blue balls are 1D $\mZ_2$-TSCs and a Kramers pair of Majorana zero modes, respectively.
			(d) The equivalence between a gapped patchwork constructed by 1D $\mZ_2$-TSCs and the vacuum.   
			(e) Boundary-gapped patchworks for the cell decomposition in (a). All yellow planes are 2D $\mZ_2$-TSCs.
			(f) The equivalence between a gapped patchwork by 2D $\mZ_2$-TSCs and the vacuum. 
			The cell decomposition assumed in panels (d) and (f) is different from the one described in (b).
		}
	\end{center}
\end{figure}

\textit{Space group $P\bar{1}$---}
Let us illustrate the above steps [other than step (E)] through discussing space group $P\bar{1}$, generated by translations and inversion, as an example. %, with the even-parity pairing.
The AU is chosen as the gray region in Fig.~\ref{fig1}(a) with $0\leq x \leq 1/2$, $0\leq y < 1$, and $0\leq z < 1$.
A cell decomposition is shown in Fig.~\ref{fig1}(b). %, where there are symmetry-inequivalent four $2$-cells, seven $1$-cells, and eight $0$-cells.
The interior of AU gives a $3$-cell. Its face on $x=1/2$ (i.e., $0\leq y < 1$ and $0\leq z < 1$) contains inversion centers [red points in Fig.~\ref{fig1}(b)] and every point on the face is mapped to another point on the same face by inversion.
Hence the face should be divided into two $2$-cells, one of which is shown by a green triangle in Fig.~\ref{fig1}(b).
Similarly, edges of $2$-cells that contain inversion centers must be decomposed into two $1$-cells, one of which is shown by a thick line in Fig.~\ref{fig1}(b). This is the end of step (A).
Step (B) is common among almost all space groups. Since the symmetry of all $1$- and $2$-cells is $\calGint$ only, these cells host 1D $\mZ_2$-TSCs and 2D $\mZ_2$-TSCs.

Now we move on to step (C). We first consider 1D $\mZ_2$-TSCs on $1$-cells.
There are seven symmetry-inequivalent $1$-cells, generating in $2^7$ patterns of patchworks.
Note that endpoints of $1$-cells are inversion centers. Hence, every TSC on a $1$-cell meets with its inversion copy at endpoints.
However, the two Majorana-Kramers zero modes sitting at the inversion center cannot be gapped, because they have different inversion eigenvalues. This point can be understood by noting that the inversion is represented by $\tau_x$ (Pauli matrix) whose eigenvalues are $\pm1$ [see Fig.~\ref{fig1}(c)].
Therefore, in the current cell decomposition,  there are no boundary-gapped patchworks composed of 1D TSCs.
One might choose a different cell decomposition and put 1D $\mZ_2$-TSCs in such a way that four Majorana-Kramers zero modes meet at each inversion center to make a boundary-gapped patchwork as illustrated in Fig.~\ref{fig1}(d). However, such a configuration can be deformed into the vacuum. We conclude that $E_{1,-1}^{\infty}=0$ regardless of the choice of the cell decomposition.

We next consider patchworks composed of 2D $\mZ_2$-TSCs.
There are four symmetry-inequivalent $2$-cells.
We consider configurations obtained by placing a 2D TSC on one of them [see in Fig.~\ref{fig1}(e)]. Since two helical edge modes always meet on each of boundary $1$-cells, they can be gapped in pairs. Thus all of these configurations are boundary-gapped patchworks.

Step (D) is the novel part of the classification. As we elaborate below, these boundary-gapped patchworks constructed from 2D $\mZ_2$-TSCs are not fully gapped because vortex zero modes survive at inversion centers.
To get rid of vortex zero modes, an even number of 2D $\mZ_2$-TSCs must be crossing at each inversion center.
However, such a combination of four planes in Fig.~\ref{fig1}(e) does not exist in this cell decomposition.
When a different cell decomposition is assumed, such a configuration may be realized, but then it can be smoothly deformed into the vacuum as shown in Fig.~\ref{fig1}(f).
Thus we conclude $E_{2,-2}^{\infty}=0$ regardless of the choice of the cell decomposition.

\section{Superconducting vortices}
\textit{Continuum model of vortices.---}
In the presence of a vortex (with the unit winding) in the superconducting order parameter, 2D $p_x \pm i p_y$ superconductors have a Majorana zero mode at the core of the vortex~\cite{PhysRevB.61.10267,PhysRevB.73.014505,PhysRevLett.98.010506,PhysRevLett.99.037001, PhysRevB.75.212509}. 
The zero mode can be described by a continuum model %of 2D $p_x \pm i p_y$ superconductor
\begin{align}
	\label{eq:pxipy}
	H_{\pm}(\bm{r}; \Delta) &= \begin{pmatrix}
		-\frac{\nabla^2}{2m} - \mu(\bm{r})  & \Delta(\bm{r})\left(\frac{1}{i}\partial_x \pm i\frac{1}{i}\partial_y \right)\\
		\Delta^*(\bm{r})\left(\frac{1}{i}\partial_x \mp i\frac{1}{i}\partial_y \right) & \frac{\nabla^2}{2m} + \mu(\bm{r})
	\end{pmatrix},
\end{align}
where PHS is represented by $\tau_x$. The superconducting gap function $\Delta(\bm{r})$ and the chemical potential $\mu(\bm{r})$ vary slowly in space. 
To implement TRS with $\mathcal{T}^2=-1$, we introduce a spin degree of freedom and use $H_{+}(\bm{r}; \Delta)$ for the spin-up component and $H_{-}(\bm{r}; \Delta^*)$ for the spin-down component. The resulting Hamiltonian reads
\begin{align}
	\label{eq:DIII-SC}
	&H_{\text{DIII}}(\bm{r}; \Delta)
	= H_{+}(\bm{r}; \Delta) \otimes\begin{pmatrix}
		1& 0 \\
		0 & 0
	\end{pmatrix}
	+H_{-}(\bm{r}; \Delta^*)\otimes\begin{pmatrix}
		0& 0 \\
		0 & 1
	\end{pmatrix}.
\end{align}
PHS and TRS are represented by $\tau_xs_0$ and $i\tau_z s_y$, respectively. In these expressions, the second matrix refers to the spin space and $s_0$ and $s_{i}$'s are another set of Pauli matrices.

Here we argue that a 2D inversion symmetric TSC  always has a vortex with an odd-integer winding at the inversion center and consequently has a Kramers pair of Majorana zero modes  at the core of a vortex.
Inversion symmetry is represented by $\tau_0$ in $H_{\pm}(\bm{r}; \Delta)$ and by $\tau_0s_0$ in $H_{\text{DIII}}(\bm{r}; \Delta)$.
We assume the form $\Delta(\bm{r})=\Delta_0(r) e^{i (\alpha+n\theta)}$ and $\mu(\bm{r})=\mu(r)$, where $r$ and $\theta$ are polar coordinates and $\alpha \in \mathbb{R}$.
From the symmetry condition $\tau_0 H_{\pm}(\bm{r}; \Delta) = H_{\pm}(-\bm{r}; \Delta)\tau_0$, we find $\Delta(-\bm{r}) = -\Delta(\bm{r})$, which implies that $n$ is an odd integer $2m-1\ (m \in \mZ)$. Following 
in Refs.~\cite{PhysRevB.61.10267,PhysRevB.73.014505,PhysRevLett.98.010506,PhysRevLett.99.037001, PhysRevB.75.212509}, we obtain the wavefunction of the vortex zero mode
[see Appendix~\ref{app:vortex} for the derivation]
\begin{align}
	\phi^{\alpha}_{\pm, m}(r,\theta) &\propto e^{-\int^r \frac{\mu(r')}{\vert \Delta_0 (r') \vert} dr'}(u,
	u^*)^T,\\
	u&=e^{\tfrac{i}{2} (\pm\alpha + \pi/2)}e^{\pm i m \theta}.
\end{align}
It satisfies $H_{\pm}(\bm{r}; \Delta)\phi^{\alpha}_{\pm, m}(r,\theta)=\bm{0}$ and $\tau_0\phi^{\alpha}_{\pm, m}(r,\theta)=(-1)^m\phi^{\alpha}_{\pm, m}(r,\theta+\pi)$, which implies that the inversion eigenvalue of the vortex zero mode is $(-1)^m$.
In the time-reversal symmetric case, a Kramers pair of Majorana zero modes emerges at the vortex core, whose
wavefunctions are
\begin{align}
	\Phi^{\alpha}_{m,\uparrow}(r, \theta) &= \phi_{+, m}^{\alpha}(r,\theta)\otimes\begin{pmatrix}
		1\\0
	\end{pmatrix},\\
	\Phi^{\alpha}_{m,\downarrow}(r, \theta) &=	\phi_{-, -m}^{-\alpha}(r,\theta)\otimes\begin{pmatrix}
		0\\1
	\end{pmatrix}.
\end{align}
This completes the proof.

\textit{Space groups $P2/c$ and $P4/m$.---}
The patchwork in Fig.~\ref{fig1}(f) was fully gapped but trivial. Here we discuss additional crystalline symmetries sometimes make it nontrivial
or gapless due to vortex zero modes. As examples, we consider space groups $P2/c$ (No.~13) and $P4/m$ (No.~83). 
The configuration remains a boundary-gapped patchwork even for these space groups. 
Space group $P2/c$ has an additional glide symmetry, which prohibits the deformation process illustrated in Fig.~\ref{fig1}(f) and makes this patchwork nontrivial.

In contrast, space group $P4/m$ contains fourfold rotation $C_{4}^{z}$ along the $z$-axis in addition to translations and inversion in $P\bar{1}$.
A patchwork in Fig.~\ref{fig1}(f) cannot be obtained from the vacuum in the presence of this symmetry. 
However, as shown in the following discussions, this state is gapless due to the vortex zero modes at each inversion center protected by rotation eigenvalues.

Let us focus on a neighborhood of an inversion center, for example, the one at the origin $(0,0,0)$ [see Fig.~\ref{fig3}(a)].
Then, the Hamiltonian is described by
\begin{align}
	H(\br) &= \begin{pmatrix}
		H^{(1)}_{\text{DIII}}(z, x; \Delta)\delta(y) & 0 \\
		0 & H^{(2)}_{\text{DIII}}(z, y; \Delta)\delta(x)
	\end{pmatrix},\\
	U(C_{4}^{z}) &= \begin{pmatrix}
		0 & \tau_0s_0 \\
		i \tau_0 s_y & 0
	\end{pmatrix},
\end{align}
where $H^{(i = 1,2)}_{\text{DIII}}(z, x_i; \Delta) = H_{\text{DIII}}( \bm{r} = (z, x_i); \Delta)$ in Eq.~\eqref{eq:DIII-SC} 
and $U(C_{4}^{z})$ is a unitary representation of the fourfold rotation.
Note that $C_{4}^{z}$ transforms $H^{(1)}_{\text{DIII}}(z, x; \Delta)$ into $H^{(2)}_{\text{DIII}}(z, y; \Delta)$ and $H^{(2)}_{\text{DIII}}(z, y; \Delta)$ into $H^{(1)}_{\text{DIII}}(z, -x; \Delta^*)$. 
The wavefunctions of vortex zero modes are $\Psi^{\alpha}_{m, s, 1} \equiv \Phi^{\alpha}_{m, s}(r, \theta)\otimes (1,0)^{T}$ and $\Psi^{\alpha}_{m, s, 2} \equiv\Phi^{\alpha}_{m, s}(r, \theta)\otimes (0,1)^{T}$, where the latter vectors describe the degree of freedom of two planes. 
They are transformed as
\begin{align}
	&U(C_{4}^{z})(\Psi^{\alpha}_{m, \uparrow, 1}, \Psi^{\alpha}_{m, \downarrow, 1},\Psi^{\alpha}_{m, \uparrow, 2}, \Psi^{\alpha}_{m, \downarrow, 2}) \nonumber\\
	&= (\Psi^{\alpha}_{m, \uparrow, 1}, \Psi^{\alpha}_{m, \downarrow, 1},\Psi^{-\alpha}_{-m, \uparrow, 2}, \Psi^{-\alpha}_{-m, \downarrow, 2})
	\begin{pmatrix}
		0 & 0 & 1 & 0 \\
		0 & 0 & 0 & 1 \\
		0 & 1 & 0 & 0\\
		-1 & 0 & 0 & 0
	\end{pmatrix}.
\end{align}
This implies that the vortex zero modes have all different eigenvalues $e^{i q \frac{\pi}{4}}\ (q=1,3,5,7)$. Hence, these four Majorana-Kramers pairs cannot be gapped among them.

The same number of Majorana-Kramers pairs can be obtained by attaching two 1D TSCs along the rotation axis ($z$-axis) such that these two interchange under inversion symmetry as illustrated in Fig 1(c). However, Majorana-Kramers pairs constructed this way have both $\pm1$ inversion eigenvalues.
In contrast, all the four vortex zero modes have the inversion eigenvalue $(-1)^m$. Therefore, they cannot be gapped by attaching these 1D chains.

\begin{figure}[t]
	\begin{center}
		\includegraphics[width=0.6\columnwidth]{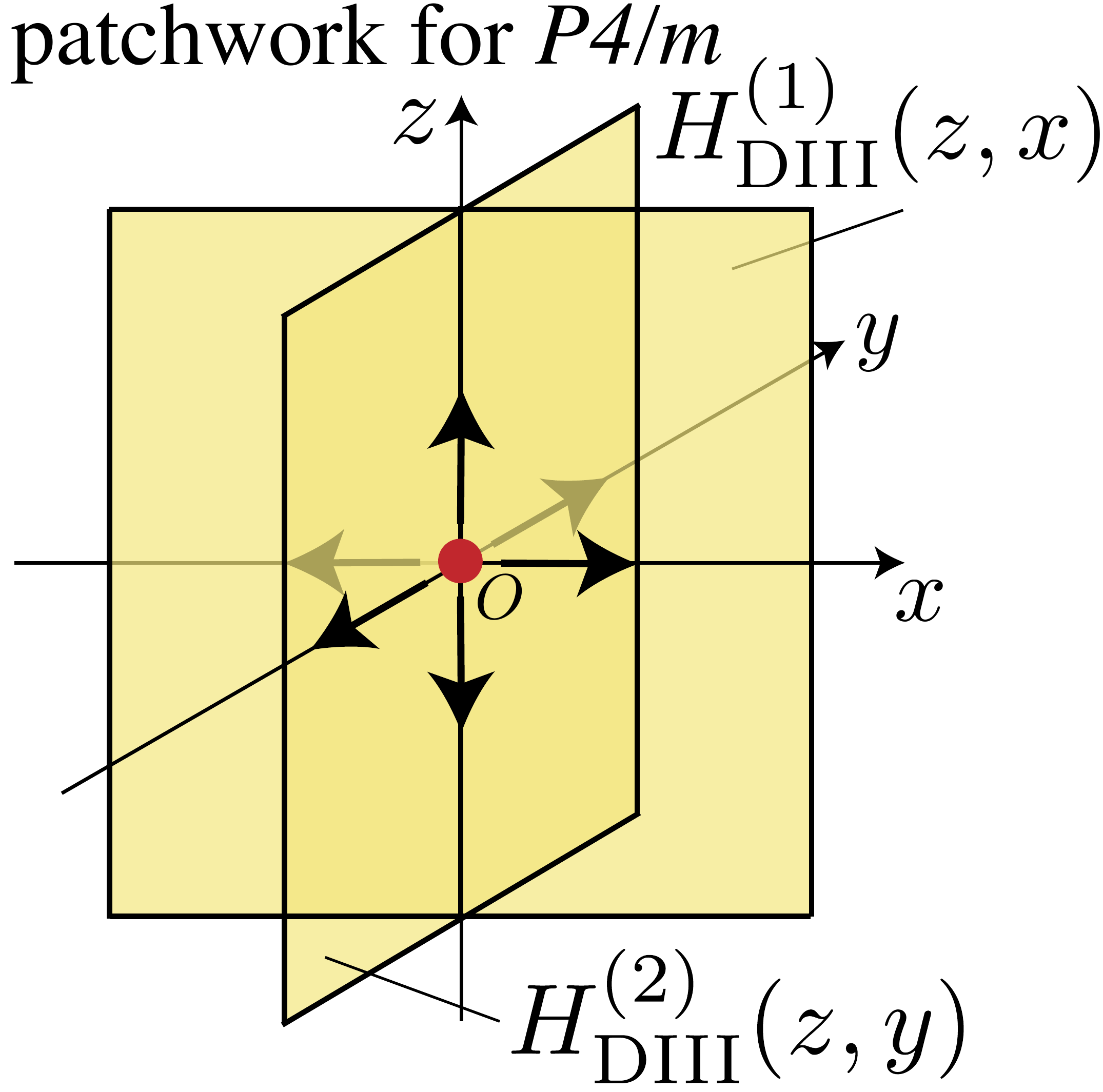}
		\caption{\label{fig2}\textbf{Illustration of patchworks in space group $P4/m$.}
			A boundary-gapped patchwork crossing two 2D TSCs near $O=(0,0,0)$. A vortex is present at $O$.
		}
	\end{center}
\end{figure}

\begin{figure}[t]
	\begin{center}
		\includegraphics[width=0.99\columnwidth]{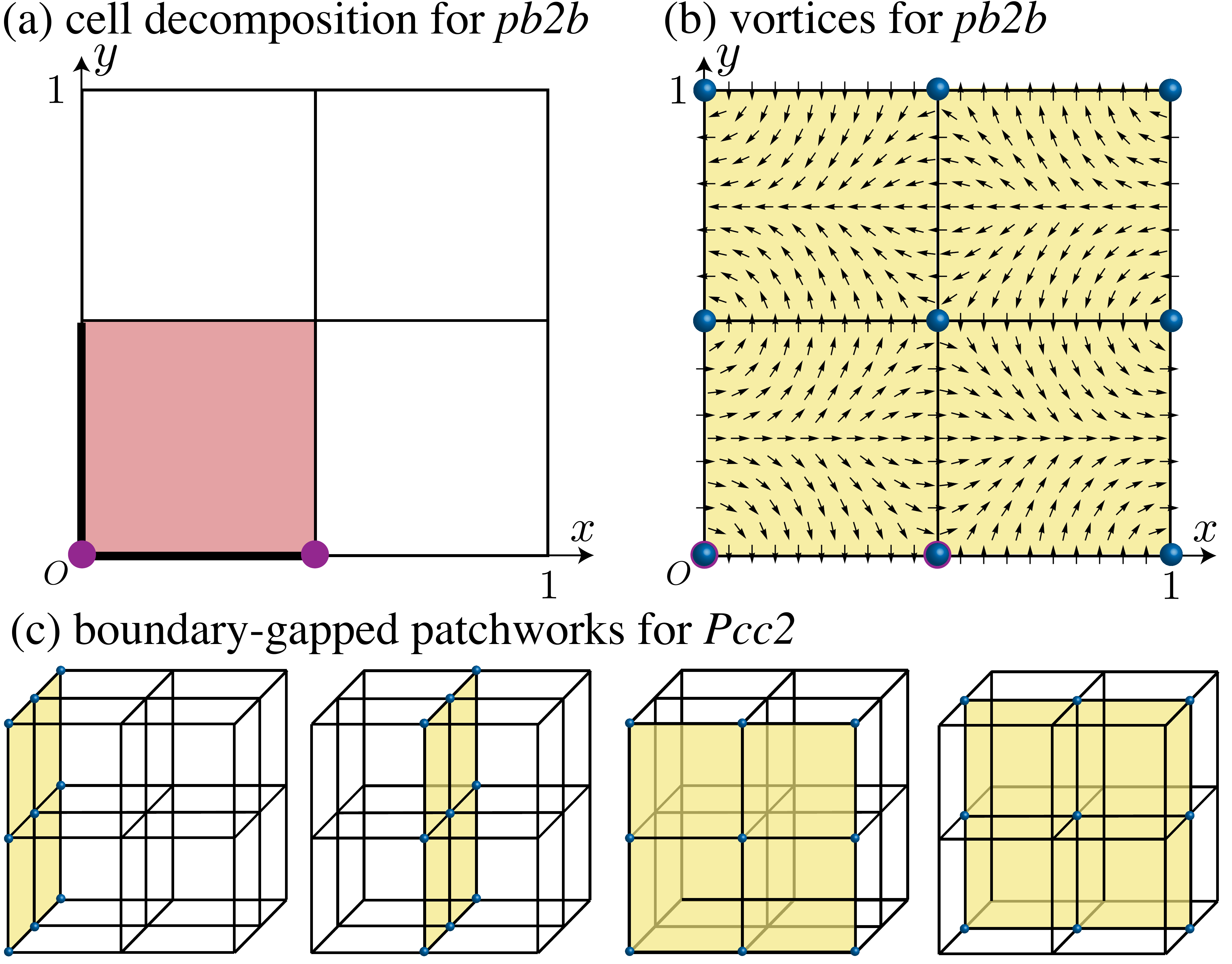}
		\caption{\label{fig3}\textbf{Illustration of patchworks in space group $Pcc2$.}
			(a) A cell decomposition for layer group $pb2b$. Symmetry-inequivalent $0$-cells, $1$-cells, and $2$-cells are represented by solid circles, black bold lines segments, and the colored face, respectively. 
			(b) An example of vortex configuration for $pb2b$. A vortex emerge at each of $0$-cells (blue balls).
			Arrows represent the phase of $\Delta(\bm{r})$. %The vorticity $m$ can be computed from the winding of $\Phi(\bm{r})$.
			(c) Boundary-gapped patchworks for space group $Pcc2$. Note that each plane is symmetric under symmetries in $pb2b$.
		}
	\end{center}
\end{figure}

\textit{Space group $Pcc2$.---}
It is tempting to think that such obstructions due to vortex zero modes happen only when inversion symmetry is present. 
However, this is untrue. Vortex zero modes can also be enforced by combinations of symmetries, not including inversion.
As an example, we consider space group $Pcc2$ generated by translation along $x$-direction, twofold rotation along $z$-axis $C_{2z}$, and glide $G_x:(x,y,z)\rightarrow (-x,y,z+1/2)$.

We focus on a 2D layer at $x=0$, which has layer group $pb2b$ (No.~30).
Our cell decomposition is shown in Fig.~\ref{fig3}(a). 
We introduce a 2D TSCs described by Eq.~\eqref{eq:DIII-SC} into each $2$-cell.
To respect the symmetry $U(C_{2y}) = i \tau_zs_x$ and $U(G_z) = i\tau_z s_z$, the gap function $\Delta(\bm{r})$ must satisfy $\Delta(C_{2y}\bm{r}) = \Delta^{*}(\bm{r})$ and $\Delta(G_{z}\bm{r}) = -\Delta(\bm{r})$, which implies the presence of vortices with odd-integer winding. We additionally require that all zero modes can exist only at 0-cells. An example of configurations satisfying all of these requirements is shown in Fig.~\ref{fig3}(b). 
Therefore, all the four boundary-gapped patchworks for space group $Pcc2$ in Fig 3(c) have vortex zero modes. A fully gapped patchwork can be constructed by stacking these four, which results in $E_{2,-2}^{\infty}=\mathbb{Z}_2$.

Vortices can also be enforced in 2D systems with other layer group symmetries, e.g., $pm2_1b$ (No.~28). After applying the above discussions to various symmetry settings, we obtain fully gapped patchworks $E_{2,-2}^{\infty}$ in all rod groups, layer groups, and space groups with conventional pairing symmetries. 
In Appendix~\ref{app:tabSG}-\ref{app:tabRG}, we show classification results of $E_{p,-p}^{\infty}$ in all symmetry settings we considered.
In addition, we tabulate patchworks that corresponds to generators of $E_{1,-1}^{\infty}$ and $E_{2,-2}^{\infty}$ in Supplemental Materials (SM).

\section{Final classification}
%\textit{Final classification.---}
As discussed in overview, we have to solve the group extension problem to obtain the final classification.
In our problem, we have to find an abelian group $X$ that satisfies
\begin{align}
	&X/E_{1,-1}^{\infty}\simeq E_{2,-2}^{\infty}.\label{extentionprob}
	%&X/Y\simeq E_{3,-3}^{\infty}=\mathbb{Z}\text{ or }0
\end{align}
There can be multiple possibilities but the correct one can be identified by a physical argument.
We analyze mass terms of the Dirac Hamiltonian that describes the phase obtained by stacking two copies of a generator of $E_{2,-2}^{\infty}$.
The group extension is trivial if the uniform mass is allowed in the Dirac Hamiltonian; otherwise, it is nontrivial. 
Given $X$, the final classification is given by $X\oplus E_{3,-3}^{\infty}$. 

\begin{figure}[t]
	\begin{center}
		\includegraphics[width=0.85\columnwidth]{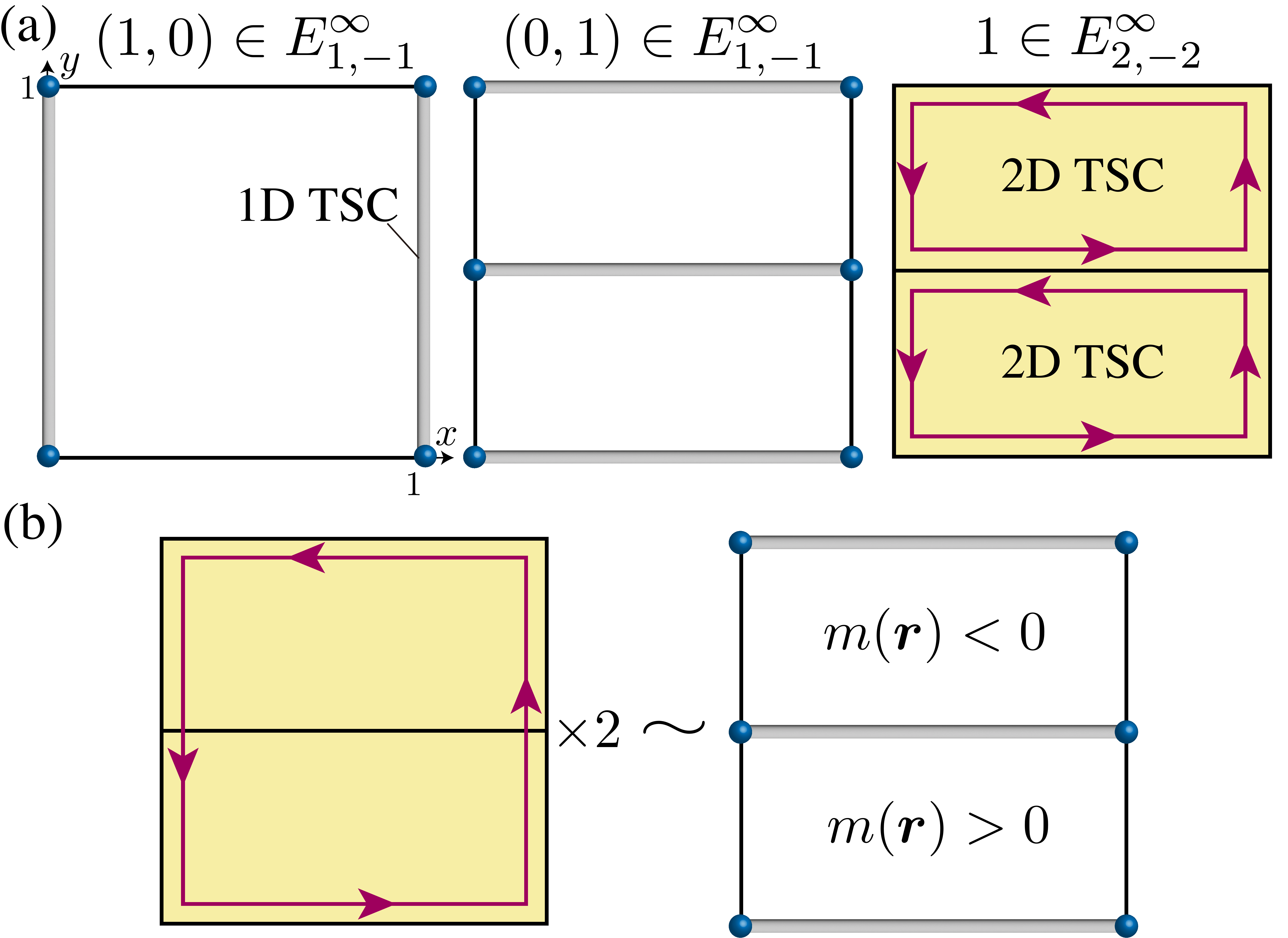}
		\caption{\label{fig4}\textbf{Illustration of the group extension problem in $pb$.}
			(a) All boundary-gapped patchworks, which are generators of $ E_{1,-1}^{\infty}$ and $ E_{2,-2}^{\infty}$. 
			(b) Relation between the stacked 2D $\mZ_2$-TSCs and a boundary-gapped patchwork composed of 1D $\mZ_2$-TSCs.
			Here, $m(\br)$ denotes the mass that will partially gap out the edge states.
		}
	\end{center}
\end{figure}

To demonstrate how to solve Eq.~\eqref{extentionprob}, here we discuss layer group $pb$ as an example. This group is generated by glide symmetry $G_x:(x,y,z)\rightarrow (-x,y+1/2,z)$ in addition to translation in $y$. After performing the real-space classification, we obtain $E_{1,-1}^{\infty} = (\mZ_2)^2$ and $E_{2,-2}^{\infty} = \mZ_2$. Their generators are illustrated in Fig.~\ref{fig4}(a).
The generator of $E_{2,-2}^{\infty}$ is described by the Dirac Hamiltonian~\eqref{eq:DIII-SC} with a constant $\Delta(\bm{r})$.
Then, we ask if the patchwork constructed by stacking two copies of the generator of $E_{2,-2}^{\infty}$ is a nontrivial element of $E_{1,-1}^{\infty}$.
By analyzing the mass term, we find that the stacked patchwork is still nontrivial, and it falls into %is equivalent to a patchwork corresponding to 
$(0,1) \in  E_{1,-1}^{\infty} =(\mZ_2)^2$.
This result indicates that the group extension is nontrivial and that the final classification is $\mZ_2 \times \mZ_4$ [see SM for more details].

Another class  of examples is layer group $pn\ (n = 2, 4, 6)$, which contains $n$-fold rotation symmetry. From the real-space classification, we have $E_{1,-1}^{\infty} = (\mZ_2)^{4-n/2}$ and $E_{2,-2}^{\infty} = \mZ_2$.
Unlike the case of layer group $pb$, the group extension is trivial, and the final classification is $E_{1,-1}^{\infty}\oplus E_{2,-2}^{\infty}=(\mZ_2)^{5-n/2}$ [see SM for more details].

The one-by-one analysis of mass terms is laborious. However, for 159 out of 230 space groups, we succeeded in sidestepping this issue by computing the final classification using AHSS in momentum space~\cite{K-AHSS}.
It should be emphasized that the real-space classifications are still crucial even when the momentum space classifications are available.
This is because the classifications in momentum space is not usually informative on the nature of the boundary modes in each entry of the classification.
In Appendix~\ref{app:tabSG}-\ref{app:tabRG}, we also show the final classifications for all rod groups, all layer groups, and 159 space groups.

\section{Conclusion}
%\textit{Conclusion \& Perspective.---}
In this work, by analyzing the wavefunctions of vortex zero modes and studying their symmetry representations, we succeeded in deriving obstructions to constructing gapped phases of superconductors. 
This enabled us to systematically perform the real-space classification for superconductors with conventional pairing symmetries in all rod, layer, and space groups. 
We found that topologically nontrivial superconducting phases exist in 199 out of 230 space groups (61 centrosymmetric and 138 noncentrosymmetric space groups). 
In addition, our results for all rod groups and layer groups are complete $K$-theoretic classifications. 159 space groups are also complete. 
We leave the group extention problem in the remaining 71 space groups as future work.

In search for materials that realize TSCs, superconductors with unconventional pairing symmetries have been actively investigated. Our work opens up a new direction of the search focusing on the conventional pairing symmetries that are majority in real materials.
	
	\begin{acknowledgments}
		We thank Hoi Chun Po for invaluable discussions in previous collaborations. 
		SO was supported by the ANRI Fellowship and KAKENHI Grant No.~JP20J21692 from the Japan Society for the Promotion of Science (JSPS)
		KS was supported by JST CREST Grant No.~JPMJCR19T2 and JST PRESTO Grant No.~JPMJPR18L4.
		HW was supported by JST PRESTO Grant No.~JPMJPR18LA, JSPS KAKENHI Grants No. JP20H01825, and No. JP21H01789.
				
		\textit{Note added.}---Recently, Ref.~\cite{Wire_Fang} appeared, which is based on a similar idea and discusses only 2D topological crystalline phases constructed only by 1D TSCs in wallpaper groups. 
	\end{acknowledgments}
	\bibliography{ref}

\begin{thebibliography}{78}%
\makeatletter
\providecommand \@ifxundefined [1]{%
 \@ifx{#1\undefined}
}%
\providecommand \@ifnum [1]{%
 \ifnum #1\expandafter \@firstoftwo
 \else \expandafter \@secondoftwo
 \fi
}%
\providecommand \@ifx [1]{%
 \ifx #1\expandafter \@firstoftwo
 \else \expandafter \@secondoftwo
 \fi
}%
\providecommand \natexlab [1]{#1}%
\providecommand \enquote  [1]{``#1''}%
\providecommand \bibnamefont  [1]{#1}%
\providecommand \bibfnamefont [1]{#1}%
\providecommand \citenamefont [1]{#1}%
\providecommand \href@noop [0]{\@secondoftwo}%
\providecommand \href [0]{\begingroup \@sanitize@url \@href}%
\providecommand \@href[1]{\@@startlink{#1}\@@href}%
\providecommand \@@href[1]{\endgroup#1\@@endlink}%
\providecommand \@sanitize@url [0]{\catcode `\\12\catcode `\$12\catcode
  `\&12\catcode `\#12\catcode `\^12\catcode `\_12\catcode `\%12\relax}%
\providecommand \@@startlink[1]{}%
\providecommand \@@endlink[0]{}%
\providecommand \url  [0]{\begingroup\@sanitize@url \@url }%
\providecommand \@url [1]{\endgroup\@href {#1}{\urlprefix }}%
\providecommand \urlprefix  [0]{URL }%
\providecommand \Eprint [0]{\href }%
\providecommand \doibase [0]{https://doi.org/}%
\providecommand \selectlanguage [0]{\@gobble}%
\providecommand \bibinfo  [0]{\@secondoftwo}%
\providecommand \bibfield  [0]{\@secondoftwo}%
\providecommand \translation [1]{[#1]}%
\providecommand \BibitemOpen [0]{}%
\providecommand \bibitemStop [0]{}%
\providecommand \bibitemNoStop [0]{.\EOS\space}%
\providecommand \EOS [0]{\spacefactor3000\relax}%
\providecommand \BibitemShut  [1]{\csname bibitem#1\endcsname}%
\let\auto@bib@innerbib\@empty
%</preamble>
\bibitem [{\citenamefont {Elliott}\ and\ \citenamefont
  {Franz}(2015)}]{RevModPhys.87.137}%
  \BibitemOpen
  \bibfield  {author} {\bibinfo {author} {\bibfnamefont {S.~R.}\ \bibnamefont
  {Elliott}}\ and\ \bibinfo {author} {\bibfnamefont {M.}~\bibnamefont
  {Franz}},\ }\bibfield  {title} {\bibinfo {title} {{Colloquium: Majorana
  fermions in nuclear, particle, and solid-state physics}},\ }\href
  {https://doi.org/10.1103/RevModPhys.87.137} {\bibfield  {journal} {\bibinfo
  {journal} {Rev. Mod. Phys.}\ }\textbf {\bibinfo {volume} {87}},\ \bibinfo
  {pages} {137} (\bibinfo {year} {2015})}\BibitemShut {NoStop}%
\bibitem [{\citenamefont {Schnyder}\ \emph {et~al.}(2008)\citenamefont
  {Schnyder}, \citenamefont {Ryu}, \citenamefont {Furusaki},\ and\
  \citenamefont {Ludwig}}]{PhysRevB.78.195125}%
  \BibitemOpen
  \bibfield  {author} {\bibinfo {author} {\bibfnamefont {A.~P.}\ \bibnamefont
  {Schnyder}}, \bibinfo {author} {\bibfnamefont {S.}~\bibnamefont {Ryu}},
  \bibinfo {author} {\bibfnamefont {A.}~\bibnamefont {Furusaki}},\ and\
  \bibinfo {author} {\bibfnamefont {A.~W.~W.}\ \bibnamefont {Ludwig}},\
  }\bibfield  {title} {\bibinfo {title} {{Classification of topological
  insulators and superconductors in three spatial dimensions}},\ }\href
  {https://doi.org/10.1103/PhysRevB.78.195125} {\bibfield  {journal} {\bibinfo
  {journal} {Phys. Rev. B}\ }\textbf {\bibinfo {volume} {78}},\ \bibinfo
  {pages} {195125} (\bibinfo {year} {2008})}\BibitemShut {NoStop}%
\bibitem [{\citenamefont {Kitaev}(2009)}]{Kitaev_bott}%
  \BibitemOpen
  \bibfield  {author} {\bibinfo {author} {\bibfnamefont {A.}~\bibnamefont
  {Kitaev}},\ }\bibfield  {title} {\bibinfo {title} {Periodic table for
  topological insulators and superconductors},\ }\href
  {https://aip.scitation.org/doi/abs/10.1063/1.3149495} {\bibfield  {journal}
  {\bibinfo  {journal} {AIP Conference Proceedings}\ }\textbf {\bibinfo
  {volume} {1134}},\ \bibinfo {pages} {22} (\bibinfo {year}
  {2009})}\BibitemShut {NoStop}%
\bibitem [{\citenamefont {Ryu}\ \emph {et~al.}(2010)\citenamefont {Ryu},
  \citenamefont {Schnyder}, \citenamefont {Furusaki},\ and\ \citenamefont
  {Ludwig}}]{Ryu_2010}%
  \BibitemOpen
  \bibfield  {author} {\bibinfo {author} {\bibfnamefont {S.}~\bibnamefont
  {Ryu}}, \bibinfo {author} {\bibfnamefont {A.~P.}\ \bibnamefont {Schnyder}},
  \bibinfo {author} {\bibfnamefont {A.}~\bibnamefont {Furusaki}},\ and\
  \bibinfo {author} {\bibfnamefont {A.~W.~W.}\ \bibnamefont {Ludwig}},\
  }\bibfield  {title} {\bibinfo {title} {Topological insulators and
  superconductors: tenfold way and dimensional hierarchy},\ }\href
  {https://doi.org/10.1088/1367-2630/12/6/065010} {\bibfield  {journal}
  {\bibinfo  {journal} {New Journal of Physics}\ }\textbf {\bibinfo {volume}
  {12}},\ \bibinfo {pages} {065010} (\bibinfo {year} {2010})}\BibitemShut
  {NoStop}%
\bibitem [{\citenamefont {Sigrist}\ and\ \citenamefont
  {Ueda}(1991)}]{Sigrist-Ueda}%
  \BibitemOpen
  \bibfield  {author} {\bibinfo {author} {\bibfnamefont {M.}~\bibnamefont
  {Sigrist}}\ and\ \bibinfo {author} {\bibfnamefont {K.}~\bibnamefont {Ueda}},\
  }\bibfield  {title} {\bibinfo {title} {Phenomenological theory of
  unconventional superconductivity},\ }\href
  {https://doi.org/10.1103/RevModPhys.63.239} {\bibfield  {journal} {\bibinfo
  {journal} {Rev. Mod. Phys.}\ }\textbf {\bibinfo {volume} {63}},\ \bibinfo
  {pages} {239} (\bibinfo {year} {1991})}\BibitemShut {NoStop}%
\bibitem [{\citenamefont {Fischer}\ \emph {et~al.}(2014)\citenamefont
  {Fischer}, \citenamefont {Neupert}, \citenamefont {Platt}, \citenamefont
  {Schnyder}, \citenamefont {Hanke}, \citenamefont {Goryo}, \citenamefont
  {Thomale},\ and\ \citenamefont {Sigrist}}]{PhysRevB.89.020509}%
  \BibitemOpen
  \bibfield  {author} {\bibinfo {author} {\bibfnamefont {M.~H.}\ \bibnamefont
  {Fischer}}, \bibinfo {author} {\bibfnamefont {T.}~\bibnamefont {Neupert}},
  \bibinfo {author} {\bibfnamefont {C.}~\bibnamefont {Platt}}, \bibinfo
  {author} {\bibfnamefont {A.~P.}\ \bibnamefont {Schnyder}}, \bibinfo {author}
  {\bibfnamefont {W.}~\bibnamefont {Hanke}}, \bibinfo {author} {\bibfnamefont
  {J.}~\bibnamefont {Goryo}}, \bibinfo {author} {\bibfnamefont
  {R.}~\bibnamefont {Thomale}},\ and\ \bibinfo {author} {\bibfnamefont
  {M.}~\bibnamefont {Sigrist}},\ }\bibfield  {title} {\bibinfo {title} {{Chiral
  $d$-wave superconductivity in SrPtAs}},\ }\href
  {https://doi.org/10.1103/PhysRevB.89.020509} {\bibfield  {journal} {\bibinfo
  {journal} {Phys. Rev. B}\ }\textbf {\bibinfo {volume} {89}},\ \bibinfo
  {pages} {020509} (\bibinfo {year} {2014})}\BibitemShut {NoStop}%
\bibitem [{\citenamefont {Fu}(2014)}]{PhysRevB.90.100509}%
  \BibitemOpen
  \bibfield  {author} {\bibinfo {author} {\bibfnamefont {L.}~\bibnamefont
  {Fu}},\ }\bibfield  {title} {\bibinfo {title} {{Odd-parity topological
  superconductor with nematic order: Application to
  {C}u$_{x}${B}i$_{2}${S}e$_{3}$}},\ }\href
  {https://doi.org/10.1103/PhysRevB.90.100509} {\bibfield  {journal} {\bibinfo
  {journal} {Phys. Rev. B}\ }\textbf {\bibinfo {volume} {90}},\ \bibinfo
  {pages} {100509(R)} (\bibinfo {year} {2014})}\BibitemShut {NoStop}%
\bibitem [{\citenamefont {Sasaki}\ \emph {et~al.}(2011)\citenamefont {Sasaki},
  \citenamefont {Kriener}, \citenamefont {Segawa}, \citenamefont {Yada},
  \citenamefont {Tanaka}, \citenamefont {Sato},\ and\ \citenamefont
  {Ando}}]{PhysRevLett.107.217001}%
  \BibitemOpen
  \bibfield  {author} {\bibinfo {author} {\bibfnamefont {S.}~\bibnamefont
  {Sasaki}}, \bibinfo {author} {\bibfnamefont {M.}~\bibnamefont {Kriener}},
  \bibinfo {author} {\bibfnamefont {K.}~\bibnamefont {Segawa}}, \bibinfo
  {author} {\bibfnamefont {K.}~\bibnamefont {Yada}}, \bibinfo {author}
  {\bibfnamefont {Y.}~\bibnamefont {Tanaka}}, \bibinfo {author} {\bibfnamefont
  {M.}~\bibnamefont {Sato}},\ and\ \bibinfo {author} {\bibfnamefont
  {Y.}~\bibnamefont {Ando}},\ }\bibfield  {title} {\bibinfo {title}
  {{Topological Superconductivity in
  ${\mathrm{Cu}}_{x}{\mathrm{Bi}}_{2}{\mathrm{Se}}_{3}$}},\ }\href
  {https://doi.org/10.1103/PhysRevLett.107.217001} {\bibfield  {journal}
  {\bibinfo  {journal} {Phys. Rev. Lett.}\ }\textbf {\bibinfo {volume} {107}},\
  \bibinfo {pages} {217001} (\bibinfo {year} {2011})}\BibitemShut {NoStop}%
\bibitem [{\citenamefont {Yanase}(2016)}]{PhysRevB.94.174502}%
  \BibitemOpen
  \bibfield  {author} {\bibinfo {author} {\bibfnamefont {Y.}~\bibnamefont
  {Yanase}},\ }\bibfield  {title} {\bibinfo {title} {{Nonsymmorphic Weyl
  superconductivity in ${\mathrm{UPt}}_{3}$ based on ${E}_{2u}$
  representation}},\ }\href {https://doi.org/10.1103/PhysRevB.94.174502}
  {\bibfield  {journal} {\bibinfo  {journal} {Phys. Rev. B}\ }\textbf {\bibinfo
  {volume} {94}},\ \bibinfo {pages} {174502} (\bibinfo {year}
  {2016})}\BibitemShut {NoStop}%
\bibitem [{\citenamefont {Yanase}\ and\ \citenamefont
  {Shiozaki}(2017)}]{PhysRevB.95.224514}%
  \BibitemOpen
  \bibfield  {author} {\bibinfo {author} {\bibfnamefont {Y.}~\bibnamefont
  {Yanase}}\ and\ \bibinfo {author} {\bibfnamefont {K.}~\bibnamefont
  {Shiozaki}},\ }\bibfield  {title} {\bibinfo {title} {{M\"obius topological
  superconductivity in ${\mathrm{UPt}}_{3}$}},\ }\href
  {https://doi.org/10.1103/PhysRevB.95.224514} {\bibfield  {journal} {\bibinfo
  {journal} {Phys. Rev. B}\ }\textbf {\bibinfo {volume} {95}},\ \bibinfo
  {pages} {224514} (\bibinfo {year} {2017})}\BibitemShut {NoStop}%
\bibitem [{\citenamefont {Kawakami}\ \emph {et~al.}(2018)\citenamefont
  {Kawakami}, \citenamefont {Okamura}, \citenamefont {Kobayashi},\ and\
  \citenamefont {Sato}}]{PhysRevX.8.041026}%
  \BibitemOpen
  \bibfield  {author} {\bibinfo {author} {\bibfnamefont {T.}~\bibnamefont
  {Kawakami}}, \bibinfo {author} {\bibfnamefont {T.}~\bibnamefont {Okamura}},
  \bibinfo {author} {\bibfnamefont {S.}~\bibnamefont {Kobayashi}},\ and\
  \bibinfo {author} {\bibfnamefont {M.}~\bibnamefont {Sato}},\ }\bibfield
  {title} {\bibinfo {title} {{Topological Crystalline Materials of $J=3/2$
  Electrons: Antiperovskites, Dirac Points, and High Winding Topological
  Superconductivity}},\ }\href {https://doi.org/10.1103/PhysRevX.8.041026}
  {\bibfield  {journal} {\bibinfo  {journal} {Phys. Rev. X}\ }\textbf {\bibinfo
  {volume} {8}},\ \bibinfo {pages} {041026} (\bibinfo {year}
  {2018})}\BibitemShut {NoStop}%
\bibitem [{\citenamefont {Daido}\ \emph {et~al.}(2019)\citenamefont {Daido},
  \citenamefont {Yoshida},\ and\ \citenamefont
  {Yanase}}]{PhysRevLett.122.227001}%
  \BibitemOpen
  \bibfield  {author} {\bibinfo {author} {\bibfnamefont {A.}~\bibnamefont
  {Daido}}, \bibinfo {author} {\bibfnamefont {T.}~\bibnamefont {Yoshida}},\
  and\ \bibinfo {author} {\bibfnamefont {Y.}~\bibnamefont {Yanase}},\
  }\bibfield  {title} {\bibinfo {title} {{${\mathbb{Z}}_{4}$ Topological
  Superconductivity in UCoGe}},\ }\href
  {https://doi.org/10.1103/PhysRevLett.122.227001} {\bibfield  {journal}
  {\bibinfo  {journal} {Phys. Rev. Lett.}\ }\textbf {\bibinfo {volume} {122}},\
  \bibinfo {pages} {227001} (\bibinfo {year} {2019})}\BibitemShut {NoStop}%
\bibitem [{\citenamefont {Ishizuka}\ \emph {et~al.}(2019)\citenamefont
  {Ishizuka}, \citenamefont {Sumita}, \citenamefont {Daido},\ and\
  \citenamefont {Yanase}}]{PhysRevLett.123.217001}%
  \BibitemOpen
  \bibfield  {author} {\bibinfo {author} {\bibfnamefont {J.}~\bibnamefont
  {Ishizuka}}, \bibinfo {author} {\bibfnamefont {S.}~\bibnamefont {Sumita}},
  \bibinfo {author} {\bibfnamefont {A.}~\bibnamefont {Daido}},\ and\ \bibinfo
  {author} {\bibfnamefont {Y.}~\bibnamefont {Yanase}},\ }\bibfield  {title}
  {\bibinfo {title} {{Insulator-Metal Transition and Topological
  Superconductivity in ${\mathrm{UTe}}_{2}$ from a First-Principles
  Calculation}},\ }\href {https://doi.org/10.1103/PhysRevLett.123.217001}
  {\bibfield  {journal} {\bibinfo  {journal} {Phys. Rev. Lett.}\ }\textbf
  {\bibinfo {volume} {123}},\ \bibinfo {pages} {217001} (\bibinfo {year}
  {2019})}\BibitemShut {NoStop}%
\bibitem [{\citenamefont {Shang}\ \emph {et~al.}(2020)\citenamefont {Shang},
  \citenamefont {Smidman}, \citenamefont {Wang}, \citenamefont {Chang},
  \citenamefont {Baines}, \citenamefont {Lee}, \citenamefont {Nie},
  \citenamefont {Pang}, \citenamefont {Xie}, \citenamefont {Jiang},
  \citenamefont {Shi}, \citenamefont {Medarde}, \citenamefont {Shiroka},\ and\
  \citenamefont {Yuan}}]{PhysRevLett.124.207001}%
  \BibitemOpen
  \bibfield  {author} {\bibinfo {author} {\bibfnamefont {T.}~\bibnamefont
  {Shang}}, \bibinfo {author} {\bibfnamefont {M.}~\bibnamefont {Smidman}},
  \bibinfo {author} {\bibfnamefont {A.}~\bibnamefont {Wang}}, \bibinfo {author}
  {\bibfnamefont {L.-J.}\ \bibnamefont {Chang}}, \bibinfo {author}
  {\bibfnamefont {C.}~\bibnamefont {Baines}}, \bibinfo {author} {\bibfnamefont
  {M.~K.}\ \bibnamefont {Lee}}, \bibinfo {author} {\bibfnamefont {Z.~Y.}\
  \bibnamefont {Nie}}, \bibinfo {author} {\bibfnamefont {G.~M.}\ \bibnamefont
  {Pang}}, \bibinfo {author} {\bibfnamefont {W.}~\bibnamefont {Xie}}, \bibinfo
  {author} {\bibfnamefont {W.~B.}\ \bibnamefont {Jiang}}, \bibinfo {author}
  {\bibfnamefont {M.}~\bibnamefont {Shi}}, \bibinfo {author} {\bibfnamefont
  {M.}~\bibnamefont {Medarde}}, \bibinfo {author} {\bibfnamefont
  {T.}~\bibnamefont {Shiroka}},\ and\ \bibinfo {author} {\bibfnamefont {H.~Q.}\
  \bibnamefont {Yuan}},\ }\bibfield  {title} {\bibinfo {title} {{Simultaneous
  Nodal Superconductivity and Time-Reversal Symmetry Breaking in the
  Noncentrosymmetric Superconductor CaPtAs}},\ }\href
  {https://doi.org/10.1103/PhysRevLett.124.207001} {\bibfield  {journal}
  {\bibinfo  {journal} {Phys. Rev. Lett.}\ }\textbf {\bibinfo {volume} {124}},\
  \bibinfo {pages} {207001} (\bibinfo {year} {2020})}\BibitemShut {NoStop}%
\bibitem [{\citenamefont {Hsu}\ \emph {et~al.}(2020)\citenamefont {Hsu},
  \citenamefont {Cole}, \citenamefont {Zhang},\ and\ \citenamefont
  {Sau}}]{PhysRevLett.125.097001}%
  \BibitemOpen
  \bibfield  {author} {\bibinfo {author} {\bibfnamefont {Y.-T.}\ \bibnamefont
  {Hsu}}, \bibinfo {author} {\bibfnamefont {W.~S.}\ \bibnamefont {Cole}},
  \bibinfo {author} {\bibfnamefont {R.-X.}\ \bibnamefont {Zhang}},\ and\
  \bibinfo {author} {\bibfnamefont {J.~D.}\ \bibnamefont {Sau}},\ }\bibfield
  {title} {\bibinfo {title} {{Inversion-Protected Higher-Order Topological
  Superconductivity in Monolayer ${\mathrm{WTe}}_{2}$}},\ }\href
  {https://doi.org/10.1103/PhysRevLett.125.097001} {\bibfield  {journal}
  {\bibinfo  {journal} {Phys. Rev. Lett.}\ }\textbf {\bibinfo {volume} {125}},\
  \bibinfo {pages} {097001} (\bibinfo {year} {2020})}\BibitemShut {NoStop}%
\bibitem [{\citenamefont {Nogaki}\ \emph {et~al.}(2021)\citenamefont {Nogaki},
  \citenamefont {Daido}, \citenamefont {Ishizuka},\ and\ \citenamefont
  {Yanase}}]{PhysRevResearch.3.L032071}%
  \BibitemOpen
  \bibfield  {author} {\bibinfo {author} {\bibfnamefont {K.}~\bibnamefont
  {Nogaki}}, \bibinfo {author} {\bibfnamefont {A.}~\bibnamefont {Daido}},
  \bibinfo {author} {\bibfnamefont {J.}~\bibnamefont {Ishizuka}},\ and\
  \bibinfo {author} {\bibfnamefont {Y.}~\bibnamefont {Yanase}},\ }\bibfield
  {title} {\bibinfo {title} {{Topological crystalline superconductivity in
  locally noncentrosymmetric ${\mathrm{CeRh}}_{2}{\mathrm{As}}_{2}$}},\ }\href
  {https://doi.org/10.1103/PhysRevResearch.3.L032071} {\bibfield  {journal}
  {\bibinfo  {journal} {Phys. Rev. Research}\ }\textbf {\bibinfo {volume}
  {3}},\ \bibinfo {pages} {L032071} (\bibinfo {year} {2021})}\BibitemShut
  {NoStop}%
\bibitem [{\citenamefont {Tang}\ \emph {et~al.}(2021)\citenamefont {Tang},
  \citenamefont {Ono}, \citenamefont {Wan},\ and\ \citenamefont
  {Watanabe}}]{Tang-Ono-Wan-Watanabe2021}%
  \BibitemOpen
  \bibfield  {author} {\bibinfo {author} {\bibfnamefont {F.}~\bibnamefont
  {Tang}}, \bibinfo {author} {\bibfnamefont {S.}~\bibnamefont {Ono}}, \bibinfo
  {author} {\bibfnamefont {X.}~\bibnamefont {Wan}},\ and\ \bibinfo {author}
  {\bibfnamefont {H.}~\bibnamefont {Watanabe}},\ }\href@noop {} {\bibinfo
  {title} {{High-throughput Investigations of Topological and Nodal
  Superconductors}}} (\bibinfo {year} {2021}),\ \Eprint
  {https://arxiv.org/abs/2106.11985} {arXiv:2106.11985 [cond-mat.supr-con]}
  \BibitemShut {NoStop}%
\bibitem [{\citenamefont {Sato}(2010)}]{PhysRevB.81.220504}%
  \BibitemOpen
  \bibfield  {author} {\bibinfo {author} {\bibfnamefont {M.}~\bibnamefont
  {Sato}},\ }\bibfield  {title} {\bibinfo {title} {{Topological odd-parity
  superconductors}},\ }\href {https://doi.org/10.1103/PhysRevB.81.220504}
  {\bibfield  {journal} {\bibinfo  {journal} {Phys. Rev. B}\ }\textbf {\bibinfo
  {volume} {81}},\ \bibinfo {pages} {220504(R)} (\bibinfo {year}
  {2010})}\BibitemShut {NoStop}%
\bibitem [{\citenamefont {Fu}\ and\ \citenamefont
  {Berg}(2010)}]{PhysRevLett.105.097001}%
  \BibitemOpen
  \bibfield  {author} {\bibinfo {author} {\bibfnamefont {L.}~\bibnamefont
  {Fu}}\ and\ \bibinfo {author} {\bibfnamefont {E.}~\bibnamefont {Berg}},\
  }\bibfield  {title} {\bibinfo {title} {{Odd-Parity Topological
  Superconductors: Theory and Application to {C}u$_{x}${B}i$_{2}${S}e$_{3}$}},\
  }\href {https://doi.org/10.1103/PhysRevLett.105.097001} {\bibfield  {journal}
  {\bibinfo  {journal} {Phys. Rev. Lett.}\ }\textbf {\bibinfo {volume} {105}},\
  \bibinfo {pages} {097001} (\bibinfo {year} {2010})}\BibitemShut {NoStop}%
\bibitem [{\citenamefont {Sato}(2009)}]{PhysRevB.79.214526}%
  \BibitemOpen
  \bibfield  {author} {\bibinfo {author} {\bibfnamefont {M.}~\bibnamefont
  {Sato}},\ }\bibfield  {title} {\bibinfo {title} {{Topological properties of
  spin-triplet superconductors and Fermi surface topology in the normal
  state}},\ }\href {https://doi.org/10.1103/PhysRevB.79.214526} {\bibfield
  {journal} {\bibinfo  {journal} {Phys. Rev. B}\ }\textbf {\bibinfo {volume}
  {79}},\ \bibinfo {pages} {214526} (\bibinfo {year} {2009})}\BibitemShut
  {NoStop}%
\bibitem [{\citenamefont {Ueno}\ \emph {et~al.}(2013)\citenamefont {Ueno},
  \citenamefont {Yamakage}, \citenamefont {Tanaka},\ and\ \citenamefont
  {Sato}}]{PhysRevLett.111.087002}%
  \BibitemOpen
  \bibfield  {author} {\bibinfo {author} {\bibfnamefont {Y.}~\bibnamefont
  {Ueno}}, \bibinfo {author} {\bibfnamefont {A.}~\bibnamefont {Yamakage}},
  \bibinfo {author} {\bibfnamefont {Y.}~\bibnamefont {Tanaka}},\ and\ \bibinfo
  {author} {\bibfnamefont {M.}~\bibnamefont {Sato}},\ }\bibfield  {title}
  {\bibinfo {title} {{Symmetry-Protected Majorana Fermions in Topological
  Crystalline Superconductors: Theory and Application to
  ${\mathrm{Sr}}_{2}{\mathrm{RuO}}_{4}$}},\ }\href
  {https://doi.org/10.1103/PhysRevLett.111.087002} {\bibfield  {journal}
  {\bibinfo  {journal} {Phys. Rev. Lett.}\ }\textbf {\bibinfo {volume} {111}},\
  \bibinfo {pages} {087002} (\bibinfo {year} {2013})}\BibitemShut {NoStop}%
\bibitem [{\citenamefont {Shiozaki}\ and\ \citenamefont
  {Sato}(2014)}]{PhysRevB.90.165114}%
  \BibitemOpen
  \bibfield  {author} {\bibinfo {author} {\bibfnamefont {K.}~\bibnamefont
  {Shiozaki}}\ and\ \bibinfo {author} {\bibfnamefont {M.}~\bibnamefont
  {Sato}},\ }\bibfield  {title} {\bibinfo {title} {{Topology of crystalline
  insulators and superconductors}},\ }\href
  {https://doi.org/10.1103/PhysRevB.90.165114} {\bibfield  {journal} {\bibinfo
  {journal} {Phys. Rev. B}\ }\textbf {\bibinfo {volume} {90}},\ \bibinfo
  {pages} {165114} (\bibinfo {year} {2014})}\BibitemShut {NoStop}%
\bibitem [{\citenamefont {Ando}\ and\ \citenamefont {Fu}(2015)}]{Ando-Fu}%
  \BibitemOpen
  \bibfield  {author} {\bibinfo {author} {\bibfnamefont {Y.}~\bibnamefont
  {Ando}}\ and\ \bibinfo {author} {\bibfnamefont {L.}~\bibnamefont {Fu}},\
  }\bibfield  {title} {\bibinfo {title} {{Topological Crystalline Insulators
  and Topological Superconductors: From Concepts to Materials}},\ }\href
  {https://doi.org/10.1146/annurev-conmatphys-031214-014501} {\bibfield
  {journal} {\bibinfo  {journal} {Annual Review of Condensed Matter Physics}\
  }\textbf {\bibinfo {volume} {6}},\ \bibinfo {pages} {361} (\bibinfo {year}
  {2015})}\BibitemShut {NoStop}%
\bibitem [{\citenamefont {Sato}\ and\ \citenamefont
  {Fujimoto}(2016)}]{SatoFujimoto}%
  \BibitemOpen
  \bibfield  {author} {\bibinfo {author} {\bibfnamefont {M.}~\bibnamefont
  {Sato}}\ and\ \bibinfo {author} {\bibfnamefont {S.}~\bibnamefont
  {Fujimoto}},\ }\bibfield  {title} {\bibinfo {title} {{Majorana Fermions and
  Topology in Superconductors}},\ }\href
  {https://doi.org/10.7566/JPSJ.85.072001} {\bibfield  {journal} {\bibinfo
  {journal} {J. Phys. Soc. Jpn.}\ }\textbf {\bibinfo {volume} {85}},\ \bibinfo
  {pages} {072001} (\bibinfo {year} {2016})}\BibitemShut {NoStop}%
\bibitem [{\citenamefont {Sato}\ and\ \citenamefont {Ando}(2017)}]{Sato_2017}%
  \BibitemOpen
  \bibfield  {author} {\bibinfo {author} {\bibfnamefont {M.}~\bibnamefont
  {Sato}}\ and\ \bibinfo {author} {\bibfnamefont {Y.}~\bibnamefont {Ando}},\
  }\bibfield  {title} {\bibinfo {title} {{Topological superconductors: a
  review}},\ }\href {https://doi.org/10.1088/1361-6633/aa6ac7} {\bibfield
  {journal} {\bibinfo  {journal} {Reports on Progress in Physics}\ }\textbf
  {\bibinfo {volume} {80}},\ \bibinfo {pages} {076501} (\bibinfo {year}
  {2017})}\BibitemShut {NoStop}%
\bibitem [{\citenamefont {Shiozaki}\ \emph {et~al.}(2017)\citenamefont
  {Shiozaki}, \citenamefont {Sato},\ and\ \citenamefont
  {Gomi}}]{Shiozaki-Sato-Gomi2017}%
  \BibitemOpen
  \bibfield  {author} {\bibinfo {author} {\bibfnamefont {K.}~\bibnamefont
  {Shiozaki}}, \bibinfo {author} {\bibfnamefont {M.}~\bibnamefont {Sato}},\
  and\ \bibinfo {author} {\bibfnamefont {K.}~\bibnamefont {Gomi}},\ }\bibfield
  {title} {\bibinfo {title} {Topological crystalline materials: General
  formulation, module structure, and wallpaper groups},\ }\href
  {https://doi.org/10.1103/PhysRevB.95.235425} {\bibfield  {journal} {\bibinfo
  {journal} {Phys. Rev. B}\ }\textbf {\bibinfo {volume} {95}},\ \bibinfo
  {pages} {235425} (\bibinfo {year} {2017})}\BibitemShut {NoStop}%
\bibitem [{\citenamefont {Langbehn}\ \emph {et~al.}(2017)\citenamefont
  {Langbehn}, \citenamefont {Peng}, \citenamefont {Trifunovic}, \citenamefont
  {von Oppen},\ and\ \citenamefont {Brouwer}}]{PhysRevLett.119.246401}%
  \BibitemOpen
  \bibfield  {author} {\bibinfo {author} {\bibfnamefont {J.}~\bibnamefont
  {Langbehn}}, \bibinfo {author} {\bibfnamefont {Y.}~\bibnamefont {Peng}},
  \bibinfo {author} {\bibfnamefont {L.}~\bibnamefont {Trifunovic}}, \bibinfo
  {author} {\bibfnamefont {F.}~\bibnamefont {von Oppen}},\ and\ \bibinfo
  {author} {\bibfnamefont {P.~W.}\ \bibnamefont {Brouwer}},\ }\bibfield
  {title} {\bibinfo {title} {{Reflection-Symmetric Second-Order Topological
  Insulators and Superconductors}},\ }\href
  {https://doi.org/10.1103/PhysRevLett.119.246401} {\bibfield  {journal}
  {\bibinfo  {journal} {Phys. Rev. Lett.}\ }\textbf {\bibinfo {volume} {119}},\
  \bibinfo {pages} {246401} (\bibinfo {year} {2017})}\BibitemShut {NoStop}%
\bibitem [{\citenamefont {Khalaf}(2018)}]{PhysRevB.97.205136}%
  \BibitemOpen
  \bibfield  {author} {\bibinfo {author} {\bibfnamefont {E.}~\bibnamefont
  {Khalaf}},\ }\bibfield  {title} {\bibinfo {title} {{Higher-order topological
  insulators and superconductors protected by inversion symmetry}},\ }\href
  {https://doi.org/10.1103/PhysRevB.97.205136} {\bibfield  {journal} {\bibinfo
  {journal} {Phys. Rev. B}\ }\textbf {\bibinfo {volume} {97}},\ \bibinfo
  {pages} {205136} (\bibinfo {year} {2018})}\BibitemShut {NoStop}%
\bibitem [{\citenamefont {Geier}\ \emph {et~al.}(2018)\citenamefont {Geier},
  \citenamefont {Trifunovic}, \citenamefont {Hoskam},\ and\ \citenamefont
  {Brouwer}}]{PhysRevB.97.205135}%
  \BibitemOpen
  \bibfield  {author} {\bibinfo {author} {\bibfnamefont {M.}~\bibnamefont
  {Geier}}, \bibinfo {author} {\bibfnamefont {L.}~\bibnamefont {Trifunovic}},
  \bibinfo {author} {\bibfnamefont {M.}~\bibnamefont {Hoskam}},\ and\ \bibinfo
  {author} {\bibfnamefont {P.~W.}\ \bibnamefont {Brouwer}},\ }\bibfield
  {title} {\bibinfo {title} {Second-order topological insulators and
  superconductors with an order-two crystalline symmetry},\ }\href
  {https://doi.org/10.1103/PhysRevB.97.205135} {\bibfield  {journal} {\bibinfo
  {journal} {Phys. Rev. B}\ }\textbf {\bibinfo {volume} {97}},\ \bibinfo
  {pages} {205135} (\bibinfo {year} {2018})}\BibitemShut {NoStop}%
\bibitem [{\citenamefont {Ono}\ and\ \citenamefont
  {Watanabe}(2018)}]{Ono-Watanabe2018}%
  \BibitemOpen
  \bibfield  {author} {\bibinfo {author} {\bibfnamefont {S.}~\bibnamefont
  {Ono}}\ and\ \bibinfo {author} {\bibfnamefont {H.}~\bibnamefont {Watanabe}},\
  }\bibfield  {title} {\bibinfo {title} {{Unified understanding of symmetry
  indicators for all internal symmetry classes}},\ }\href
  {https://doi.org/10.1103/PhysRevB.98.115150} {\bibfield  {journal} {\bibinfo
  {journal} {Phys. Rev. B}\ }\textbf {\bibinfo {volume} {98}},\ \bibinfo
  {pages} {115150} (\bibinfo {year} {2018})}\BibitemShut {NoStop}%
\bibitem [{\citenamefont {Trifunovic}\ and\ \citenamefont
  {Brouwer}(2019)}]{PhysRevX.9.011012}%
  \BibitemOpen
  \bibfield  {author} {\bibinfo {author} {\bibfnamefont {L.}~\bibnamefont
  {Trifunovic}}\ and\ \bibinfo {author} {\bibfnamefont {P.~W.}\ \bibnamefont
  {Brouwer}},\ }\bibfield  {title} {\bibinfo {title} {Higher-order
  bulk-boundary correspondence for topological crystalline phases},\ }\href
  {https://doi.org/10.1103/PhysRevX.9.011012} {\bibfield  {journal} {\bibinfo
  {journal} {Phys. Rev. X}\ }\textbf {\bibinfo {volume} {9}},\ \bibinfo {pages}
  {011012} (\bibinfo {year} {2019})}\BibitemShut {NoStop}%
\bibitem [{\citenamefont {Cornfeld}\ and\ \citenamefont
  {Chapman}(2019)}]{Cornfeld-Chapman}%
  \BibitemOpen
  \bibfield  {author} {\bibinfo {author} {\bibfnamefont {E.}~\bibnamefont
  {Cornfeld}}\ and\ \bibinfo {author} {\bibfnamefont {A.}~\bibnamefont
  {Chapman}},\ }\bibfield  {title} {\bibinfo {title} {Classification of
  crystalline topological insulators and superconductors with point group
  symmetries},\ }\href {https://doi.org/10.1103/PhysRevB.99.075105} {\bibfield
  {journal} {\bibinfo  {journal} {Phys. Rev. B}\ }\textbf {\bibinfo {volume}
  {99}},\ \bibinfo {pages} {075105} (\bibinfo {year} {2019})}\BibitemShut
  {NoStop}%
\bibitem [{\citenamefont {Ono}\ \emph {et~al.}(2019)\citenamefont {Ono},
  \citenamefont {Yanase},\ and\ \citenamefont
  {Watanabe}}]{Ono-Yanase-Watanabe2019}%
  \BibitemOpen
  \bibfield  {author} {\bibinfo {author} {\bibfnamefont {S.}~\bibnamefont
  {Ono}}, \bibinfo {author} {\bibfnamefont {Y.}~\bibnamefont {Yanase}},\ and\
  \bibinfo {author} {\bibfnamefont {H.}~\bibnamefont {Watanabe}},\ }\bibfield
  {title} {\bibinfo {title} {{Symmetry indicators for topological
  superconductors}},\ }\href {https://doi.org/10.1103/PhysRevResearch.1.013012}
  {\bibfield  {journal} {\bibinfo  {journal} {Phys. Rev. Res.}\ }\textbf
  {\bibinfo {volume} {1}},\ \bibinfo {pages} {013012} (\bibinfo {year}
  {2019})}\BibitemShut {NoStop}%
\bibitem [{\citenamefont {Skurativska}\ \emph {et~al.}(2020)\citenamefont
  {Skurativska}, \citenamefont {Neupert},\ and\ \citenamefont
  {Fischer}}]{Skurativska2019}%
  \BibitemOpen
  \bibfield  {author} {\bibinfo {author} {\bibfnamefont {A.}~\bibnamefont
  {Skurativska}}, \bibinfo {author} {\bibfnamefont {T.}~\bibnamefont
  {Neupert}},\ and\ \bibinfo {author} {\bibfnamefont {M.~H.}\ \bibnamefont
  {Fischer}},\ }\bibfield  {title} {\bibinfo {title} {Atomic limit and
  inversion-symmetry indicators for topological superconductors},\ }\href
  {https://doi.org/10.1103/PhysRevResearch.2.013064} {\bibfield  {journal}
  {\bibinfo  {journal} {Phys. Rev. Research}\ }\textbf {\bibinfo {volume}
  {2}},\ \bibinfo {pages} {013064} (\bibinfo {year} {2020})}\BibitemShut
  {NoStop}%
\bibitem [{\citenamefont {Ono}\ \emph {et~al.}(2020)\citenamefont {Ono},
  \citenamefont {Po},\ and\ \citenamefont {Watanabe}}]{Ono-Po-Watanabe2020}%
  \BibitemOpen
  \bibfield  {author} {\bibinfo {author} {\bibfnamefont {S.}~\bibnamefont
  {Ono}}, \bibinfo {author} {\bibfnamefont {H.~C.}\ \bibnamefont {Po}},\ and\
  \bibinfo {author} {\bibfnamefont {H.}~\bibnamefont {Watanabe}},\ }\bibfield
  {title} {\bibinfo {title} {{Refined symmetry indicators for topological
  superconductors in all space groups}},\ }\href
  {https://advances.sciencemag.org/content/6/18/eaaz8367} {\bibfield  {journal}
  {\bibinfo  {journal} {Science Advances}\ }\textbf {\bibinfo {volume} {6}},\
  \bibinfo {pages} {eaaz8367} (\bibinfo {year} {2020})}\BibitemShut {NoStop}%
\bibitem [{\citenamefont {{Ahn}}\ and\ \citenamefont {{Yang}}(2020)}]{Ahn2019}%
  \BibitemOpen
  \bibfield  {author} {\bibinfo {author} {\bibfnamefont {J.}~\bibnamefont
  {{Ahn}}}\ and\ \bibinfo {author} {\bibfnamefont {B.-J.}\ \bibnamefont
  {{Yang}}},\ }\bibfield  {title} {\bibinfo {title} {{Higher-order topological
  superconductivity of spin-polarized fermions}},\ }\href
  {https://doi.org/10.1103/PhysRevResearch.2.012060} {\bibfield  {journal}
  {\bibinfo  {journal} {Phys. Rev. Research}\ }\textbf {\bibinfo {volume}
  {2}},\ \bibinfo {pages} {012060} (\bibinfo {year} {2020})}\BibitemShut
  {NoStop}%
\bibitem [{\citenamefont {Geier}\ \emph {et~al.}(2020)\citenamefont {Geier},
  \citenamefont {Brouwer},\ and\ \citenamefont {Trifunovic}}]{SI_Luka}%
  \BibitemOpen
  \bibfield  {author} {\bibinfo {author} {\bibfnamefont {M.}~\bibnamefont
  {Geier}}, \bibinfo {author} {\bibfnamefont {P.~W.}\ \bibnamefont {Brouwer}},\
  and\ \bibinfo {author} {\bibfnamefont {L.}~\bibnamefont {Trifunovic}},\
  }\bibfield  {title} {\bibinfo {title} {{Symmetry-based indicators for
  topological Bogoliubov--de Gennes Hamiltonians}},\ }\href
  {https://doi.org/10.1103/PhysRevB.101.245128} {\bibfield  {journal} {\bibinfo
   {journal} {Phys. Rev. B}\ }\textbf {\bibinfo {volume} {101}},\ \bibinfo
  {pages} {245128} (\bibinfo {year} {2020})}\BibitemShut {NoStop}%
\bibitem [{\citenamefont {Tiwari}\ \emph {et~al.}(2020)\citenamefont {Tiwari},
  \citenamefont {Jahin},\ and\ \citenamefont
  {Wang}}]{PhysRevResearch.2.043300}%
  \BibitemOpen
  \bibfield  {author} {\bibinfo {author} {\bibfnamefont {A.}~\bibnamefont
  {Tiwari}}, \bibinfo {author} {\bibfnamefont {A.}~\bibnamefont {Jahin}},\ and\
  \bibinfo {author} {\bibfnamefont {Y.}~\bibnamefont {Wang}},\ }\bibfield
  {title} {\bibinfo {title} {Chiral dirac superconductors: Second-order and
  boundary-obstructed topology},\ }\href
  {https://doi.org/10.1103/PhysRevResearch.2.043300} {\bibfield  {journal}
  {\bibinfo  {journal} {Phys. Rev. Research}\ }\textbf {\bibinfo {volume}
  {2}},\ \bibinfo {pages} {043300} (\bibinfo {year} {2020})}\BibitemShut
  {NoStop}%
\bibitem [{\citenamefont {Ono}\ \emph {et~al.}(2021)\citenamefont {Ono},
  \citenamefont {Po},\ and\ \citenamefont {Shiozaki}}]{Ono-Po-Shiozaki2020}%
  \BibitemOpen
  \bibfield  {author} {\bibinfo {author} {\bibfnamefont {S.}~\bibnamefont
  {Ono}}, \bibinfo {author} {\bibfnamefont {H.~C.}\ \bibnamefont {Po}},\ and\
  \bibinfo {author} {\bibfnamefont {K.}~\bibnamefont {Shiozaki}},\ }\bibfield
  {title} {\bibinfo {title} {{$\mathbb{Z}_2$-enriched symmetry indicators for
  topological superconductors in the 1651 magnetic space groups}},\ }\href
  {https://link.aps.org/doi/10.1103/PhysRevResearch.3.023086} {\bibfield
  {journal} {\bibinfo  {journal} {Phys. Rev. Res.}\ }\textbf {\bibinfo {volume}
  {3}},\ \bibinfo {pages} {023086} (\bibinfo {year} {2021})}\BibitemShut
  {NoStop}%
\bibitem [{\citenamefont {Huang}\ and\ \citenamefont
  {Hsu}(2021)}]{huang2020faithful}%
  \BibitemOpen
  \bibfield  {author} {\bibinfo {author} {\bibfnamefont {S.-J.}\ \bibnamefont
  {Huang}}\ and\ \bibinfo {author} {\bibfnamefont {Y.-T.}\ \bibnamefont
  {Hsu}},\ }\bibfield  {title} {\bibinfo {title} {{Faithful derivation of
  symmetry indicators: A case study for topological superconductors with
  time-reversal and inversion symmetries}},\ }\href
  {https://doi.org/10.1103/PhysRevResearch.3.013243} {\bibfield  {journal}
  {\bibinfo  {journal} {Phys. Rev. Research}\ }\textbf {\bibinfo {volume}
  {3}},\ \bibinfo {pages} {013243} (\bibinfo {year} {2021})}\BibitemShut
  {NoStop}%
\bibitem [{\citenamefont {Cornfeld}\ and\ \citenamefont
  {Carmeli}(2021)}]{PhysRevResearch.3.013052}%
  \BibitemOpen
  \bibfield  {author} {\bibinfo {author} {\bibfnamefont {E.}~\bibnamefont
  {Cornfeld}}\ and\ \bibinfo {author} {\bibfnamefont {S.}~\bibnamefont
  {Carmeli}},\ }\bibfield  {title} {\bibinfo {title} {Tenfold topology of
  crystals: Unified classification of crystalline topological insulators and
  superconductors},\ }\href {https://doi.org/10.1103/PhysRevResearch.3.013052}
  {\bibfield  {journal} {\bibinfo  {journal} {Phys. Rev. Research}\ }\textbf
  {\bibinfo {volume} {3}},\ \bibinfo {pages} {013052} (\bibinfo {year}
  {2021})}\BibitemShut {NoStop}%
\bibitem [{\citenamefont {Ono}\ and\ \citenamefont
  {Shiozaki}(2022)}]{Ono-Shiozaki2022}%
  \BibitemOpen
  \bibfield  {author} {\bibinfo {author} {\bibfnamefont {S.}~\bibnamefont
  {Ono}}\ and\ \bibinfo {author} {\bibfnamefont {K.}~\bibnamefont {Shiozaki}},\
  }\bibfield  {title} {\bibinfo {title} {{Symmetry-Based Approach to
  Superconducting Nodes: Unification of Compatibility Conditions and Gapless
  Point Classifications}},\ }\href {https://doi.org/10.1103/PhysRevX.12.011021}
  {\bibfield  {journal} {\bibinfo  {journal} {Phys. Rev. X}\ }\textbf {\bibinfo
  {volume} {12}},\ \bibinfo {pages} {011021} (\bibinfo {year}
  {2022})}\BibitemShut {NoStop}%
\bibitem [{\citenamefont {Chen}\ \emph {et~al.}(2022)\citenamefont {Chen},
  \citenamefont {Huang}, \citenamefont {Hsu},\ and\ \citenamefont
  {Wei}}]{PhysRevB.105.094518}%
  \BibitemOpen
  \bibfield  {author} {\bibinfo {author} {\bibfnamefont {Y.}~\bibnamefont
  {Chen}}, \bibinfo {author} {\bibfnamefont {S.-J.}\ \bibnamefont {Huang}},
  \bibinfo {author} {\bibfnamefont {Y.-T.}\ \bibnamefont {Hsu}},\ and\ \bibinfo
  {author} {\bibfnamefont {T.-C.}\ \bibnamefont {Wei}},\ }\bibfield  {title}
  {\bibinfo {title} {Topological invariants beyond symmetry indicators:
  Boundary diagnostics for twofold rotationally symmetric superconductors},\
  }\href {https://doi.org/10.1103/PhysRevB.105.094518} {\bibfield  {journal}
  {\bibinfo  {journal} {Phys. Rev. B}\ }\textbf {\bibinfo {volume} {105}},\
  \bibinfo {pages} {094518} (\bibinfo {year} {2022})}\BibitemShut {NoStop}%
\bibitem [{\citenamefont {Scammell}\ \emph {et~al.}(2021)\citenamefont
  {Scammell}, \citenamefont {Ingham}, \citenamefont {Geier},\ and\
  \citenamefont {Li}}]{2111.07252}%
  \BibitemOpen
  \bibfield  {author} {\bibinfo {author} {\bibfnamefont {H.~D.}\ \bibnamefont
  {Scammell}}, \bibinfo {author} {\bibfnamefont {J.}~\bibnamefont {Ingham}},
  \bibinfo {author} {\bibfnamefont {M.}~\bibnamefont {Geier}},\ and\ \bibinfo
  {author} {\bibfnamefont {T.}~\bibnamefont {Li}},\ }\href
  {https://arxiv.org/abs/2111.07252} {\bibinfo {title} {{Intrinsic first and
  higher-order topological superconductivity in a doped topological
  insulator}}} (\bibinfo {year} {2021})\BibitemShut {NoStop}%
\bibitem [{\citenamefont {Le}\ \emph {et~al.}(2021)\citenamefont {Le},
  \citenamefont {Yang}, \citenamefont {Cui}, \citenamefont {Schnyder},\ and\
  \citenamefont {Chiu}}]{2108.04534}%
  \BibitemOpen
  \bibfield  {author} {\bibinfo {author} {\bibfnamefont {C.}~\bibnamefont
  {Le}}, \bibinfo {author} {\bibfnamefont {Z.}~\bibnamefont {Yang}}, \bibinfo
  {author} {\bibfnamefont {F.}~\bibnamefont {Cui}}, \bibinfo {author}
  {\bibfnamefont {A.~P.}\ \bibnamefont {Schnyder}},\ and\ \bibinfo {author}
  {\bibfnamefont {C.-K.}\ \bibnamefont {Chiu}},\ }\href
  {https://arxiv.org/abs/2108.04534} {\bibinfo {title} {{Generalized Fermion
  Doubling Theorems: Classification of 2D Nodal Systems in Terms of Wallpaper
  Groups}}} (\bibinfo {year} {2021})\BibitemShut {NoStop}%
\bibitem [{\citenamefont {Simon}\ \emph {et~al.}(2021)\citenamefont {Simon},
  \citenamefont {Geier},\ and\ \citenamefont {Brouwer}}]{2109.02664}%
  \BibitemOpen
  \bibfield  {author} {\bibinfo {author} {\bibfnamefont {S.}~\bibnamefont
  {Simon}}, \bibinfo {author} {\bibfnamefont {M.}~\bibnamefont {Geier}},\ and\
  \bibinfo {author} {\bibfnamefont {P.~W.}\ \bibnamefont {Brouwer}},\ }\href
  {https://arxiv.org/abs/2109.02664} {\bibinfo {title} {{Higher-order
  topological semimetals and nodal superconductors with an order-two
  crystalline symmetry}}} (\bibinfo {year} {2021})\BibitemShut {NoStop}%
\bibitem [{\citenamefont {Timm}\ and\ \citenamefont
  {Bhattacharya}(2021)}]{PhysRevB.104.094529}%
  \BibitemOpen
  \bibfield  {author} {\bibinfo {author} {\bibfnamefont {C.}~\bibnamefont
  {Timm}}\ and\ \bibinfo {author} {\bibfnamefont {A.}~\bibnamefont
  {Bhattacharya}},\ }\bibfield  {title} {\bibinfo {title} {Symmetry, nodal
  structure, and bogoliubov fermi surfaces for nonlocal pairing},\ }\href
  {https://doi.org/10.1103/PhysRevB.104.094529} {\bibfield  {journal} {\bibinfo
   {journal} {Phys. Rev. B}\ }\textbf {\bibinfo {volume} {104}},\ \bibinfo
  {pages} {094529} (\bibinfo {year} {2021})}\BibitemShut {NoStop}%
\bibitem [{\citenamefont {Lo}\ \emph {et~al.}(2022)\citenamefont {Lo},
  \citenamefont {Po},\ and\ \citenamefont {Nevidomskyy}}]{PhysRevB.105.104501}%
  \BibitemOpen
  \bibfield  {author} {\bibinfo {author} {\bibfnamefont {C.~F.~B.}\
  \bibnamefont {Lo}}, \bibinfo {author} {\bibfnamefont {H.~C.}\ \bibnamefont
  {Po}},\ and\ \bibinfo {author} {\bibfnamefont {A.~H.}\ \bibnamefont
  {Nevidomskyy}},\ }\bibfield  {title} {\bibinfo {title} {Inherited topological
  superconductivity in two-dimensional dirac semimetals},\ }\href
  {https://doi.org/10.1103/PhysRevB.105.104501} {\bibfield  {journal} {\bibinfo
   {journal} {Phys. Rev. B}\ }\textbf {\bibinfo {volume} {105}},\ \bibinfo
  {pages} {104501} (\bibinfo {year} {2022})}\BibitemShut {NoStop}%
\bibitem [{\citenamefont {Yu}\ \emph {et~al.}(2022)\citenamefont {Yu},
  \citenamefont {Chen},\ and\ \citenamefont {Das~Sarma}}]{PhysRevB.105.104515}%
  \BibitemOpen
  \bibfield  {author} {\bibinfo {author} {\bibfnamefont {J.}~\bibnamefont
  {Yu}}, \bibinfo {author} {\bibfnamefont {Y.-A.}\ \bibnamefont {Chen}},\ and\
  \bibinfo {author} {\bibfnamefont {S.}~\bibnamefont {Das~Sarma}},\ }\bibfield
  {title} {\bibinfo {title} {Euler-obstructed cooper pairing: Nodal
  superconductivity and hinge majorana zero modes},\ }\href
  {https://doi.org/10.1103/PhysRevB.105.104515} {\bibfield  {journal} {\bibinfo
   {journal} {Phys. Rev. B}\ }\textbf {\bibinfo {volume} {105}},\ \bibinfo
  {pages} {104515} (\bibinfo {year} {2022})}\BibitemShut {NoStop}%
\bibitem [{\citenamefont {Shiozaki}(2022)}]{CC-Shiozaki}%
  \BibitemOpen
  \bibfield  {author} {\bibinfo {author} {\bibfnamefont {K.}~\bibnamefont
  {Shiozaki}},\ }\bibfield  {title} {\bibinfo {title} {{The classification of
  surface states of topological insulators and superconductors with magnetic
  point group symmetry}},\ }\href {https://doi.org/10.1093/ptep/ptep026}
  {\bibfield  {journal} {\bibinfo  {journal} {Progress of Theoretical and
  Experimental Physics}\ }\textbf {\bibinfo {volume} {2022}} (\bibinfo {year}
  {2022})},\ \bibinfo {note} {04A104}\BibitemShut {NoStop}%
\bibitem [{\citenamefont {Zhang}\ \emph {et~al.}(2021)\citenamefont {Zhang},
  \citenamefont {Ren},\ and\ \citenamefont {Fang}}]{Surface_Fang}%
  \BibitemOpen
  \bibfield  {author} {\bibinfo {author} {\bibfnamefont {Z.}~\bibnamefont
  {Zhang}}, \bibinfo {author} {\bibfnamefont {J.}~\bibnamefont {Ren}},\ and\
  \bibinfo {author} {\bibfnamefont {C.}~\bibnamefont {Fang}},\ }\href@noop {}
  {\bibinfo {title} {{Classification of intrinsic topological superconductors
  jointly protected by time-reversal and point-group symmetries in three
  dimensions}}} (\bibinfo {year} {2021}),\ \Eprint
  {https://arxiv.org/abs/2109.14629} {arXiv:2109.14629 [cond-mat.str-el]}
  \BibitemShut {NoStop}%
\bibitem [{\citenamefont {Sato}\ and\ \citenamefont
  {Fujimoto}(2009)}]{PhysRevB.79.094504}%
  \BibitemOpen
  \bibfield  {author} {\bibinfo {author} {\bibfnamefont {M.}~\bibnamefont
  {Sato}}\ and\ \bibinfo {author} {\bibfnamefont {S.}~\bibnamefont
  {Fujimoto}},\ }\bibfield  {title} {\bibinfo {title} {{Topological phases of
  noncentrosymmetric superconductors: Edge states, Majorana fermions, and
  non-Abelian statistics}},\ }\href
  {https://doi.org/10.1103/PhysRevB.79.094504} {\bibfield  {journal} {\bibinfo
  {journal} {Phys. Rev. B}\ }\textbf {\bibinfo {volume} {79}},\ \bibinfo
  {pages} {094504} (\bibinfo {year} {2009})}\BibitemShut {NoStop}%
\bibitem [{\citenamefont {Qin}\ \emph {et~al.}(2022)\citenamefont {Qin},
  \citenamefont {Fang}, \citenamefont {Zhang},\ and\ \citenamefont
  {Hu}}]{Fe-based_TSC}%
  \BibitemOpen
  \bibfield  {author} {\bibinfo {author} {\bibfnamefont {S.}~\bibnamefont
  {Qin}}, \bibinfo {author} {\bibfnamefont {C.}~\bibnamefont {Fang}}, \bibinfo
  {author} {\bibfnamefont {F.-C.}\ \bibnamefont {Zhang}},\ and\ \bibinfo
  {author} {\bibfnamefont {J.}~\bibnamefont {Hu}},\ }\bibfield  {title}
  {\bibinfo {title} {{Topological Superconductivity in an Extended $s$-Wave
  Superconductor and Its Implication to Iron-Based Superconductors}},\ }\href
  {https://doi.org/10.1103/PhysRevX.12.011030} {\bibfield  {journal} {\bibinfo
  {journal} {Phys. Rev. X}\ }\textbf {\bibinfo {volume} {12}},\ \bibinfo
  {pages} {011030} (\bibinfo {year} {2022})}\BibitemShut {NoStop}%
\bibitem [{\citenamefont {Qi}\ \emph {et~al.}(2010)\citenamefont {Qi},
  \citenamefont {Hughes},\ and\ \citenamefont {Zhang}}]{PhysRevB.81.134508}%
  \BibitemOpen
  \bibfield  {author} {\bibinfo {author} {\bibfnamefont {X.-L.}\ \bibnamefont
  {Qi}}, \bibinfo {author} {\bibfnamefont {T.~L.}\ \bibnamefont {Hughes}},\
  and\ \bibinfo {author} {\bibfnamefont {S.-C.}\ \bibnamefont {Zhang}},\
  }\bibfield  {title} {\bibinfo {title} {{Topological invariants for the Fermi
  surface of a time-reversal-invariant superconductor}},\ }\href
  {https://doi.org/10.1103/PhysRevB.81.134508} {\bibfield  {journal} {\bibinfo
  {journal} {Phys. Rev. B}\ }\textbf {\bibinfo {volume} {81}},\ \bibinfo
  {pages} {134508} (\bibinfo {year} {2010})}\BibitemShut {NoStop}%
\bibitem [{\citenamefont {Sato}\ \emph {et~al.}(2011)\citenamefont {Sato},
  \citenamefont {Tanaka}, \citenamefont {Yada},\ and\ \citenamefont
  {Yokoyama}}]{PhysRevB.83.224511}%
  \BibitemOpen
  \bibfield  {author} {\bibinfo {author} {\bibfnamefont {M.}~\bibnamefont
  {Sato}}, \bibinfo {author} {\bibfnamefont {Y.}~\bibnamefont {Tanaka}},
  \bibinfo {author} {\bibfnamefont {K.}~\bibnamefont {Yada}},\ and\ \bibinfo
  {author} {\bibfnamefont {T.}~\bibnamefont {Yokoyama}},\ }\bibfield  {title}
  {\bibinfo {title} {{Topology of Andreev bound states with flat dispersion}},\
  }\href {https://doi.org/10.1103/PhysRevB.83.224511} {\bibfield  {journal}
  {\bibinfo  {journal} {Phys. Rev. B}\ }\textbf {\bibinfo {volume} {83}},\
  \bibinfo {pages} {224511} (\bibinfo {year} {2011})}\BibitemShut {NoStop}%
\bibitem [{\citenamefont {Karoubi}(2008)}]{Karoubi}%
  \BibitemOpen
  \bibfield  {author} {\bibinfo {author} {\bibfnamefont {M.}~\bibnamefont
  {Karoubi}},\ }\href
  {https://link.springer.com/book/10.1007/978-3-540-79890-3} {\emph {\bibinfo
  {title} {{K-Theory: An Introduction}}}}\ (\bibinfo  {publisher} {Springer
  Berlin, Heidelberg},\ \bibinfo {year} {2008})\BibitemShut {NoStop}%
\bibitem [{\citenamefont {Freed}\ and\ \citenamefont
  {Moore}(2013)}]{Freed2013}%
  \BibitemOpen
  \bibfield  {author} {\bibinfo {author} {\bibfnamefont {D.~S.}\ \bibnamefont
  {Freed}}\ and\ \bibinfo {author} {\bibfnamefont {G.~W.}\ \bibnamefont
  {Moore}},\ }\bibfield  {title} {\bibinfo {title} {{Twisted Equivariant
  Matter}},\ }\href {https://doi.org/10.1007/s00023-013-0236-x} {\bibfield
  {journal} {\bibinfo  {journal} {Annales Henri Poincar{\'e}}\ }\textbf
  {\bibinfo {volume} {14}},\ \bibinfo {pages} {1927} (\bibinfo {year}
  {2013})}\BibitemShut {NoStop}%
\bibitem [{\citenamefont {Shiozaki}\ \emph
  {et~al.}(2018{\natexlab{a}})\citenamefont {Shiozaki}, \citenamefont {Sato},\
  and\ \citenamefont {Gomi}}]{K-AHSS}%
  \BibitemOpen
  \bibfield  {author} {\bibinfo {author} {\bibfnamefont {K.}~\bibnamefont
  {Shiozaki}}, \bibinfo {author} {\bibfnamefont {M.}~\bibnamefont {Sato}},\
  and\ \bibinfo {author} {\bibfnamefont {K.}~\bibnamefont {Gomi}},\ }\href@noop
  {} {\bibinfo {title} {{Atiyah-Hirzebruch Spectral Sequence in Band Topology:
  General Formalism and Topological Invariants for 230 Space Groups}}}
  (\bibinfo {year} {2018}{\natexlab{a}}),\ \Eprint
  {https://arxiv.org/abs/1802.06694} {arXiv:1802.06694 [cond-mat.str-el]}
  \BibitemShut {NoStop}%
\bibitem [{\citenamefont {Song}\ \emph {et~al.}(2017)\citenamefont {Song},
  \citenamefont {Huang}, \citenamefont {Fu},\ and\ \citenamefont
  {Hermele}}]{TC_PRX}%
  \BibitemOpen
  \bibfield  {author} {\bibinfo {author} {\bibfnamefont {H.}~\bibnamefont
  {Song}}, \bibinfo {author} {\bibfnamefont {S.-J.}\ \bibnamefont {Huang}},
  \bibinfo {author} {\bibfnamefont {L.}~\bibnamefont {Fu}},\ and\ \bibinfo
  {author} {\bibfnamefont {M.}~\bibnamefont {Hermele}},\ }\bibfield  {title}
  {\bibinfo {title} {{Topological Phases Protected by Point Group Symmetry}},\
  }\href {https://doi.org/10.1103/PhysRevX.7.011020} {\bibfield  {journal}
  {\bibinfo  {journal} {Phys. Rev. X}\ }\textbf {\bibinfo {volume} {7}},\
  \bibinfo {pages} {011020} (\bibinfo {year} {2017})}\BibitemShut {NoStop}%
\bibitem [{\citenamefont {Huang}\ \emph {et~al.}(2017)\citenamefont {Huang},
  \citenamefont {Song}, \citenamefont {Huang},\ and\ \citenamefont
  {Hermele}}]{TC_PRB}%
  \BibitemOpen
  \bibfield  {author} {\bibinfo {author} {\bibfnamefont {S.-J.}\ \bibnamefont
  {Huang}}, \bibinfo {author} {\bibfnamefont {H.}~\bibnamefont {Song}},
  \bibinfo {author} {\bibfnamefont {Y.-P.}\ \bibnamefont {Huang}},\ and\
  \bibinfo {author} {\bibfnamefont {M.}~\bibnamefont {Hermele}},\ }\bibfield
  {title} {\bibinfo {title} {Building crystalline topological phases from
  lower-dimensional states},\ }\href
  {https://doi.org/10.1103/PhysRevB.96.205106} {\bibfield  {journal} {\bibinfo
  {journal} {Phys. Rev. B}\ }\textbf {\bibinfo {volume} {96}},\ \bibinfo
  {pages} {205106} (\bibinfo {year} {2017})}\BibitemShut {NoStop}%
\bibitem [{\citenamefont {Xiong}(2018)}]{Xiong_2018}%
  \BibitemOpen
  \bibfield  {author} {\bibinfo {author} {\bibfnamefont {C.~Z.}\ \bibnamefont
  {Xiong}},\ }\bibfield  {title} {\bibinfo {title} {{Minimalist approach to the
  classification of symmetry protected topological phases}},\ }\href
  {https://doi.org/10.1088/1751-8121/aae0b1} {\bibfield  {journal} {\bibinfo
  {journal} {Journal of Physics A: Mathematical and Theoretical}\ }\textbf
  {\bibinfo {volume} {51}},\ \bibinfo {pages} {445001} (\bibinfo {year}
  {2018})}\BibitemShut {NoStop}%
\bibitem [{\citenamefont {Song}\ \emph
  {et~al.}(2020{\natexlab{a}})\citenamefont {Song}, \citenamefont {Fang},\ and\
  \citenamefont {Qi}}]{R-AHSS_Song}%
  \BibitemOpen
  \bibfield  {author} {\bibinfo {author} {\bibfnamefont {Z.}~\bibnamefont
  {Song}}, \bibinfo {author} {\bibfnamefont {C.}~\bibnamefont {Fang}},\ and\
  \bibinfo {author} {\bibfnamefont {Y.}~\bibnamefont {Qi}},\ }\bibfield
  {title} {\bibinfo {title} {{Real-space recipes for general topological
  crystalline states}},\ }\href {https://doi.org/10.1038/s41467-020-17685-5}
  {\bibfield  {journal} {\bibinfo  {journal} {Nature Communications}\ }\textbf
  {\bibinfo {volume} {11}},\ \bibinfo {pages} {4197} (\bibinfo {year}
  {2020}{\natexlab{a}})}\BibitemShut {NoStop}%
\bibitem [{\citenamefont {Shiozaki}\ \emph
  {et~al.}(2018{\natexlab{b}})\citenamefont {Shiozaki}, \citenamefont {Xiong},\
  and\ \citenamefont {Gomi}}]{R-AHSS_Shiozaki}%
  \BibitemOpen
  \bibfield  {author} {\bibinfo {author} {\bibfnamefont {K.}~\bibnamefont
  {Shiozaki}}, \bibinfo {author} {\bibfnamefont {C.~Z.}\ \bibnamefont
  {Xiong}},\ and\ \bibinfo {author} {\bibfnamefont {K.}~\bibnamefont {Gomi}},\
  }\href@noop {} {\bibinfo {title} {{Generalized homology and Atiyah-Hirzebruch
  spectral sequence in crystalline symmetry protected topological phenomena}}}
  (\bibinfo {year} {2018}{\natexlab{b}}),\ \Eprint
  {https://arxiv.org/abs/1810.00801} {arXiv:1810.00801 [cond-mat.str-el]}
  \BibitemShut {NoStop}%
\bibitem [{\citenamefont {{Song}}\ \emph {et~al.}(2019)\citenamefont {{Song}},
  \citenamefont {{Huang}}, \citenamefont {{Qi}}, \citenamefont {{Fang}},\ and\
  \citenamefont {{Hermele}}}]{TC_AII}%
  \BibitemOpen
  \bibfield  {author} {\bibinfo {author} {\bibfnamefont {Z.}~\bibnamefont
  {{Song}}}, \bibinfo {author} {\bibfnamefont {S.-J.}\ \bibnamefont {{Huang}}},
  \bibinfo {author} {\bibfnamefont {Y.}~\bibnamefont {{Qi}}}, \bibinfo {author}
  {\bibfnamefont {C.}~\bibnamefont {{Fang}}},\ and\ \bibinfo {author}
  {\bibfnamefont {M.}~\bibnamefont {{Hermele}}},\ }\bibfield  {title} {\bibinfo
  {title} {{Topological states from topological crystals}},\ }\href
  {https://advances.sciencemag.org/content/5/12/eaax2007} {\bibfield  {journal}
  {\bibinfo  {journal} {Science Advances}\ }\textbf {\bibinfo {volume} {5}},\
  \bibinfo {pages} {eaax2007} (\bibinfo {year} {2019})}\BibitemShut {NoStop}%
\bibitem [{\citenamefont {Else}\ and\ \citenamefont
  {Thorngren}(2019)}]{defect_network}%
  \BibitemOpen
  \bibfield  {author} {\bibinfo {author} {\bibfnamefont {D.~V.}\ \bibnamefont
  {Else}}\ and\ \bibinfo {author} {\bibfnamefont {R.}~\bibnamefont
  {Thorngren}},\ }\bibfield  {title} {\bibinfo {title} {{Crystalline
  topological phases as defect networks}},\ }\href
  {https://doi.org/10.1103/PhysRevB.99.115116} {\bibfield  {journal} {\bibinfo
  {journal} {Phys. Rev. B}\ }\textbf {\bibinfo {volume} {99}},\ \bibinfo
  {pages} {115116} (\bibinfo {year} {2019})}\BibitemShut {NoStop}%
\bibitem [{\citenamefont {Zhang}\ \emph {et~al.}(2020)\citenamefont {Zhang},
  \citenamefont {Wang}, \citenamefont {Yang}, \citenamefont {Qi},\ and\
  \citenamefont {Gu}}]{PhysRevB.101.100501}%
  \BibitemOpen
  \bibfield  {author} {\bibinfo {author} {\bibfnamefont {J.-H.}\ \bibnamefont
  {Zhang}}, \bibinfo {author} {\bibfnamefont {Q.-R.}\ \bibnamefont {Wang}},
  \bibinfo {author} {\bibfnamefont {S.}~\bibnamefont {Yang}}, \bibinfo {author}
  {\bibfnamefont {Y.}~\bibnamefont {Qi}},\ and\ \bibinfo {author}
  {\bibfnamefont {Z.-C.}\ \bibnamefont {Gu}},\ }\bibfield  {title} {\bibinfo
  {title} {Construction and classification of point-group symmetry-protected
  topological phases in two-dimensional interacting fermionic systems},\ }\href
  {https://doi.org/10.1103/PhysRevB.101.100501} {\bibfield  {journal} {\bibinfo
   {journal} {Phys. Rev. B}\ }\textbf {\bibinfo {volume} {101}},\ \bibinfo
  {pages} {100501} (\bibinfo {year} {2020})}\BibitemShut {NoStop}%
\bibitem [{\citenamefont {Rasmussen}\ and\ \citenamefont
  {Lu}(2020)}]{PhysRevB.101.085137}%
  \BibitemOpen
  \bibfield  {author} {\bibinfo {author} {\bibfnamefont {A.}~\bibnamefont
  {Rasmussen}}\ and\ \bibinfo {author} {\bibfnamefont {Y.-M.}\ \bibnamefont
  {Lu}},\ }\bibfield  {title} {\bibinfo {title} {Classification and
  construction of higher-order symmetry-protected topological phases of
  interacting bosons},\ }\href {https://doi.org/10.1103/PhysRevB.101.085137}
  {\bibfield  {journal} {\bibinfo  {journal} {Phys. Rev. B}\ }\textbf {\bibinfo
  {volume} {101}},\ \bibinfo {pages} {085137} (\bibinfo {year}
  {2020})}\BibitemShut {NoStop}%
\bibitem [{\citenamefont {Song}\ \emph
  {et~al.}(2020{\natexlab{b}})\citenamefont {Song}, \citenamefont {Xiong},\
  and\ \citenamefont {Huang}}]{PhysRevB.101.165129}%
  \BibitemOpen
  \bibfield  {author} {\bibinfo {author} {\bibfnamefont {H.}~\bibnamefont
  {Song}}, \bibinfo {author} {\bibfnamefont {C.~Z.}\ \bibnamefont {Xiong}},\
  and\ \bibinfo {author} {\bibfnamefont {S.-J.}\ \bibnamefont {Huang}},\
  }\bibfield  {title} {\bibinfo {title} {Bosonic crystalline symmetry protected
  topological phases beyond the group cohomology proposal},\ }\href
  {https://doi.org/10.1103/PhysRevB.101.165129} {\bibfield  {journal} {\bibinfo
   {journal} {Phys. Rev. B}\ }\textbf {\bibinfo {volume} {101}},\ \bibinfo
  {pages} {165129} (\bibinfo {year} {2020}{\natexlab{b}})}\BibitemShut
  {NoStop}%
\bibitem [{\citenamefont {Peng}\ \emph {et~al.}(2021)\citenamefont {Peng},
  \citenamefont {Jiang}, \citenamefont {Fang}, \citenamefont {Weng},\ and\
  \citenamefont {Fang}}]{TC_MSG}%
  \BibitemOpen
  \bibfield  {author} {\bibinfo {author} {\bibfnamefont {B.}~\bibnamefont
  {Peng}}, \bibinfo {author} {\bibfnamefont {Y.}~\bibnamefont {Jiang}},
  \bibinfo {author} {\bibfnamefont {Z.}~\bibnamefont {Fang}}, \bibinfo {author}
  {\bibfnamefont {H.}~\bibnamefont {Weng}},\ and\ \bibinfo {author}
  {\bibfnamefont {C.}~\bibnamefont {Fang}},\ }\href@noop {} {\bibinfo {title}
  {Topological classification and diagnosis in magnetically ordered electronic
  materials}} (\bibinfo {year} {2021}),\ \Eprint
  {https://arxiv.org/abs/2102.12645} {arXiv:2102.12645 [cond-mat.mes-hall]}
  \BibitemShut {NoStop}%
\bibitem [{\citenamefont {Hahn}(2006)}]{ITC}%
  \BibitemOpen
  \bibinfo {editor} {\bibfnamefont {T.}~\bibnamefont {Hahn}},\ ed.,\ \href@noop
  {} {\emph {\bibinfo {title} {{International Tables for Crystallography}}}},\
  \bibinfo {edition} {5th}\ ed.,\ Vol.\ \bibinfo {volume} {A: Space-group
  symmetry}\ (\bibinfo  {publisher} {Springer},\ \bibinfo {year}
  {2006})\BibitemShut {NoStop}%
\bibitem [{\citenamefont {Okuma}\ \emph {et~al.}(2019)\citenamefont {Okuma},
  \citenamefont {Sato},\ and\ \citenamefont {Shiozaki}}]{PhysRevB.99.085127}%
  \BibitemOpen
  \bibfield  {author} {\bibinfo {author} {\bibfnamefont {N.}~\bibnamefont
  {Okuma}}, \bibinfo {author} {\bibfnamefont {M.}~\bibnamefont {Sato}},\ and\
  \bibinfo {author} {\bibfnamefont {K.}~\bibnamefont {Shiozaki}},\ }\bibfield
  {title} {\bibinfo {title} {Topological classification under nonmagnetic and
  magnetic point group symmetry: Application of real-space atiyah-hirzebruch
  spectral sequence to higher-order topology},\ }\href
  {https://doi.org/10.1103/PhysRevB.99.085127} {\bibfield  {journal} {\bibinfo
  {journal} {Phys. Rev. B}\ }\textbf {\bibinfo {volume} {99}},\ \bibinfo
  {pages} {085127} (\bibinfo {year} {2019})}\BibitemShut {NoStop}%
\bibitem [{\citenamefont {Read}\ and\ \citenamefont
  {Green}(2000)}]{PhysRevB.61.10267}%
  \BibitemOpen
  \bibfield  {author} {\bibinfo {author} {\bibfnamefont {N.}~\bibnamefont
  {Read}}\ and\ \bibinfo {author} {\bibfnamefont {D.}~\bibnamefont {Green}},\
  }\bibfield  {title} {\bibinfo {title} {Paired states of fermions in two
  dimensions with breaking of parity and time-reversal symmetries and the
  fractional quantum hall effect},\ }\href
  {https://doi.org/10.1103/PhysRevB.61.10267} {\bibfield  {journal} {\bibinfo
  {journal} {Phys. Rev. B}\ }\textbf {\bibinfo {volume} {61}},\ \bibinfo
  {pages} {10267} (\bibinfo {year} {2000})}\BibitemShut {NoStop}%
\bibitem [{\citenamefont {Stone}\ and\ \citenamefont
  {Chung}(2006)}]{PhysRevB.73.014505}%
  \BibitemOpen
  \bibfield  {author} {\bibinfo {author} {\bibfnamefont {M.}~\bibnamefont
  {Stone}}\ and\ \bibinfo {author} {\bibfnamefont {S.-B.}\ \bibnamefont
  {Chung}},\ }\bibfield  {title} {\bibinfo {title} {Fusion rules and vortices
  in ${p}_{x}+i{p}_{y}$ superconductors},\ }\href
  {https://doi.org/10.1103/PhysRevB.73.014505} {\bibfield  {journal} {\bibinfo
  {journal} {Phys. Rev. B}\ }\textbf {\bibinfo {volume} {73}},\ \bibinfo
  {pages} {014505} (\bibinfo {year} {2006})}\BibitemShut {NoStop}%
\bibitem [{\citenamefont {Tewari}\ \emph
  {et~al.}(2007{\natexlab{a}})\citenamefont {Tewari}, \citenamefont
  {Das~Sarma}, \citenamefont {Nayak}, \citenamefont {Zhang},\ and\
  \citenamefont {Zoller}}]{PhysRevLett.98.010506}%
  \BibitemOpen
  \bibfield  {author} {\bibinfo {author} {\bibfnamefont {S.}~\bibnamefont
  {Tewari}}, \bibinfo {author} {\bibfnamefont {S.}~\bibnamefont {Das~Sarma}},
  \bibinfo {author} {\bibfnamefont {C.}~\bibnamefont {Nayak}}, \bibinfo
  {author} {\bibfnamefont {C.}~\bibnamefont {Zhang}},\ and\ \bibinfo {author}
  {\bibfnamefont {P.}~\bibnamefont {Zoller}},\ }\bibfield  {title} {\bibinfo
  {title} {{Quantum Computation using Vortices and Majorana Zero Modes of a
  ${p}_{x}+i{p}_{y}$ Superfluid of Fermionic Cold Atoms}},\ }\href
  {https://doi.org/10.1103/PhysRevLett.98.010506} {\bibfield  {journal}
  {\bibinfo  {journal} {Phys. Rev. Lett.}\ }\textbf {\bibinfo {volume} {98}},\
  \bibinfo {pages} {010506} (\bibinfo {year} {2007}{\natexlab{a}})}\BibitemShut
  {NoStop}%
\bibitem [{\citenamefont {Tewari}\ \emph
  {et~al.}(2007{\natexlab{b}})\citenamefont {Tewari}, \citenamefont
  {Das~Sarma},\ and\ \citenamefont {Lee}}]{PhysRevLett.99.037001}%
  \BibitemOpen
  \bibfield  {author} {\bibinfo {author} {\bibfnamefont {S.}~\bibnamefont
  {Tewari}}, \bibinfo {author} {\bibfnamefont {S.}~\bibnamefont {Das~Sarma}},\
  and\ \bibinfo {author} {\bibfnamefont {D.-H.}\ \bibnamefont {Lee}},\
  }\bibfield  {title} {\bibinfo {title} {{Index Theorem for the Zero Modes of
  Majorana Fermion Vortices in Chiral $p$-Wave Superconductors}},\ }\href
  {https://doi.org/10.1103/PhysRevLett.99.037001} {\bibfield  {journal}
  {\bibinfo  {journal} {Phys. Rev. Lett.}\ }\textbf {\bibinfo {volume} {99}},\
  \bibinfo {pages} {037001} (\bibinfo {year} {2007}{\natexlab{b}})}\BibitemShut
  {NoStop}%
\bibitem [{\citenamefont {Gurarie}\ and\ \citenamefont
  {Radzihovsky}(2007)}]{PhysRevB.75.212509}%
  \BibitemOpen
  \bibfield  {author} {\bibinfo {author} {\bibfnamefont {V.}~\bibnamefont
  {Gurarie}}\ and\ \bibinfo {author} {\bibfnamefont {L.}~\bibnamefont
  {Radzihovsky}},\ }\bibfield  {title} {\bibinfo {title} {Zero modes of
  two-dimensional chiral $p$-wave superconductors},\ }\href
  {https://doi.org/10.1103/PhysRevB.75.212509} {\bibfield  {journal} {\bibinfo
  {journal} {Phys. Rev. B}\ }\textbf {\bibinfo {volume} {75}},\ \bibinfo
  {pages} {212509} (\bibinfo {year} {2007})}\BibitemShut {NoStop}%
\bibitem [{\citenamefont {Peng}\ \emph {et~al.}(2022)\citenamefont {Peng},
  \citenamefont {Weng},\ and\ \citenamefont {Fang}}]{Wire_Fang}%
  \BibitemOpen
  \bibfield  {author} {\bibinfo {author} {\bibfnamefont {B.}~\bibnamefont
  {Peng}}, \bibinfo {author} {\bibfnamefont {H.}~\bibnamefont {Weng}},\ and\
  \bibinfo {author} {\bibfnamefont {C.}~\bibnamefont {Fang}},\ }\href@noop {}
  {\bibinfo {title} {{Wire Construction of Topological Crystalline
  Superconductors}}} (\bibinfo {year} {2022}),\ \Eprint
  {https://arxiv.org/abs/2205.13397} {arXiv:2205.13397 [cond-mat.supr-con]}
  \BibitemShut {NoStop}%
\bibitem [{\citenamefont {Bradley}\ and\ \citenamefont
  {Cracknell}(1972)}]{Bradley}%
  \BibitemOpen
  \bibfield  {author} {\bibinfo {author} {\bibfnamefont {C.~J.}\ \bibnamefont
  {Bradley}}\ and\ \bibinfo {author} {\bibfnamefont {A.~P.}\ \bibnamefont
  {Cracknell}},\ }\href
  {https://global.oup.com/academic/product/the-mathematical-theory-of-symmetry-in-solids-9780199582587?cc=jp&lang=en&#}
  {\emph {\bibinfo {title} {{The Mathematical Theory of Symmetry in Solids}}}}\
  (\bibinfo  {publisher} {Oxford University Press},\ \bibinfo {year}
  {1972})\BibitemShut {NoStop}%
\end{thebibliography}%
	
	\clearpage
	\appendix
	\onecolumngrid
	\section{Vortex zero modes in two-dimensional TSCs}
	\label{app:vortex}
		%In this appendix, we show that two-dimensional topological superconductors sometimes have vortex zero modes in their bulk when some crystalline symmetries prohibit the existence of nontrivial topological invariants. 
		In this appendix, we provide more detailed discussions on vortex zero modes in two-dimensional topological superconductors (TSCs).
		First, we discuss the vortex zero modes in $p_x \pm i p_y$ superconductor and derive wavefunctions of them. 
		Next, we generalize the discussions for $p_x \pm i p_y$ superconductor to those for time-reversal symmetric TSCs in two-dimension. Also, we analyze symmetry representations of the wavefunctions. 
	
	\subsection{$p_x \pm i p_y$ superconductor with inversion symmetry}
	\subsubsection{Model and wavefunctions of vortex zero modes}
	Following Refs.~\cite{PhysRevB.61.10267,PhysRevB.75.212509,SatoFujimoto}, we here show that the $p_x \pm i p_y$ superconductor has a vortex zero mode when the vortex has an odd-integer winding.
	%\subsection{Model and wavefunctions of vortex zero modes}
	Let us consider the following continuum model in the $xy$-plane:
	\begin{align}
		\label{eq:pxipy_app}
		H_{\pm}(\bm{r}; \Delta) &= \begin{pmatrix}
			-\frac{\nabla^2}{2m} - \mu(\bm{r})  & \Delta(\bm{r})\left(\frac{1}{i}\partial_x \pm i\frac{1}{i}\partial_y \right)\\
			\Delta^*(\bm{r})\left(\frac{1}{i}\partial_x \mp i\frac{1}{i}\partial_y \right) & \frac{\nabla^2}{2m} + \mu(\bm{r})
		\end{pmatrix},\\
		\label{eq:PHS1}
		u(\calC)&= \tau_x, 
	\end{align}
	which describes the {$p_x \pm i p_y$ superconductor. 
	Here, $\calC$ denotes PHS, and  the superconducting gap function $\Delta(\bm{r})$ and the chemical potential $\mu(\bm{r})$ vary slowly in space.
	It is convenient to use the polar coordinate $(r, \theta)$ such that $(x, y) = (r \cos\theta, r \sin\theta)$.
	In the polar coordinate, the Hamiltonian is described by
	\begin{align}
		\label{eq:pxipy_app2}
		H_{\pm}(\bm{r}; \Delta) &= \begin{pmatrix}
			-\frac{\nabla^2}{2m} - \mu(\bm{r})  & -i \Delta(\bm{r})e^{\pm i \theta}\left(\partial_r \pm i \frac{1}{r}\partial_\theta \right)\\
			i \Delta^*(\bm{r})e^{\mp i \theta}\left(\partial_r \mp i \frac{1}{r}\partial_\theta \right) & \frac{\nabla^2}{2m} + \mu(\bm{r})
		\end{pmatrix}.
	\end{align}

		Next, we consider the Bogoliubov--de Gennes (BdG) equation to obtain the wavefunction of vortex zero mode
		\begin{align}
			\label{eq:BdGeq}
			\begin{pmatrix}
				- \mu(r) & -i\Delta(\bm{r})(\partial_x \pm i \partial_y) \\
				i\Delta^*(\bm{r})(\partial_x \mp i \partial_y) & \mu(r)\\ 
			\end{pmatrix}
			\begin{pmatrix}
				u \\ v
			\end{pmatrix} = 
			\begin{pmatrix}
				0 \\ 0
			\end{pmatrix}.
		\end{align}
	We here assume that $\mu(\br) = \mu(r)$ such that $\mu(r) \rightarrow \mu_0 > 0$ as $r\rightarrow \infty$ and $\mu(r) \rightarrow -\mu_0$ as $r\rightarrow 0$.
		According to \cite{PhysRevB.61.10267}, we consider a vortex as a small circular edge with vacuum (vanishing density) at the center and study Majorana zero modes near the phase transition $(\mu(r) = 0)$ so that the kinetic term $-\nabla^2/2m$ is omitted.
	Since $v = u^*$ due to PHS, the equation is rewritten as	
		\begin{align}
			\label{eq:polar_BdG-1}
			-\mu(\bm{r})u -i \Delta(\bm{r})e^{\pm i \theta}\left(\partial_r \pm i \frac{1}{r}\partial_\theta \right)u^* &= 0.
			%-\mu(\bm{r})v +i \Delta^*(\bm{r})e^{\mp i \theta}\left(\partial_r \mp i \frac{1}{r}\partial_\theta \right)u &= 0.
		\end{align}
		
		Let us start with the case $\Delta(\bm{r}) = \vert \Delta_0 (r) \vert e^{\pm i \alpha}e^{\mp i \theta}$, where $\alpha$ is a constant. 
		This form indicates that the vortex is present in $p_x\pm i p_y$ superconductor and that its vorticity is $-1$. 
		For this case, the equation becomes a simpler form
		\begin{align}
			-\frac{\mu(r)}{\vert \Delta_0 (r) \vert}u(r) &= e^{i (\pm\alpha + \pi/2)}\partial_r u^*(r).
			%-\frac{\mu(r)}{\vert \Delta_0 (r) \vert}[e^{-\tfrac{i}{2} (\pm\alpha + \pi/2)}u(r)] &= \partial_r [e^{-\tfrac{i}{2} (\pm\alpha + \pi/2)}u(r)]^*.
		\end{align}
		A solution of this equation is proportional to $e^{\tfrac{i}{2} (\pm\alpha + \pi/2)}e^{-\int^r \frac{\mu(r')}{\vert \Delta_0 (r') \vert} dr'}$~\cite{SatoFujimoto}. Thus, we obtain the wavefunction of this vortex zero mode
		\begin{align}
			\label{eq:wave-1}
			\phi(r,\theta) &\propto \begin{pmatrix}
				e^{\tfrac{i}{2} (\pm \alpha + \pi/2)}\\
				e^{-\tfrac{i}{2} (\pm\alpha + \pi/2)}
			\end{pmatrix}e^{-\int^r \frac{\mu(r')}{\vert \Delta_0 (r') \vert} dr'}.
		\end{align}
	
		Next, we argue that the vortex with an even-integer winding $2m$ can be eliminated by the gauge transformation
		\begin{align}
			\tilde{u} = e^{-i m \theta}u,\quad \tilde{v} = e^{i m \theta}v.
		\end{align}
		Then, the equation is rewritten as
		\begin{align}
			e^{i m \theta}\left[-\mu(r)\tilde{u} - i\Delta_0(r)e^{\pm i \theta}\left(\partial_r \pm i \frac{1}{r}\partial_\theta \right)\tilde{v} \right] \pm e^{i(2m)\theta}\Delta_0(r)e^{\pm i \theta}\frac{1}{r}\partial_\theta[e^{-im\theta}]\tilde{v} = E e^{i m \theta} \tilde{u}.
		\end{align}
		As discussed in Ref.~\cite{PhysRevB.75.212509}, we can consider a smooth deformation in such a way that the second term in the left-hand side vanishes. 
		After the deformation, we find that the equation becomes the equation for vortex-free systems in which $\Delta(\bm{r}) = \vert \Delta_0 (r) \vert$.
		Therefore, we conclude that the vortex with an even-integer winding does not exhibit zero modes. 
		
		Using this property, we find the wavefunction of this vortex zero mode with its vorticity $2m-1\ (m \in \mathbb{Z})$,
		\begin{align}
			\phi_{\pm, m}(r,\theta) &\propto \begin{pmatrix}
				e^{\tfrac{i}{2} (\pm\alpha + \pi/2)}e^{\pm i m \theta}\\
				e^{-\tfrac{i}{2} (\pm\alpha + \pi/2)}e^{\mp i m \theta}
			\end{pmatrix}e^{-\int^r \frac{\mu(r')}{\vert \Delta_0 (r') \vert} dr'}.
		\end{align}

		\subsubsection{Symmetries}
		Here, we discuss symmetries of $H_{+}$ and $H_{-}$ in Eq.~\eqref{eq:pxipy_app} and relationship between them. 
		$H_{\pm}$ has even-parity inversion symmetry $u(I) = \tau_0$ and $n$-fold rotation symmetry $u(C_{n}^{z}) = e^{i \tfrac{\pi}{n}\tau_z}(n=2,3,4,6)$, which satisfy
		\begin{align}
			\label{eq:even-inversion}
			u(I)H_{\pm}(\bm{r}; \Delta) &= H_{\pm}(-\bm{r}; \Delta)u(I);\\
			\label{eq:even-rotation}
			u(C_{n}^{z})H_{+}(\bm{r}; \Delta) &= H_{+}(C_{n}^{z}\bm{r}; \Delta)u(C_{n}^{z});\\
			\label{eq:even-rotation2}
			[u(C_{n}^{z})]^*H_{-}(\bm{r}; \Delta) &= H_{-}(C_{n}^{z}\bm{r}; \Delta)[u(C_{n}^{z})]^*.
		\end{align}
		From Eq.~\eqref{eq:even-inversion}, we find $\Delta(-\bm{r}) = -\Delta(\bm{r})$, which indicates the vorticity is odd. That is, inversion symmetry in even-parity pairing superconductors enforce the emergence of a vortex zero mode at the vortex cores. On the other hand, $\Delta(C_{n}^{z}\bm{r}) = \Delta(\bm{r})$ for $C_{n}^{z}$, which implies that the winding of a vortex is even. 
		
		Furthermore, $H_{+}$ and $H_{-}$ are related to each other as follows:
		\begin{align}
			\label{eq:rel1}
			\tau_z [H_{+}(\bm{r}; \Delta)]^{*}\tau_z &= H_{-}(\bm{r}; \Delta^*);\\
			\label{eq:rel2}
			H_{+}(\bm{r}; \Delta) &= H_{-}(C_{2}^{x}\bm{r}; \Delta),
		\end{align}
		which lead to time-reversal and two-fold rotation along the $x$-axis, as discussed in the next section.

		\subsection{Two-dimensional topological superconductor with time-reversal symmetries}
		\subsubsection{Model and wavefunctions of vortex zero modes}
		Here, we discuss two-dimensional topological superconductors with time-reversal symmetry. While the $p_x + i p_y$ and $p_x - i p_y$ superconductors are not invariant under time-reversal symmetry, they are interchanged by the time-reversal operation. As discussed in the main text, by stacking these two, we can construct a time-reversal symmetric superconductor. The Hamiltonian is described by
		\begin{align}
			\label{eq:DIII-SC_app}
			H_{\text{DIII}}(\bm{r}; \Delta) %&= H_{p_x + i p_y}(\bm{r}, \bm{k}; \Delta) \oplus H_{p_x - i p_y}(\bm{r}, \bm{k}; \Delta^*), \\
			&= \begin{pmatrix}
				H_{+}(\bm{r}; \Delta) & 0 \\
				0 & H_{-}(\bm{r}; \Delta^*)
			\end{pmatrix}_s,\\
			U(\calC) &= \begin{pmatrix}
				\tau_x & 0 \\
				0 & \tau_x
			\end{pmatrix}_s =\tau_xs_0,
			%\calT &= i\tau_z s_y K
		\end{align}
		where we introduce a spin degree of freedom and use $H_{+}(\bm{r}; \Delta)$ for the spin-up component and $H_{-}(\bm{r}; \Delta^*)$ for the spin-down component. In the following, $U(g)$ denotes a unitary representation of $g$.
		From Eq.~\eqref{eq:rel1}, we find time-reversal symmetry 
		\begin{align}
			U(\calT) = i\tau_zs_y
		\end{align}
		such that 
		\begin{align}
			U(\calT)[H_{\text{DIII}}(\bm{r}; \Delta)]^* &= H_{\text{DIII}}(\bm{r}; \Delta)U(\calT).
		\end{align}
	
		Correspondingly, the wavefunctions of vortex zero modes are
		\begin{align}
			\label{eq:DIII-vortex}
			\Phi^{\alpha}_{m,\uparrow}(r, \theta) &= \phi_{+, m}^{\alpha}(r,\theta)\otimes\begin{pmatrix}
				1\\0
			\end{pmatrix}_s \propto \begin{pmatrix}
			e^{\tfrac{i}{2} (\alpha + \pi/2)}e^{ im\theta}\\
			e^{-\tfrac{i}{2} (\alpha + \pi/2)}e^{- im\theta}\
		\end{pmatrix}_\tau\otimes\begin{pmatrix}
		1\\0 
	\end{pmatrix}_s e^{-\int^r \frac{\mu(r')}{\vert \Delta_0 (r') \vert} dr'},\\
			\Phi^{\alpha}_{m,\downarrow}(r, \theta) &=	\phi_{-, -m}^{-\alpha}(r,\theta)\otimes\begin{pmatrix}
				0\\1
			\end{pmatrix}_s \propto \begin{pmatrix}
			e^{\tfrac{i}{2} (-\alpha + \pi/2)}e^{- im\theta}\\
			e^{-\tfrac{i}{2} (-\alpha + \pi/2)}e^{im\theta}\
		\end{pmatrix}_\tau\otimes\begin{pmatrix}
		0\\1
	\end{pmatrix}_se^{-\int^r \frac{\mu(r')}{\vert \Delta_0 (r') \vert} dr'}.
		\end{align}

		\subsection{Symmetry representations of Hamiltonian}
		Here, we find symmetry representations of $H_{\text{DIII}}$. In the following discussions, we consider the case where the pairing symmetry is trivial, i.e., the representation of the superconducting order parameter under point group symmetries is trivial. 
		
		From Eqs.~\eqref{eq:even-inversion} and \eqref{eq:even-rotation}, we find that even-parity inversion and 
		$n$-fold rotation symmetry are represented by
		\begin{align}
			U(I) &= \begin{pmatrix}
				\tau_0 & 0 \\
				0 & \tau_0
			\end{pmatrix}_s =\tau_0s_0,\\
			U(C_{n}^{z})  &= \begin{pmatrix}
				e^{i \tfrac{\pi}{n}\tau_z} & 0 \\
				0 & e^{-i \tfrac{\pi}{n}\tau_z}
			\end{pmatrix}_s =  e^{i \tfrac{\pi}{n}\tau_z s_z}
		\end{align}
		such that 
		\begin{align}
			\label{eq:symm_inv}
			U(I)H_{\text{DIII}}(\bm{r}; \Delta) &= H_{\text{DIII}}(-\bm{r}; -\Delta)U(I),\\
			\label{eq:symm_rotation_z}
			U(C_{n}^{z})H_{\text{DIII}}(\bm{r}; \Delta) &= H_{\text{DIII}}(C_{n}^{z}\bm{r}; \Delta)U(C_{n}^{z}).
			%U(\calC)[U(C_{n}^{z})]^* &= U(C_{n}^{z})U(\calC).
		\end{align}
		Equations~\eqref{eq:symm_inv} and~\eqref{eq:symm_rotation_z} suggest that $\Delta(-\bm{r}) = -\Delta(\bm{r})$ and $\Delta(C_{n}^{z}\bm{r}) = \Delta(\bm{r})$. 
		
		Equation \eqref{eq:rel2} implies that two-fold rotation along the $x$-axis interchange $H_{\pm}$. Then, it is easy to find the two-fold rotation symmetry
		\begin{align}
			\label{eq:C2x-1}
			U(C_{2}^{x}) &= \begin{pmatrix}
				0 & \mathds{1} \\
				-\mathds{1} & 0
			\end{pmatrix}_s =is_y
		\end{align}
		such that
		\begin{align}
			\label{eq:symm_rotation_x}
			U(C_{2}^{x})H_{\text{DIII}}(\bm{r}; \Delta) &= H_{\text{DIII}}(C_{2}^{x}\bm{r}; \Delta^*)U(C_{2}^{x}).
			%U(\calC)[U(C_{2}^{x})]^* &= U(C_{2}^{x})U(\calC).
		\end{align}
		Equation \eqref{eq:symm_rotation_x} implies that $\Delta(C_{2}^{x}\bm{r}) = [\Delta(\bm{r})]^*$. 
		
		Other symmetries are constructed by $I, C_{n}^{z},$ and $C_{2}^{x}$. For example, two-fold rotation symmetry along [110] (denoted by $C_{2}^{[110]} = C_{4}^{z}C_{2}^{x}$) is defined by $U(C_{2}^{[110]}) \equiv U(C_{4}^{z})U(C_{2}^{x})$. Then, $\Delta(C_{2}^{[110]}\bm{r}) = [\Delta(\bm{r})]^*$. We summarize the representations and transformations of $\Delta(\bm{r})$ for all symmetries in Table~\ref{tab:rep}. 
		
		Although we always use Eq.~(\ref{eq:C2x-1}) in our comprehensive calculations, it is also possible to consider a different representation of $C_{2}^{x}$
		\begin{align}
			\label{eq:C2x-2}
			U(C_{2}^{x}) &= is_x\tau_z,
		\end{align}
		which suggests that $\Delta(C_{2}^{x}\bm{r}) = -[\Delta(\bm{r})]^*$. 
		See Table~\ref{tab:rep2} for other operations in this choice.
		
		\begin{table}[H]
			\begin{center}
				\caption{\label{tab:rep}\textbf{Summary of representations and transformations of $\Delta(\bm{r})\propto e^{i \varphi(\br)}$ in the choice of Eq.~\eqref{eq:C2x-1}.}
				}
				\begin{tabular}{c|c|c|c}
					\hline
					$g$ & $U(g)$ & transformation of $\Delta(\bm{r})$&$\alpha$ at $g\bm{r} = \bm{r}$\\
					\hline\hline
					$I$ & $\tau_0s_0$ & $\Delta(-\bm{r}) = -\Delta(\bm{r})$&$-$\\
					$C_{n}^{z}$ & $e^{i \tfrac{\pi}{n}\tau_z s_z}$ & $\Delta(C_{n}^{z}\bm{r}) = \Delta(\bm{r})$& $-$\\%$\varphi(\br)  \in (-\pi, \pi]$\\
					$C_{2}^{x}$ & $is_y$ & $\Delta(C_{2}^{x}\bm{r}) = [\Delta(\bm{r})]^*$&$\varphi(\br)  \in \{0, \pi\}$\\
					$S_{n}^{z}=IC_{n}^{z}$ & $e^{i \tfrac{\pi}{n}\tau_z s_z}$ & $\Delta(S_{n}^{z}\bm{r}) = -\Delta(\bm{r})$&$-$\\
					%$M^{z}$ & $e^{i \tfrac{\pi}{n}\tau_z s_z}$ & $\Delta(M^{z}\bm{r}) = -\Delta(\bm{r})$\\
					$C_{2}^{d}=C_{n}^{z}C_{2}^{x}$ & $e^{i \tfrac{\pi}{n}\tau_z s_z}is_y$ & $\Delta(C_{2}^{d}\bm{r}) = [\Delta(\bm{r})]^*$&$\varphi(\br)  \in \{0, \pi\}$\\
					$M^{d}=IC_{2}^{d}$ & $e^{i \tfrac{\pi}{n}\tau_z s_z}is_y$ & $\Delta(M^{d}\bm{r}) = -[\Delta(\bm{r})]^*$&$\varphi(\br)  \in \{-\tfrac{\pi}{2}, \tfrac{\pi}{2}\}$\\
					\hline
				\end{tabular}
			\end{center}
		\end{table}
	
			\begin{table}[H]
		\begin{center}
			\caption{\label{tab:rep2}\textbf{Summary of representations and transformations of $\Delta(\bm{r}) \propto e^{i \varphi(\br)}$ in the choice of Eq.~\eqref{eq:C2x-2}.}
			}
			\begin{tabular}{c|c|c|c}
				\hline
				$g$ & $U(g)$ & transformation of $\Delta(\bm{r})$&$ \varphi(\br)$ at $g\bm{r} = \bm{r}$\\
				\hline\hline
				$I$ & $\tau_0s_0$ & $\Delta(-\bm{r}) = -\Delta(\bm{r})$&$-$\\
				$C_{n}^{z}$ & $e^{i \tfrac{\pi}{n}\tau_z s_z}$ & $\Delta(C_{n}^{z}\bm{r}) = \Delta(\bm{r})$&$-$\\
				$C_{2}^{x}$ & $is_x\tau_z$ & $\Delta(C_{2}^{x}\bm{r}) = -[\Delta(\bm{r})]^*$&$ \varphi(\br) \in \{0, \pi\}$\\
				$S_{n}^{z}=IC_{n}^{z}$ & $e^{i \tfrac{\pi}{n}\tau_z s_z}$ & $\Delta(S_{n}^{z}\bm{r}) = -\Delta(\bm{r})$&$-$\\
				$C_{2}^{d}=C_{n}^{z}C_{2}^{x}$ & $e^{i \tfrac{\pi}{n}\tau_z s_z}is_x\tau_z$ & $\Delta(C_{2}^{d}\bm{r}) = -[\Delta(\bm{r})]^*$&$ \varphi(\br) \in \{0, \pi\}$\\
				$M^{d}=IC_{2}^{d}$ & $e^{i \tfrac{\pi}{n}\tau_z s_z}is_x\tau_z$ & $\Delta(M^{d}\bm{r}) = [\Delta(\bm{r})]^*$&$ \varphi(\br) \in \{-\tfrac{\pi}{2}, \tfrac{\pi}{2}\}$\\
				\hline
			\end{tabular}
		\end{center}
	\end{table}

		\subsection{Symmetry representations of vortex zero modes}
		Here, we discuss representations of vortex zero modes in Eq.~\eqref{eq:DIII-vortex}. For each element, the representations are obtained by
		\begin{align}
			U(g)\Phi_{m}(r, \theta)  &= \Phi_{m}(r, g(\theta))U_{\text{vortex}}(g) ,
		\end{align}
		where $g(\theta)$ is the angle after the transformation by $g$. For example, $I(\theta) = \theta + \pi$. In Table \ref{tab:rep_vortex}, we tabulate $g(\theta)$ and $U_{\text{vortex}}(g)$ for an arbitrary $g$. 
		
		\begin{table}[H]
			\begin{center}
				\caption{\label{tab:rep_vortex}\textbf{Summary of representations of vortex zero modes for all symmetries in the choice of Eq.~\eqref{eq:C2x-1}}.
				}
				\begin{tabular}{c|c|c}
					\hline
					$g$ & $g(\theta)$ & $U_{\text{vortex}}(g)$ \\
					\hline\hline
					$I$ & $\theta + \pi$ & $(-1)^m \sigma_0$ \\
					$C_{n}^{z}$ & $\theta + \frac{2\pi}{n}$ & $\cos \frac{(2m-1)\pi}{n}\sigma_0$ \\
					$C_{2}^{x}$ & $-\theta$ & $\cos\alpha\ i\sigma_y$ \\
					$S_{n}^{z}$ & $\theta + \frac{2\pi}{n} + \pi$ & $(-1)^m\cos \frac{(2m-1)\pi}{n}\sigma_0$ \\
					$C_{2}^{d}$ & $-\theta + \frac{2\pi}{n}$ & $\cos \frac{(2m-1)\pi + n \alpha}{n}\ i\sigma_y$ \\
					$M^{d}$ & $-\theta+ \frac{2\pi}{n} + \pi$ & $(-1)^m\cos \frac{(2m-1)\pi + n \alpha}{n}\ i\sigma_y$ \\
					$\calT$ & $\theta$& $\sigma_y$ \\
					$\calC$ & $\theta$& $\sigma_0$ \\
					\hline
				\end{tabular}
			\end{center}
		\end{table}	
	\clearpage
	
	\twocolumngrid
	\section{Atiyah-Hirzebruch spectral sequence in real space}
	\label{app:RAHSS}
	In this section, we briefly review Atiyah-Hirzebruch spectral sequence for superconductors, which enables us to systematically perform the real-space classification. More detailed discussions are found in Ref.~\cite{R-AHSS_Shiozaki}.
	
	\subsection{Symmetry and representations}
	Suppose that the system of interest in the normal phase is invariant under symmetries in a magnetic space group (MSG) $\calM = \calG + \calA$, where $\calG$ and $\calA$ denote space-group and anti-unitary parts of $\calM$, respectively. 
	Translation group $T$, which is composed of lattice translations, is always a subgroup of $\calM$. 
	While we consider only the case of type II MSGs for which $\calA = \calG \calT$ ($\calT$: TRS) in this work, the following discussions except for Sec.~\ref{app:d232} are applicable to any MSG. 
	A point $\bx$ is transformed into $g\bx = p_g\bx+\bm{t}_g$ by an element $g\in \calM$, where $p_g$ is an element of $\text{O}(3)$ and $\bm{t}_g$ is a lattice translation or a fractional translation.

	We now move on to the action of spatial symmetries on electrons. A symmetry $g \in \calG$ transforms the fermionic
	 creation operator $\creation_{\sigma, \bR}\ (\sigma=1, 2, \cdots, N_{\text{orb}})$ into
	\begin{align}
		\label{eq:symm_real}
		\hat{g}\creation_{\sigma,\bm{R}}\hat{g}^{-1} = \sum_{\sigma'}\creation_{\sigma',g_{\sigma'}(\bm{R})}[U(g)]_{\sigma'\sigma}, 
	\end{align}
	where $U(g)$ is a $N_{\text{orb}}$-dimensional unitary matrix encoding transformation properties of the degree of freedom in unit cell. 
	It should be noted that $g_{\sigma'}(\bm{R}) \neq p_g\bR + \bt_g$ in general. 
	By performing Fourier transformation, we also obtain
	\begin{align}
		\hat{g}\creation_{\sigma,\bm{k}}\hat{g}^{-1} &= \sum_{\sigma'}\creation_{\sigma',p_g\bm{k}}[U_{\bk}(g)]_{\sigma'\sigma}, 
	\end{align}
	where $U_{\bk}(g)$ is a unitary representation in momentum space.
	It should be emphasized that $U(g)$ and $U_{\bk}(g)$ are projective representations such that 
	\begin{align}
		U(g) U(g') &= z_{g,g'}U(gg'),\\
		U_{g'\bk}(g) U_{\bk}(g') &= z_{g,g'}U_{\bk}(gg'),
	\end{align}
	where $z_{g,g'}\in \text{U}(1)$ is a projective factor of $\calG$. For spinless systems, we can always choose $z_{g, g'} = +1$ for $g,g' \in \calG$.

	In this work, we always consider translation-invariant superconductors which can be described by the mean-field theory
	\begin{align}
		\label{eq:BdG}
		\hat{H} &= \hat{h}+\hat{\Delta} \\
		\hat{h} &= \sum_{\bm{R}, \delta\bm{R}}\sum_{\sigma, \sigma'} \left[\hat{c}^{\dagger}_{\sigma, \bm{R}+\delta\bm{R}} [h_{\delta\bm{R}}]_{\sigma \sigma'}\hat{c}_{\sigma', \bm{R}}\right]\nonumber\\
		&= \sum_{\bm{k}}\sum_{\sigma, \sigma'} \left[\hat{c}^{\dagger}_{\sigma, \bk} [h_{\bk}]_{\sigma \sigma'}\hat{c}_{\sigma', \bm{k}}  \right],\\
		\hat{\Delta} &=\sum_{\bm{R}, \delta\bm{R}}\sum_{\sigma, \sigma'} \left[ \hat{c}^{\dagger}_{\sigma, \bm{R}+\delta\bm{R}}[\Delta_{\delta\bm{R}}]_{\sigma \sigma'}\hat{c}^{\dagger}_{\sigma', \bm{R}} + \text{h.c.} \right]\nonumber\\
		&= \sum_{\bm{k}}\sum_{\sigma, \sigma'} \left[\hat{c}^{\dagger}_{\sigma, \bk}[\Delta_{\bk}]_{\sigma \sigma'}\hat{c}^{\dagger}_{\sigma', -\bm{k}} + \text{h.c.} \right],
	\end{align}
	where $\hat{h}$ and $\hat{\Delta}$ are the Hamiltonian in the normal phase and the superconducting order parameter.
	To hold $\hat{g}\hat{H}\hat{g}^{-1} = \hat{H}$, $U_{\bk}(g)$ must satisfy
	\begin{align}
		\label{eq:symm_rel}
		U_{\bk}(g)h_{\bk}U_{\bk}^{\dagger}(g) &= h_{p_g\bk}, \\
		U_{\bk}(g)\Delta_{\bk}U_{-\bk}^{T}(g) &= \chi_g \Delta_{p_g\bk}\ (\chi_g \in \text{U}(1)).
	\end{align}
	The set $\{\chi_g\}_{g\in \calG}$ represents the pairing symmetry. 
	In particular, when $\chi_g=+1$ for all $g\in\calG$, we say the pairing symmetry is conventional.
	
	It is convenient to introduce BdG Hamiltonian in momentum space
	\begin{align}
		%\label{eq:BdG_Ham}
		\hat{H}&\simeq (\hat{\bm{c}}_{\bk}^{\dagger}\  \hat{\bm{c}}_{-\bk} )H^{\text{BdG}}_{\bk} \begin{pmatrix}
			\hat{\bm{c}}_{\bk} \\
			\hat{\bm{c}}_{-\bk}^{\dagger} 
		\end{pmatrix}\\
		%(\hat{\bm{c}}_{\bk}\  \hat{\bm{c}}_{-\bk}^{\dagger} )^T\\
		H_{\bk}^{\text{BdG}}&=\begin{pmatrix}
			h_{\bk}& \Delta_{\bk} \\
			\Delta_{\bk}^{\dagger} & -h_{-\bm{k}}^{*}
		\end{pmatrix},\\
	U_{\bk}^{\text{BdG}}(g)&= \begin{pmatrix}
		U_{\bk}(g)& 0 \\
		0 & \chi_gU_{-\bk}^{*}(g)
	\end{pmatrix},
	\end{align}
	where $\hat{\bm{c}}_{\bk}^{\dagger} = (\creation_{1\bm{k}}, \creation_{2\bm{k}}, \cdots, \creation_{N_{\text{orb}}\bm{k}})$ and $U^{\text{BdG}}_{\bk}(g)H^{\text{BdG}}_{\bk}U^{\text{BdG}}{\bk}^{\dagger}(g) = H^{\text{BdG}}_{p_g\bk}. $
	In this expression, one can find an additional symmetry
	\begin{align}
		U(\calC) &= \begin{pmatrix}
			0& \mathds{1}\\
			\mathds{1} & 0
		\end{pmatrix},
	\end{align}
	where $U(\calC)$ is a unitary representation of PHS $\calC$ that satisfies $U(\calC)H_{\bk}^{*} = -H_{-\bk}U(\calC)$. 	
	In the presence of PHS $\calC$, the full symmetry group $G$ is decomposed into the following four parts
	\begin{align}
		G &= \calM + \calM \calC \nonumber\\
		&= \calG + \calA + \mathcal{P} + \mathcal{J},
	\end{align}
	where $\mathcal{P}=\calG\calC$ and $\mathcal{J}=\calA\calC$ are sets of particle-hole like and chiral like symmetries.
	In particular, for type II MSGs, $G = \calG + \calG\calT +  \calG\calC +  \calG\Gamma$ ($\Gamma$: chiral symmetry).
	
	\subsection{Cell decomposition}
	\label{app:cell}
	In the remaining subsections, we focus only on real space. 
	Here, we provide a formal discussion on \textit{cell decomposition}. We introduce a series of subspace of $\mathbb{R}^3$ such that
	\begin{align}
		X_0 \subset X_1 \subset \cdots \subset X_3 =\mathbb{R}^3,
	\end{align}
	where $X_p$ is a $p$-dimensional subspace of $\mathbb{R}^3$ and referred to as $p$-skelton. 
	Each $X_p$ is invariant under $G$, i.e., $g\bx \in X_p$ for $\forall g \in G$ if and only if $\bx \in X_p$. 
	Such a series is called \textit{$G$-symmetric filtration}. 
	We remark that the choice of filtration is not unique. 
	
	To obtain a $G$-symmetric filtration, we introduce a cell decomposition. 
	As discussed in the main text, we first find an asymmetric unit (AU) and decompose the asymmetric unit into the set of $p$-cells $\{D^{p}_{i}\}_i$.
	In the following, $D^{p}_{i}$ denote the $i$-th $p$-cell $(p=0,1,2,\text{and}\ 3)$. 
	Next, by taking copies of these $p$-cells throughout the entire 3D space, we define the entire set of $p$-cells by
	\begin{align}
		\calC_p := \bigcup_{i} \bigcup_{g \in G} D_{g(i)}^{p},
	\end{align}
	where $D_{g(i)}^{p} = gD^{p}_{i}$.
	
	In this construction, each $p$-cell satisfies the following conditions~\cite{K-AHSS,R-AHSS_Shiozaki,Ono-Shiozaki2022}: 
	\begin{enumerate}
		\setlength{\itemsep}{-2pt}
		\item[(i)] Any two $p$-cells in $\calC_p$ do not overlap.
		\item[(ii)] Any point in a $p$-cell $D_{i}^{p}$ is invariant under symmetries or transformed into points in different $p$-cells by symmetries.
		\item[(iii)] For $p\geq 1$, the boundary of a $p$-cell $D_{i}^{p}$, denoted by $\partial D_{i}^{p}$, is composed of $(p-1)$-cells.
	\end{enumerate}
	Also, we define orientations of $p$-cell ($p\geq 1$) in a symmetric manner.
	Finally, the $p$-skelton $X_p$ is obtained by
	\begin{align}
		X_0 = \calC_0, X_p= \calC_p \cup X_{p-1}\ \ (p \geq 1).
	\end{align}
	
	\subsection{$E^1$-pages: Building blocks of topological superconductors}
	\label{app:E1}
	As discussed in Ref.~\cite{R-AHSS_Shiozaki}, $E^1$-pages are defined by $K$-homology as
	\begin{align}
		\label{eq:E1}
		E_{p, -n}^{1} &:= K_{p-n}^{G}(X_p , X_{p-1}),\\
		&= \prod_{j} K_{p-n}^{G_{D_{j}^{p}}}\left(D_{j}^p, \partial D_{j}^p\right).
	\end{align}
	The physical meaning of $E_{p, -n}^{1}$ is as follows.
	\begin{enumerate}
		\item[(1)] $E_{p, -p}^{1}$ is an abelian group whose entries correspond to $p$-dimensional topological superconductors defined on $p$-cells.
		\item[(2)] $E_{p-1, -p}^{1}$ is an abelian group whose entries correspond to $(p-1)$-dimensional boundary modes of $p$-dimensional topological superconductors. 
		\item[(3)] $E_{p+1, -p}^{1}$ is an abelian group whose entries correspond to generating processes of $p$-dimensional topological superconductors on $(p+1)$-cells. 
	\end{enumerate}
	To obtain $E^1$-pages, we have to (i) identify a subgroup of $G$ for each $p$-cell and (ii) determine effective Altland–Zirnbauer symmetry (EAZ) classes on $p$-cells.
	
	Let us begin by discussing (i). 
	For a point $\bx$, one can define a subgroup $\calG_{\bx}$ of $\calG$ by
	\begin{align}
		\mathcal{G}_{\bx} = \{g \in  \mathcal{G}\ |\ g\bx = \bx\}.
	\end{align}
	Similarly, one can also find a subset of $\calA, \calP$, and $\calJ$ as
	\begin{align}
		\mathcal{A}_{\bx} &= \{a \in  \mathcal{A}\ |\ a\bx = \bx\},\\
		\mathcal{P}_{\bx} &= \{c \in  \mathcal{P}\ |\ c\bx = \bx\},\\
		\mathcal{J}_{\bx} &= \{\gamma \in  \mathcal{J}\ |\ \gamma\bx = \bx\},
	\end{align}
	which result in a subgroup $G_{\bx} = \calG_{\bx} + \calA_{\bx} + \calP_{\bx} + \calJ_{\bx}$ of $G$.
	This subgroup $G_{\bx}$ is called \text{site-symmetry group}. 
	In our construction of the cell decomposition, the site-symmetry group $G_{\bx}$ at any point $\bx$ in a $p$-cell $D^p$ are in common, and then the common site-symmetry group is denoted by $G_{D^p} = \calG_{D^p} + \calA_{D^p} + \calP_{D^p} + \calJ_{D^p}$. 
	%In the same way as $\calG_{D^p}$, we define a subset $\mathcal{V}_{D^p}$ of $\mathcal{V}$ by $\mathcal{V}_{D^p} = \{g \in  \mathcal{V}| g\bx = \bx \ \text{for}\ \forall \bx \in D^p\}$, where $\mathcal{V}=\calA, \mathcal{P}, \mathcal{J}$. 

	Next, we move on to the discussions on (2). For a given the site-symmetry group $G_{D^p}$ on a $p$-cell $D^p$, we identify EAZ classes by the Wigner criteria~\cite{Bradley,K-AHSS}
	\begin{align}
		\label{eq:wigner_C}
		W^{\alpha}_{D^p}(\mathcal{P}) &=\frac{1}{\vert \mathcal{P}_{\bx}/T \vert}\sum_{c \in \calP_{\bk}/T }z_{c, c}\chi_{\bx}^{\alpha}(c^2) \in \{0, \pm 1\},\\
		W^{\alpha}_{D^p}(\calA) &=\frac{1}{\vert \calA_{\bx}/T \vert}\sum_{a \in \calA_{\bx}/T }z_{a, a}\chi_{\bx}^{\alpha}(a^2) \in \{0, \pm 1\},\\
		\label{eq:wigner_G}
		W^{\alpha}_{D^p}(\mathcal{J}) &= \frac{1}{\vert \calG_{\bx}/T \vert} \sum_{g \in \calG_{\bx}/T } \frac{z_{\gamma, \gamma^{-1}g\gamma}}{z_{g, \gamma}} [\chi^{\alpha}_{\bx}(\gamma^{-1} g \gamma)]^{*}\chi^{\alpha}_{\bx}(g)\\
		&\in\{0,1\}\nonumber.
	\end{align}
	Here,
	\begin{align}
		\label{eq:char_irrep}
		\chi_{\bx}^{\alpha}(g) =\mathrm{tr}[u_{\bx}^{\alpha}(g)]\ (g\in \calG_{\bx})
	\end{align} 
	is a character of an irreducible representation at $\bx \in D^p$,
	and $\gamma$ is a chiral like symmetry. Note that, in fact, it is enough for our purpose to consider a point $\bx$ in $D^p$. $K_{p-n}^{G_{D^{p}}}\left(D^p, \partial D^p\right)$ is found from the EAZ classes~\cite{R-AHSS_Shiozaki}.

	For conventional pairing symmetries in type II MSGs, we find that $(W^{\alpha}_{D^p}(\calA) , W^{\alpha}_{D^p}(\mathcal{P}), W^{\alpha}_{D^p}(\mathcal{J}))$ is always either $(\pm1,\mp1, 1)$ or $(0,0,1)$. That is, EAZ classes belong to one of DIII, CI, or AIII. Consequentially, $E_{0,0}^{1} = 0$.  
	
	\subsection{First differential $d^1$}
	Once we have $E^1$-pages, the next step is to find inequivalent boundary-gapped patchworks, denoted by $E_{p, -p}^{2}$.
	To achieve this, we consider the following two processes. 
	For $p\geq 1$, when we place $p$-dimensional TSCs on $p$-cells, $(p-1)$-dimensional boundary modes emerge on their boundary $(p-1)$-cells.
	This process is implemented as a map from $E_{p, -p}^{1}$ to $E_{p-1, -p}^{1}$
	\begin{align}
		d^{1}_{p,-p}: E_{p, -p}^{1} \rightarrow E_{p-1, -p}^{1}.
	\end{align}
	Then, since the boundary gapped patchworks do not have any $(p-1)$-dimensional boundary mode as discussed in the main text, the solution space $\text{Ker}\ d^{1}_{p,-p}$ is a set of boundary gapped patchworks. 

	For $p \leq 2$, we obtain $p$-dimensional TSCs on $p$-cells generated from $(p+1)$-cells of the vacuum. This is also described by a map from $E_{p+1, -p}^{1}$ to $E_{p, -p}^{1}$
	\begin{align}
		d^{1}_{p+1,-p}: E_{p+1, -p}^{1} \rightarrow E_{p, -p}^{1}.
	\end{align}
	As discussed in the main text, we identify boundary-gapped patchworks that can be deformed into each other, where the difference between them is an entry of $\text{Im}\ d^{1}_{p+1,-p}$. 
	Then, we obtain all inequivalent boundary-gapped patchworks $E_{p, -p}^{2}$ by
	\begin{align}
		E_{p, -p}^{2} := \text{Ker}\ d^{1}_{p,-p}/ \text{Im}\ d^{1}_{p+1,-p}.
	\end{align}

	On the other hand, $\text{Im}\ d^{1}_{p,-p}$ represents $(p-1)$-dimensional boundary modes obtained by placing $p$-dimensional TSCs on $p$-cells. 
	%Then, we find $(p-1)$-dimensional boundary modes that cannot be obtained from $p$-dimensional TSCs on $p$-cells. 
	Then, for $p=0$ and $n = 1$,
	\begin{align}
		E_{0, -1}^{2} := E^{1}_{0,-1}/ \text{Im}\ d^{1}_{1,-1}
	\end{align}
	is an abelian group whose entries are Majorana zero modes at $0$-cells that cannot be obtained by 1D TSCs. 

	Also, $\text{Ker}\ d^{1}_{p+1,-p}$ represents the generating processes of $p$-dimensional TSCs that do not result in $p$-dimensional TSCs on $p$-cells. Then, for $p=3$ and $n = 2$, we define an abelian group by
	\begin{align}
		E_{3, -2}^{2} := \text{Ker}\ d^{1}_{3,-2},
	\end{align}
	whose entries are the generating processes of 2D TSCs but all of 2D TSCs are annihilated on $2$-cells.

	\subsection{Higher differential $d^r$}
	As is the case of $E^1$-pages, we define a map among $E^2$-pages by 
	\begin{align}
		d^{r}_{p,-p}: E_{p, -p}^{r} \rightarrow E_{p-r, -p-r+1}^{1}.
	\end{align}
	In the following, we discuss only $d_{2,-2}^{2}, d_{3,-2}^{2}, d_{3,-3}^{2}$, and $d_{3,-3}^{3}$, which appeared in our classifications for conventional pairing symmetries. 
	
	\subsubsection{$d_{2,-2}^{2}$}
	As discussed in the main text and Appendix~\ref{app:vortex}, when we place 2D $\mZ_2$-TSCs on a plane, crystalline symmetries sometimes enforce the emergence of vorticies with odd-integer winding. As a result, vortex zero modes appear at the vortex cores. 
	This process is implemented by a map from $E_{2, -2}^{2}$ to $E_{0, -1}^{2}$
	\begin{align}
		d^{2}_{2,-2}: E_{2, -2}^{2} \rightarrow E_{0, -1}^{2}.
	\end{align}
	Therefore, $E_{2, -2}^{\infty} := \text{Ker}\ E_{2, -2}^{2}$ is an abelian group whose entries are all inequivalent fully gapped patchworks constructed by 2D TSCs. 
	
	\subsubsection{$d_{3,-2}^{2}$}
	\label{app:d232}
	%In this subsection, we discuss further trivialization of $E_{1, -1}^{2}$. 
	In general, elements in $E_{1, -1}^{2}$ are subject to further trivialization originating from $E_{3, -2}^{2}$. Although generating processes of 2D TSCs from the vacuum, corresponding to entries of $E_{3, -2}^{2}$, do not bring nontrivial 2D TSCs on $2$-cells, they can carry nontrivial 1D TSCs on $1$-cells.
	This process is represented by a map
	\begin{align}
		d^{2}_{3,-2}: E_{3, -2}^{2} \rightarrow E_{1, -1}^{2}.
	\end{align}
	To obtain all inequivalent fully gapped patchworks, we subtract $\text{Im}\ d^{2}_{3,-2}$ from $E_{1, -1}^{2}$.
	Then, 
	\begin{align}
		E_{1, -1}^{\infty} := E_{1, -1}^{2}/\text{Im}\ d^{2}_{3,-2}
	\end{align}
	is an abelian group of all inequivalent fully gapped patchworks composed of 1D TSCs.

	However, for the conventional pairing symmetries in which we are interested, we find that $d^{2}_{3,-2}$ is always trivial. This can be understood as follows. 
	As defined in Eq.~\eqref{eq:E1}, we have
	\begin{align}
		E_{3, -2}^{1} &:= K_{1}^{G}(X_3 , X_2).
	\end{align}
	From the $K$-theory point of view, this definition implies that the generating processes can be regarded as topological phases with a chiral symmetry in addition to $G$. For the conventional pairing symmetries, the symmetry class with a chiral symmetry in addition to $G$ is equivalent to class AII in which all crystalline symmetries commute with TRS. It is well known that topological insulators are compatible with all space groups, and thus any entry of $E_{3, -2}^{1}$ does not carry any nontrivial TSC on $1$-cells and $2$-cells. As a result, we find that $E_{1, -1}^{\infty} = E_{1, -1}^{2}$ due to $\text{Im}\ d^{2}_{3,-2} = 0$.
	As a sanity check, we explicitly confirmed that $E_{3, -2}^{2} = E_{3, -2}^{1}$ and $d_{3,-2}^{1}$ was trivial in all space groups.

	\subsubsection{$d_{3,-3}^{2}$ and $d_{3,-3}^{3}$}
	Here, we discuss 3D TSCs protected by chiral symmetry. The topological invariant for the 3D TSC is 3D winding number $\nu_{\text{w}}$. 
	It is known that $\nu_{\text{w}}$ is transformed as $\nu_{\text{w}} = \det p_g \chi_g \nu_{\text{w}}$, which indicates that the topological invariant is trivial when $\det p_g \neq \chi_g$. We actually find that $E_{3,-3}^{2} = 0$ for space groups that contain at least an element such that $\det p_g \neq \chi_g$.

	Then, in the following, we focus only on the case where $\det p_g = \chi_g$ for $\forall g \in \calG$.
	We argue that $d_{3,-3}^{2}$ and $d_{3,-3}^{3}$ are always trivial for the case. 
	This can be justified as follows. 
	Let us start with the Hamiltonian of $^3$He
	\begin{align}
		\label{eq:3He}
		H_{\bk} &= \left(\frac{k^2}{2m} - \mu\right)\tau_z + (\bk \cdot \bm{\sigma} (i\sigma_y))\tau_x,
	\end{align}
	where $\sigma_i$ and $\tau_i$ are Pauli matrices. This Hamiltonian is continuous translational invariant and invariant under any element in $\text{O}(3)$, i.e., invariant under the three-dimensional Euclidean group ${\rm E}(3)$. 
	Any space group can be obtained by a subgroup of ${\rm E}(3)$,} the existence of 3D TSCs is compatible with any space group.
In summary, 
\begin{align}
E_{3,-3}^{\infty} &= \begin{cases}
	0 \quad (\exists g \in \calG\text{ s.t. } \det p_g \neq \chi_g)\\
	\mathbb{Z} \quad (\det p_g = \chi_g\ \text{for } \forall g \in \calG)
\end{cases}.
\end{align}
	
%{Thus, the existence of 3D TSCs is compatible with any point group.
%Furthermore, for any space group, the symmetry group at $\Gamma = (0,0,0)$ in momentum space is always equivalent to the point group of the space group. 
%	This indicates that, when a band inversion occurs at $\Gamma$, the system can be described by Eq.~\eqref{eq:3He}. 
%	That is, }{Any space group can be obtained by a subgroup of ${\rm E}(3)$,} the existence of 3D TSCs is compatible with any space group.
%	In summary, 
%	\begin{align}
%		E_{3,-3}^{\infty} &= \begin{cases}
%			0 \quad (\exists g \in \calG\text{ s.t. } \det p_g \neq \chi_g)\\
%			\mathbb{Z} \quad (\det p_g = \chi_g\ \text{for } \forall g \in \calG)
%		\end{cases}.
%	\end{align}

%	\begin{align}
%		\nu_{\text{w}} = \frac{1}{48\pi^2}\int d^3\bk \epsilon^{ijk}\mathrm{tr}\left[\Gamma \left(H^{-1}_{\bk}\partial_i H_{\bk}\right) \left(H^{-1}_{\bk}\partial_j H_{\bk}\right) \left(H^{-1}_{\bk}\partial_k H_{\bk}\right) \right]
%	\end{align}

	\subsection{Group extension}
	In general, $\oplus_{p=0}^3E_{p,-p}^{\infty}$ does not reproduce the $K$-group result, because $E_{p,-p}^{\infty}$ $(p=0, 1,2,3)$ are not mutually independent.
	To obtain the final classification, we have to solve the group extension problem that is expressed by the following short exact sequences
	\begin{align}
			0 \rightarrow E_{0,-0}^{\infty} \rightarrow F_{1}K_0 \rightarrow  E_{1,-1}^{\infty}\rightarrow  0;\\
			\label{eq:ext_2}
			0 \rightarrow F_{1}K_0  \rightarrow   F_{2}K_0 \rightarrow  E_{2,-2}^{\infty}\rightarrow  0;\\
			0 \rightarrow F_{2}K_0 \rightarrow K_{0}^{G}(\mathbb{R}^3)  \rightarrow  E_{3,-3}^{\infty}\rightarrow  0,
	\end{align}
	or equivalently,
	\begin{align}
		F_{1}K_0 /E_{0,-0}^{\infty} \simeq  E_{1,-1}^{\infty};\\
		\label{eq:ext_2-2}
		F_{2}K_0/F_{1}K_0 \simeq  E_{2,-2}^{\infty};\\
		K_{0}^{G}(\mathbb{R}^3)/F_{2}K_0  \simeq E_{3,-3}^{\infty}.
	\end{align}
	Since $E_{0,-0}^{\infty} = 0$ and $E_{3,-3}^{\infty} = 0$ or $\mZ$ for conventional pairing symmetries, $ F_{1}K_0 \simeq E_{1,-1}^{\infty}$ and $K_{0}^{G}(\mathbb{R}^3) \simeq F_{2}K_0\oplus E_{3,-3}^{\infty}$. 
	Thus, the remaining task is to solve Eq.~\eqref{eq:ext_2}.
	Although there are usually multiple possibilities, the correct one can be identified by physical considerations.
	We will see how to find physical $F_{2}K_0$ in Eq.~\eqref{eq:ext_2} through some examples.

	\clearpage
	\section{Detailed calculations of AHSS for some examples}
	In this section, we show computation processes of AHSS in several examples. 
	In the following, we always consider conventional pairing symmetries.
	\subsection{layer group $p\bar{1}$}
	\label{app:p-1}
	We discuss layer group $p\bar{1}$, whose generator are translations and inversion symmetry $I:(x,y,z)\rightarrow (-x,-y,-z)$.

	\subsubsection{Cell decomposition and building blocks}
\begin{figure}[b]
		\begin{center}
			\includegraphics[width=0.8\columnwidth]{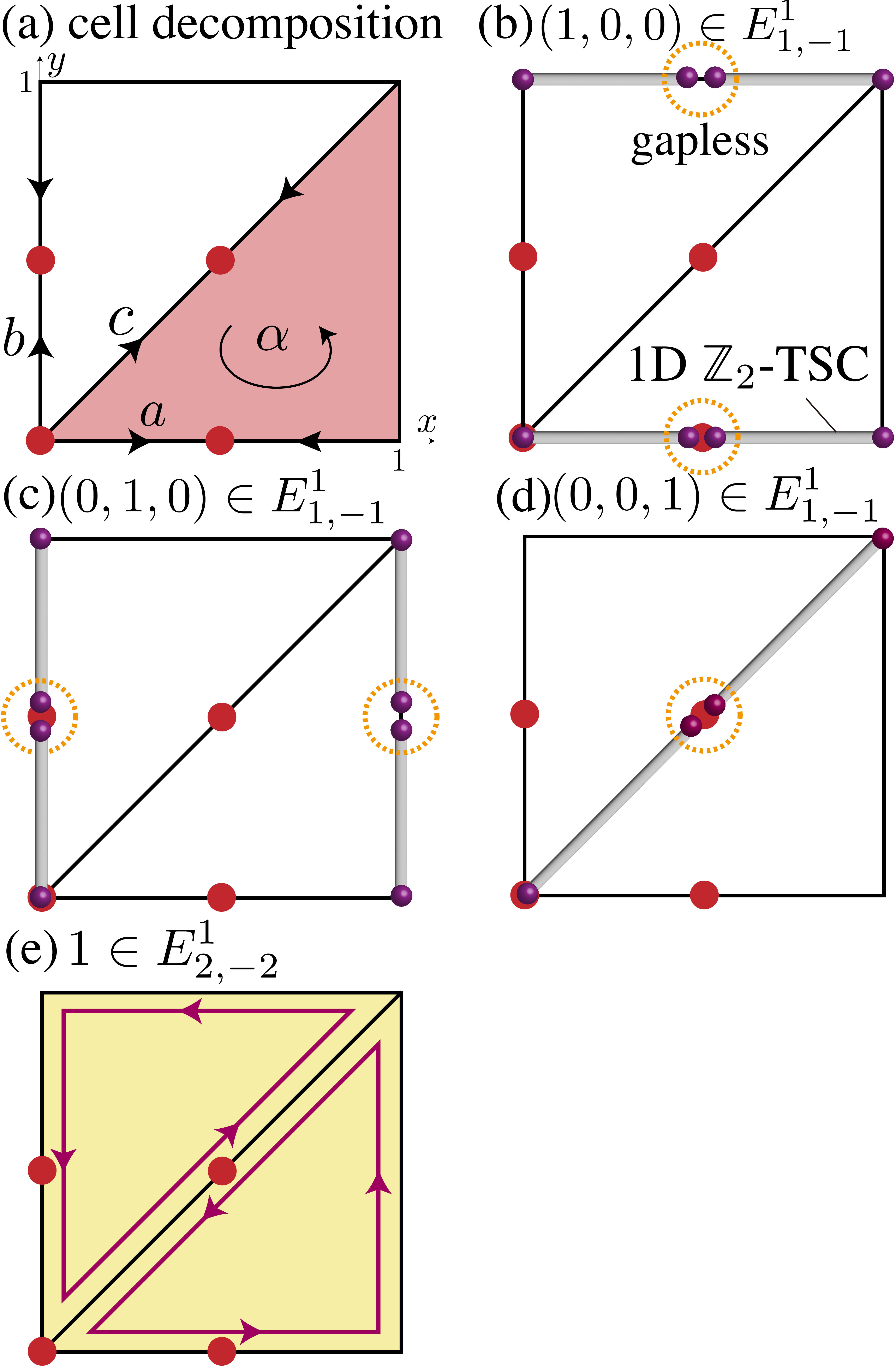}
			\caption{\label{app:fig_p-1}\textbf{Illustration of topological patchworks in $p\bar{1}$.}
				(a) Cell decomposition. 
				(b)-(e) Generators of $E_{1,-1}^{1}$ and $E_{2,-2}^{1}$. 
			}
		\end{center}
	\end{figure}

	Our cell decomposition is shown in Fig.~\ref{app:fig_p-1} (a). The shaded region in Fig.~\ref{app:fig_p-1} (a) is an AU. 
	Note that all boundary line segments of the AU should be divided into two $1$-cells since a point in a line segment is symmetry-related to another point in the same line segment.
	As a result, there are inequivalent a $2$-cell, three $1$-cells, and four $0$-cells. 
	
	Next, we identify EAZ classes for each $p$-cell. 
	For each of $0$-cells denoted by $\bx = (a/2, b/2)$ with $a,b=0\text{ or }1$, the site-symmetry group $G_{\bx} = \calG_{\bx} + \calA_{\bx} + \calP_{\bx} + \calJ_{\bx}$ is
	\begin{align}
		\calG_{\bx} &= \{e, IT_{x}^{a}T_{y}^{b}\};\\
		\calA_{\bx} &= \{\calT, IT_{x}^{a}T_{y}^{b}\calT\};\\
		\calP_{\bx} &= \{\calC, IT_{x}^{a}T_{y}^{b}\calC\};\\
		\calJ_{\bx} &= \{\Gamma, IT_{x}^{a}T_{y}^{b}\Gamma\},
	\end{align}
	where $T_{x}$ and $T_{y}$ are translations along $x$- and $y$-directions. 
	Since $z_{IT_{x}^{a}T_{y}^{b}, IT_{x}^{a}T_{y}^{b}} = -z_{\calT, \calT} = -z_{IT_{x}^{a}T_{y}^{b}\calT, IT_{x}^{a}T_{y}^{b}\calT} =  z_{\calC, \calC} =  z_{ IT_{x}^{a}T_{y}^{b}\calC,  IT_{x}^{a}T_{y}^{b}\calC} = z_{IT_{x}^{a}T_{y}^{b}, \Gamma} = z_{\Gamma, IT_{x}^{a}T_{y}^{b}} = +1$, we find that, for $\chi_{\bx}^{\pm}(e) = +1$ and $\chi_{\bx}^{\pm}(IT_{x}^{a}T_{y}^{b}) = \pm 1$ in Eq.~\eqref{eq:char_irrep},
	\begin{align}
		W^{\pm}_{\bx}(\mathcal{P}) &= \frac{1}{2}\left( 1 + 1 \right) = 1,\\
		W^{\pm}_{\bx}(\mathcal{A}) &= \frac{1}{2}\left( -1 -1 \right) = -1,\\
		W^{\pm}_{\bx}(\mathcal{J}) &= \frac{1}{2}\left( 1 + 1 \right) = 1.
	\end{align}
	This indicates that EAZ classes for each parity sector are class DIII. Therefore, $E_{0,0}^1 = 0.$
	We also have $E_{0,-1}^{1} = (\mZ_2)^8$ that is the set of Kramers pair of Majorana zero modes characterized by inversion parity at each $0$-cell.
	
	For each of $1$- and $2$-cells, the site-symmetry group is
	\begin{align}
		\calG_{\bx} &= \{e\};\\
		\calA_{\bx} &= \{\calT\};\\
		\calP_{\bx} &= \{\calC\};\\
		\calJ_{\bx} &= \{\Gamma\}.
	\end{align}
	Then, EAZ classes for them are class DIII.
	Thus, the building block TSCs on $1$-cells are 1D $\mZ_2$-TSCs, and those on $2$-cells are 2D $\mZ_2$-TSCs, which results in $E^1$-pages in Table~\ref{app:E1_p-1}.
	In Fig.~\ref{app:fig_p-1} (b)-(e), we show generators of $E_{1,-1}^{1}$ and $E_{2,-2}^{1}$.
	Also, $E_{2,-1}^{1} = \mZ_2$ and $E_{1,-2}^{1} = (\mZ_2)^3$, whose entries are generating processes of 1D $\mZ_2$-TSCs on $2$-cells and helical edge modes on $1$-cells, respectively.  
	\begin{table}[H]
		\begin{center}
			\caption{\label{app:E1_p-1}$E^1$-pages for layer group $p\bar{1}$. Here, $\star$ denotes that $E^1$-pages are not involved in the  classification.}
			\begin{tabular}{c|c|c|c}
				$n=0$ & $0$ & $\star$ & $\star$\\
				$n=1$ & ($\mZ_2)^8$ & $(\mZ_2)^3$& $\mZ_2$\\
				$n=2$ & $\star$ & $(\mZ_2)^3$& $\mZ_2$\\
				\hline
				$E_{p,-n}^{1}$ & $p=0$ & $p=1$ & $p=2$
			\end{tabular}
		\end{center}
	\end{table}
	
	\subsubsection{first differential}
	Let us start with $d_{1,-1}^{1}: E_{1,-1}^{1} \rightarrow E_{0,-1}^{1}$. 
	Recall that $E_{0,-1}^{1} = (\mZ_2)^8$ is the set of a single Majorana zero mode characterized by inversion parity at each $0$-cell.
	When we place a 1D $\mZ_2$-TSC on a $1$-cell and its inversion copy [see Fig.~\ref{app:fig_p-1}(b-d)], we find Majorana zero modes with positive and negative parities at $0$-cells of the boundaries of the 1D $\mZ_2$-TSCs.
	Thus, we get
	\begin{align}
		d_{1,-1}^{1} &= \begin{pmatrix}
			1 &  1& 1 \\
			1 &  1& 1 \\
			1 &  0& 0 \\
			1 &  0& 0 \\
			0 &  1& 0 \\
			0 &  1& 0 \\
			0 &  0& 1 \\
			0 &  0& 1 
		\end{pmatrix}
	\end{align}
	and 
	\begin{align}
		\text{Im }d_{1,-1}^{1} &= \text{span}\left\{\begin{array}{cccccccc}
			\ba^{(1)}_1 = (1 & 1 & 1 & 1 & 0 & 0 & 0 & 0)^T \\
			\ba^{(1)}_2 = (1 & 1 & 0 & 0 & 1 & 1 & 0 & 0)^T \\
			\ba^{(1)}_3 = (1 & 1 & 0 & 0 & 0 & 0 & 1 & 1)^T \\
		\end{array}\right\},\\
		E_{0,-1}^{2} &= E_{0,-1}^{1}/\text{Im }d_{1,-1}^{1} \simeq (\mZ_2)^5\\
		&= \text{span}\left\{\begin{array}{cccccccc}
			\bb^{(1)}_1 = (1 & 0 & 0 & 0 & 0 & 0 & 0 & 0)^T \\
			\bb^{(1)}_2 =(0 & 1 & 0 & 0 & 0 & 0 & 0 & 0)^T \\
			\bb^{(1)}_3 =(0 & 0 & 1 & 0 & 0 & 0 & 0 & 0)^T \\
			\bb^{(1)}_4 =(0 & 0 & 0 & 0 & 1 & 0 & 0 & 0)^T \\
			\bb^{(1)}_5 =(0 & 0 & 0 & 0 & 0 & 0 & 1 & 0)^T \\
		\end{array}\right\}.
	\end{align}
	As discussed in the previous section, $E_{0,-1}^{2}$ is an abelian group whose entries are Kramers pairs of Majorana zero modes that cannot be obtained by placing 1D $\mZ_2$-TSCs. 
	
	%We then ask if the two Majorana-Kramers zero modes at each inversion center can be gapped out. Indeed, this patchwork is gapless since inversion parities of Majorana-Kramers zero modes are different. 
	
	Next, we consider $d_{2,-1}^{1}: E_{2,-1}^{1} \rightarrow E_{1,-1}^{1}$.
	This process can be interpreted as the generating process of 1D $\mZ_2$-TSCs from a vacuum on $2$-cells [see Fig.~1(c) in the main text].
	When we generate 1D TSCs from a vacuum on $2$-cells, two 1D $\mZ_2$-TSCs always meet on each 1-cell.
	Thus, any element in $E_{1,-1}^{1}$ cannot be generated from a vacuum, i.e., $d_{2,-1}^{1} = (0, 0)^T$.
	Also, we get $E_{1,-1}^{2} = \text{Ker }d_{1,-1}^{1}/\text{Im }d_{2,-1}^{1} = 0$, which means there are no fully-gapped patchworks constructed by 1D TSCs.
	
	The similar thing happens in the computation $d_{2,-2}^{1}: E_{2,-2}^{1} \rightarrow E_{1,-2}^{1}$. Two helical edge modes always meet on 1-cells when we put 2D $\mZ_2$-TSCs on $\alpha$, and therefore $d_{2,-2}^{1} = (0, 0)^T$ and $E_{2,-2}^{2} = \text{Ker }d_{2,-2}^{1} = E_{2,-2}^{1} = \mZ_2$.
    $E^2$-pages are summarized in Table~\ref{app:E2_p-1}.
	
	\begin{table}[H]
		\begin{center}
			\caption{\label{app:E2_p-1}$E^2$-pages for layer group $p\bar{1}$. Here, $\star$ denotes that $E^2$-pages are not involved in the  classification.}
			\begin{tabular}{c|c|c|c}
				$n=0$ & $0$ & $\star$ & $\star$\\
				$n=1$ & ($\mZ_2)^5$ & $0$& $\mZ_2$\\
				$n=2$ & $\star$ & $\star$& $\mZ_2$\\
				\hline
				$E_{p,-n}^{2}$ & $p=0$ & $p=1$ & $p=2$
			\end{tabular}
		\end{center}
	\end{table}

	\subsubsection{second differential}
\begin{figure}[t]
		\begin{center}
			\includegraphics[width=0.8\columnwidth]{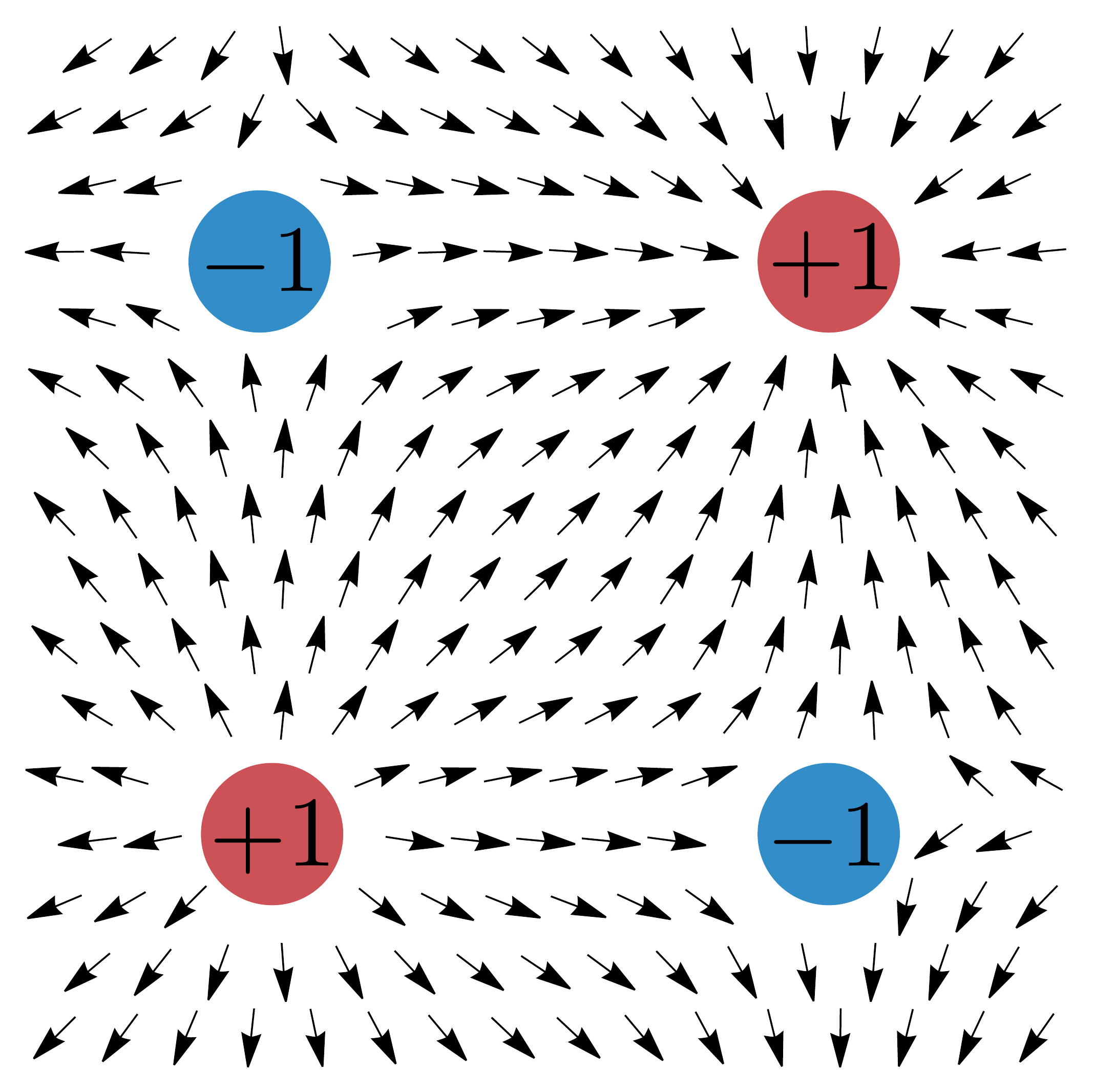}
			\caption{\label{fig:vortex_p-1}\textbf{A configuration of vortex zero modes in layer group $p\bar{1}$.} The vortex zero modes appear at inversion centers $(0,0), (1/2,0), (0,1/2)$, and $(1/2,1/2)$. Here, arrows denote the phase factor of $\Delta(\bm{r}) = \vert \Delta_0 (r) \vert e^{\pm i \alpha}e^{i \pm n \theta}$. While $n=+1\ (m=1)$ at $(0,0)$ and $(1/2,1/2)$, $n=-1\ (m=0)$ at $(1/2,0)$ and $(0,1/2)$.
			}
		\end{center}
	\end{figure}

	We discuss $d_{2,-2}^{2}:E_{2,-2}^{2}\rightarrow E_{0,-1}^{2}$, i.e.,
	we check if the entry $1\in E_{2,-2}^{2} = \mZ_2$ has to contain vortex zero modes.
	To do so, we consider the Hamiltonian $H_{\text{DIII}}(\bm{r}; \Delta)$~\eqref{eq:DIII-SC_app} with the inversion symmetry $U(I) = \tau_0s_0$.
	As discussed in Sec.~\ref{app:vortex}, the presence of even-parity inversion symmetry enforces emergence of vortex zero modes.
	A possible configuration of vortices is shown in Fig.~\ref{fig:vortex_p-1}. 
	As discussed in Sec.~\ref{app:vortex}, the inversion parities of a vortex is $(-1)^m$ $(m \text{ characterizes its vorticity by} 2m-1)$.
	For the vortex configuration, $m=1$ for vortices at $(0,0)$ and  $(1/2,1/2)$; $m=0$ for vortices at $(1/2,0)$ and  $(0,1/2)$.
	Then, $\bb^{(2)} \in E_{2,-2}^{2}$ is mapped 
	\begin{align}
		& d_{2,-2}^{2}(\bb^{(2)}) = \bb^{(1)}_1 + \bb^{(1)}_3 + \bb^{(1)}_4 + \bb^{(1)}_5 + \ba^{(1)}_3.
	\end{align}
	This implies that
	\begin{align}
		d_{2,-2}^{2} &= \begin{pmatrix}
			1 \\
			0 \\
			1 \\
			1 \\
			1 
		\end{pmatrix},\\
		E_{2,-2}^{\infty} &=\text{Ker }d_{2,-2}^{2} = 0.
	\end{align}
	It should be noted that the matrix form of $d_{2,-2}^{2}$ is the same for any vortex configuration. 
	This is because the difference of vortex configurations can eliminate by attaching 1D $\mZ_2$-TSCs, i.e., the linear combinations of $\im d_{1,-1}^{1}$, as shown in the above.
	$E^\infty$-pages are summarized in Table~\ref{app:E3_p-1}.
	Finally, we see that there are no topological phases in $p\bar{1}$.
	
	\begin{table}[H]
		\begin{center}
			\caption{\label{app:E3_p-1}$E^\infty$-pages for layer group $p\bar{1}$. Here, $\star$ denotes that $E^\infty$-pages are not involved in the  classification.}
			\begin{tabular}{c|c|c|c}
				$n=0$ & $0$ & $\star$ & $\star$\\
				$n=1$ & ($\mZ_2)^4$ & $0$& $\mZ_2$\\
				$n=2$ & $\star$ & $\star$& $0$\\
				\hline
				$E_{p,-n}^{\infty}$ & $p=0$ & $p=1$ & $p=2$
			\end{tabular}
		\end{center}
	\end{table}

	\subsection{layer group $pb$}
	\label{app:pb}
	Next, we discuss layer group $pb$, which is generated by glide symmetry $G_x:(x,y,z)\rightarrow (-x,y+1/2,z)$. 
	\subsubsection{Cell decomposition and building blocks}
	Our cell decomposition is shown in Fig.~\ref{app:fig_pb} (a). 
	An AU is the shades region in Fig.~\ref{app:fig_pb} (a). 
	The interior of AU is a $2$-cell (denoted by $\alpha$).
	Boundary line segments of $2$-cells are $1$-cells, and endpoints of $1$-cells are $0$-cells, where two of $1$-cells (denoted by $a$ and $b$) and one of $0$-cells are inequivalent cells.
	For all cells, the site symmetry groups are common
		\begin{align}
		\calG_{\bx} &= \{e\};\\
		\calA_{\bx} &= \{\calT\};\\
		\calP_{\bx} &= \{\calC\};\\
		\calJ_{\bx} &= \{\Gamma\}.
	\end{align}
	EAZ classes for them are class DIII.
	Then, there are three building block TSCs:(i) 2D $\mZ_2$-TSC on $\alpha$; (ii) 1D $\mZ_2$-TSC on $a$; (iii) 1D $\mZ_2$-TSC on $b$, which results in 
	$E_{p,-p}^{1}$ in Table~\ref{app:E1_pb}.
	As is the case of $p\bar{1}$, we find that $E_{0,-1}^{1} = \mZ_2$, $E_{2,-1}^{1} = \mZ_2$, and $E_{1,-2}^{1} = (\mZ_2)^2$.
	$E^1$-pages are summarized in Table~\ref{app:E1_pb}.
	\begin{table}[H]
		\begin{center}
			\caption{\label{app:E1_pb}$E^1$-pages for layer group $pb$. Here, $\star$ denotes that $E^1$-pages are not involved in the  classification.}
			\begin{tabular}{c|c|c|c}
				$n=0$ & $0$ & $\star$ & $\star$\\
				$n=1$ & $\mZ_2$ & $(\mZ_2)^2$& $\mZ_2$\\
				$n=2$ & $\star$ & $(\mZ_2)^2$& $\mZ_2$\\
				\hline
				$E_{p,-n}^{1}$ & $p=0$ & $p=1$ & $p=2$
			\end{tabular}
		\end{center}
	\end{table}

	\subsubsection{first differential}
	Let us begin by discussing $d_{1,-1}^{1}: E_{1,-1}^{1} \rightarrow E_{0,-1}^{1}$. 
	As discussed in Sec.~\ref{app:E1}, $E_{0,-1}^{1} = (\mZ_2)^2$ is the set of a single Majorana zero mode at the $0$-cell.
	Then, we ask how many Majorana zero modes appear when we place 1D $\mZ_2$-TSCs on $a$ and $b$.
	As shown in Fig.~\ref{app:fig_pb}, two Majoarana zero modes emerge at $O$, which can be gapped out. 
	Thus, $d_{1,-1}^{1} = (0, 0)$.
	
	The discussions on $d_{2,-1}^{1}$ and $d_{2,-2}^{1}$ for $pb$ are the completely same as those for $p\bar{1}$, i.e.,
	we have $d_{2,-1}^{1} = (0, 0)^T$ and $d_{2,-2}^{1} = (0, 0)^T$.
	Then, we get $E_{1,-1}^{2} = \text{Ker }d_{1,-1}^{1}/\text{Im }d_{2,-1}^{1} = E_{1,-1}^{1}$ and $E_{2,-2}^{2} = \text{Ker}d_{2,-2}^{1} = E_{2,-2}^{1}$. $E^2$-pages are summarized in Table~\ref{app:E2_pb}.

	\begin{table}[H]
		\begin{center}
			\caption{\label{app:E2_pb}$E^2$-pages for layer group $pb$. Again, $\star$ represents that $E^2$-pages are not involved in the  classification.}
			\begin{tabular}{c|c|c|c}
				$n=0$ & $0$ & $\star$ & $\star$\\
				$n=1$ & $\mZ_2$ & $(\mZ_2)^2$& $\mZ_2$\\
				$n=2$ & $\star$ & $\star$& $\mZ_2$\\
				\hline
				$E_{p,-n}^{2}$ & $p=0$ & $p=1$ & $p=2$
			\end{tabular}
		\end{center}
	\end{table}
	
	\subsubsection{second differential}
	We discuss $d_{2,-2}^{2}:E_{2,-2}^{2}\rightarrow E_{0,-1}^{2}$, i.e.,
	we check if the entry $1\in E_{2,-2}^{2} = \mZ_2$ has to have vortex zero modes.
	To do so, we again consider the Hamiltonian $H_{\text{DIII}}(\bm{r},\bk, \Delta)$~\eqref{eq:DIII-SC_app} with the glide symmetry
	\begin{align}
		U(G_x) &= is_x\tau_z
	\end{align}
	such that $U(G_x)H_{\text{DIII}}(\bm{r},\bk, \Delta) = H_{\text{DIII}}(G_x\bm{r}, (-k_x, k_y, k_z), \Delta^*)U(G_x)$.
	This implies that the phase of $\Delta(\bm{r})$ can be uniformly chosen as real. As a result, there does not exist any vortex, and $E_{2,-2}^{\infty}=\text{Ker }d_{2,-2}^{2} = E_{2,-2}^{2}$. 
    Finally, we have $E^\infty$-pages in Table~\ref{app:E3_pb}.
	
	\begin{table}[H]
		\begin{center}
			\caption{\label{app:E3_pb}$E^\infty$-pages for layer group $pb$. Again, $\star$ represents that $E^\infty$-pages are not involved in the  classification.}
			\begin{tabular}{c|c|c|c}
				$n=0$ & $0$ & $\star$ & $\star$\\
				$n=1$ & $\mZ_2$ & $(\mZ_2)^2$& $\mZ_2$\\
				$n=2$ & $\star$ & $\star$& $\mZ_2$\\
				\hline
				$E_{p,-n}^{\infty}$ & $p=0$ & $p=1$ & $p=2$
			\end{tabular}
		\end{center}
	\end{table}

	\subsubsection{Group extension}
\begin{figure}[t]
		\begin{center}
			\includegraphics[width=0.9\columnwidth]{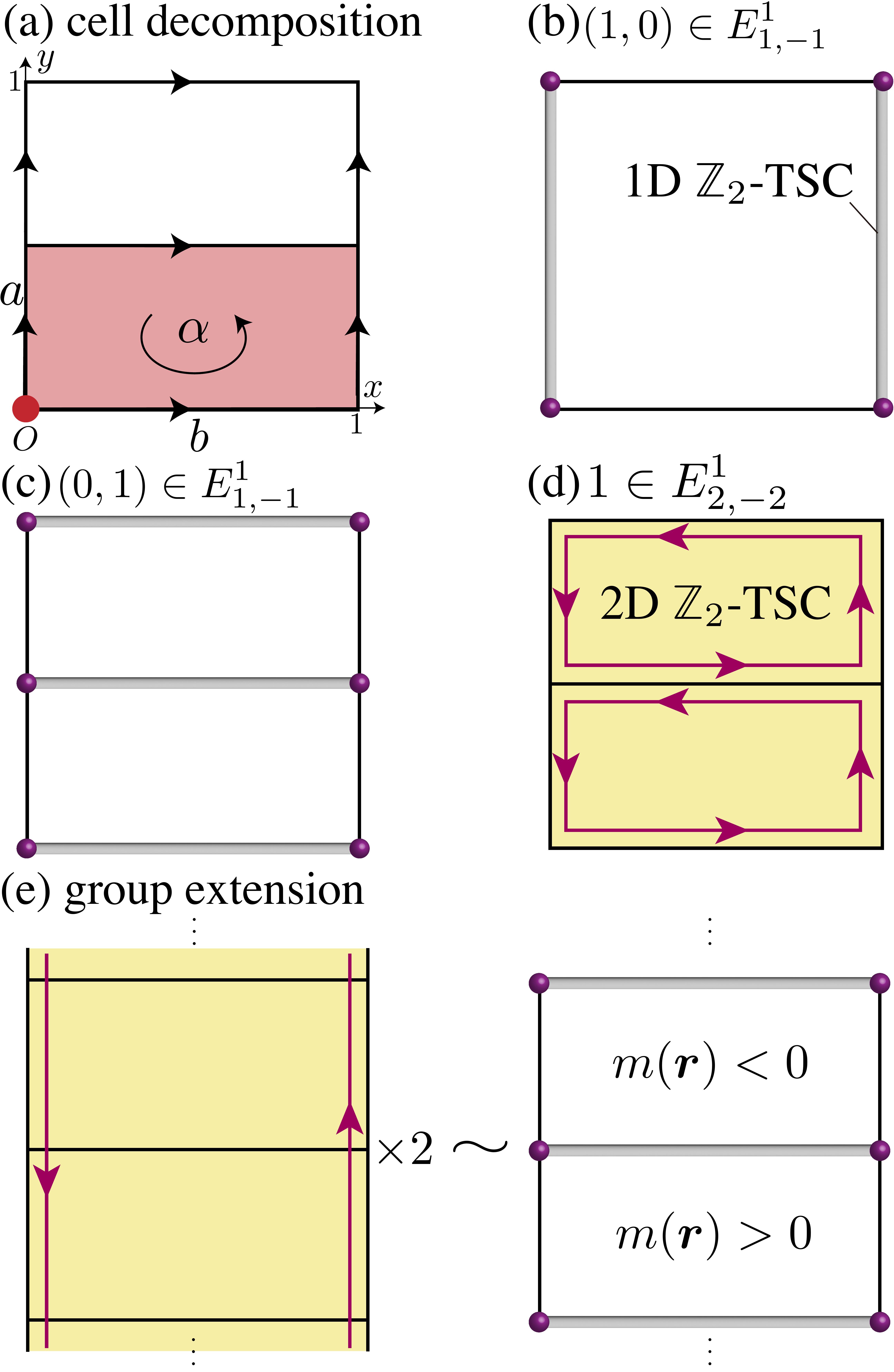}
			\caption{\label{app:fig_pb}\textbf{Illustration of topological patchworks in $pb$.}
				(a) Cell decomposition. 
				(b)-(d) All boundary-gapped patchworks, which are generators of $ E_{1,-1}^{\infty}=E_{1,-1}^{1}$ and $E_{2,-2}^{\infty}=E_{2,-2}^{1}$. 
				(e) Relation between the stacked 2D $\mZ_2$-TSCs and a boundary-gapped patchwork composed of 1D $\mZ_2$-TSCs.
			}
		\end{center}
	\end{figure}

	Once we find $E_{1,-1}^{\infty} = (\mZ_2)^2$ and $E_{2,-2}^{\infty} = \mZ_2$, we check if the final classification is $E_{1,-1}^{\infty}\oplus E_{2,-2}^{\infty} = (\mZ_2)^3$.
	To understand the group structure of $E_{1,-1}^{\infty}$ and $E_{2,-2}^{\infty}$, let us consider doubled 2D $\mZ_2$-TSCs (DTSCs), which is constructed by stacking two 2D $\mZ_2$-TSCs. 
	\begin{align}
		\label{eq:DTSC_pb}
		H &= \Delta(-i \partial_z \tau_x -i \partial_x \tau_z s_z) + m \tau_z,\\
		U(\calC) &= \mu_0\tau_x,\\
		U(\calT) &= i\mu_0s_y\tau_z,\\
		\label{eq:glide_pb}
		U(G_x) &= i\mu_0s_x\tau_z,
	\end{align}
	where $U(\calC)$ and $U(\calT)$ are unitary representations of PHS and TRS.
	The only possible $\calC$- and $\calT$-symmetric mass term is 
	\begin{align}
		\label{eq:mass_pb}
		M(\bm{r}) &= m(\bm{r})  \mu_y s_x \tau_y.
	\end{align}
	This mass term must satisfy $U_{\bm{r}}(G_x)M_{\bm{r}} = M_{G_x\bm{r}}U_{\bm{r}}(G_x)$. 
	Therefore, the coefficient $m(\bm{r}) $ is transformed as $m(-x,y+1/2,z) = -m(x,y,z)$. 
	Here, we choose $m(\br) > 0$ for $0 < y <  1/2$ and $m(\br) < 0$ for $1/2 < y <  1$.
	The transformation of $m(\br)$ implies that the domain wall exists on $y = l/2\ (l = 0, 1, 2, ...)$ lines, which results in 1D TSCs corresponding to $(0,1)\in E_{1,-1}^{\infty}$ [see Fig.~\ref{app:fig_pb} (e)].
	
	Also, we can understand this group structure from edge point of view.
	We periodically place the DTSCs along $y$-direction and impose periodic boundary condition on $y$-axis and open boundary condition on $x$-axis [see Fig.~\ref{app:fig_pb}(e)]. We introduce a vector 
	\begin{align}
		\bm{n}_{\parallel}(\bm{r}) = \begin{cases}
			(0,1)^T \quad \text{for }\ \bm{r} = (1/2, y)\\
			(0,-1)^T \quad \text{for }\  \bm{r} = (-1/2, y)\\
		\end{cases},
	\end{align} 
	which defines the directions of boundaries.

	We here project Eqs.~\eqref{eq:DTSC_pb}-\eqref{eq:mass_pb} onto the edges.	
	The edge Hamiltonian and symmetries are described by
	\begin{align}
		h_{\bm{r},\bm{k}} &= \bm{k}\cdot\bm{n}_{\parallel}(\bm{r}) \gamma_0\sigma_z,\\
		u_{\bm{r}}(\calC) &= (-1)^{1/2-x}i\gamma_0\sigma_0,\\
		u_{\bm{r}}(\calT)  &= i \gamma_0\sigma_y,\\
		u_{\bm{r}}(G_x)  &= i \gamma_0\sigma_y,\\
		\tilde{M}(\bm{r}) &= m(\bm{r})  \gamma_y\sigma_x,
	\end{align}
	where $u_{\bm{r}}(g)$ is a unitary representation for $g \in G$ and $\gamma$ denotes the flavor space of two copies of TSCs.
	Note that $G_x$ transforms a point $\bm{r}$ into $\bm{r}+(0,1/2,0)^T$, i.e., $u_{\bm{r}}(G_x)$ satisfies $u_{\bm{r}}(G_x)h_{\bm{r},k_y} = h_{G_x(\bm{r}),k_y}u_{\bm{r}}(G_x)$.
	Again, one can see that $m_{(\pm 1/2,y+1/2,z)} = -m_{(\pm 1/2,y,z)}$.
	This transformation implies that the mass term cannot fully gap out the edge modes, i.e., gapless modes are still at $y = l/2\ (l = 0, 1, 2, ...)$. Thus, we conclude that the DTSC has the same boundary signature of $(0,1)\in E_{1,-1}^{\infty}$.
	Therefore, the final classification is $\mZ_2 \times \mZ_4$. 
	
	\subsection{layer group $p4$}
	\label{app:p4}
	We move on to discussions for layer group $p4$, whose generator is fourfold rotation symmetry $C_{4}^{z}:(x,y,z)\rightarrow (y,-x,z)$. 
	\subsubsection{Cell decomposition and building blocks}
\begin{figure}[t]
		\begin{center}
			\includegraphics[width=0.9\columnwidth]{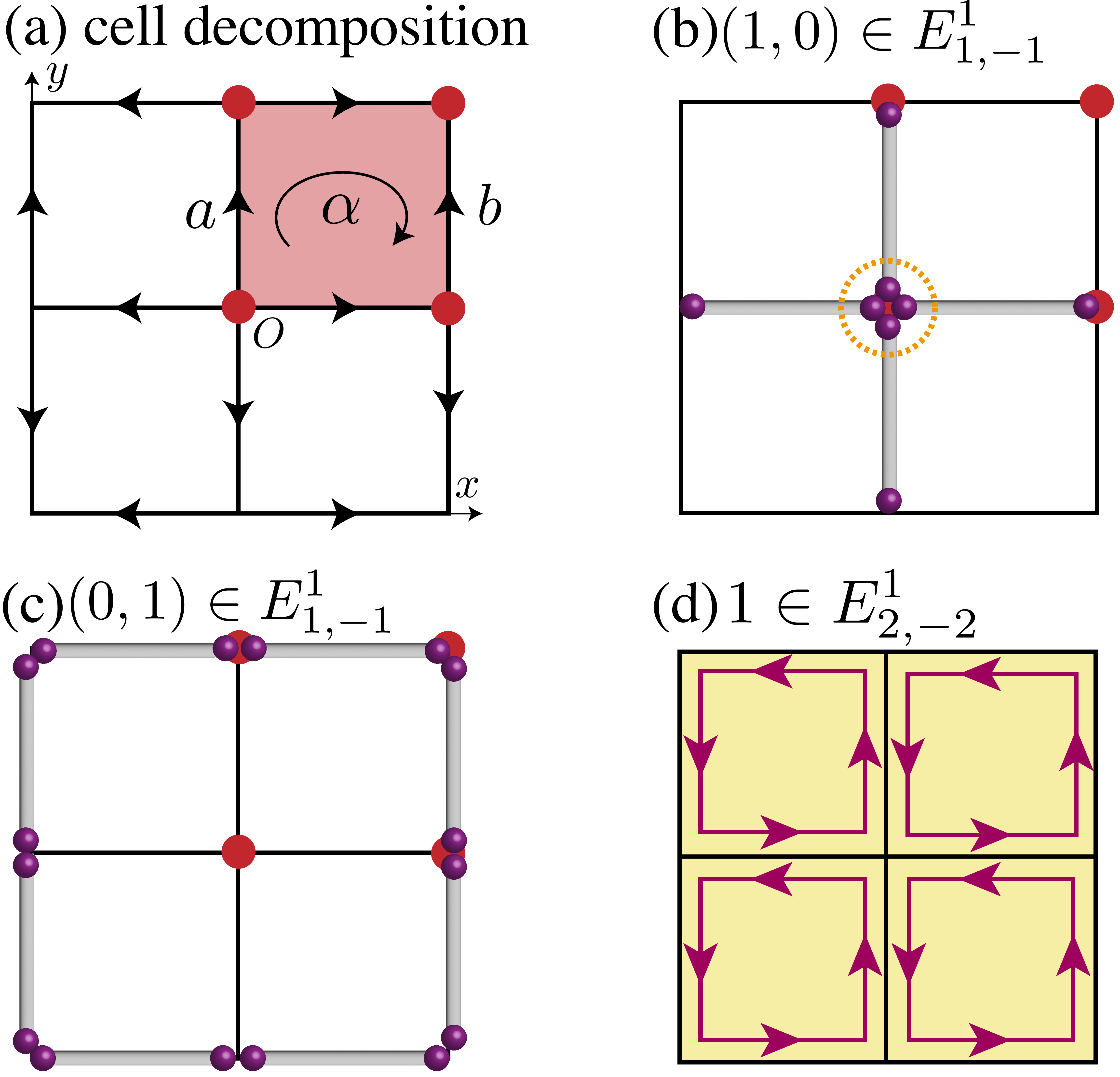}
			\caption{\label{app:fig_p4}\textbf{Illustration of topological patchworks in $p4$.}
				(a) Cell decomposition. 
				(b)-(d) All boundary-gapped patchworks, which are generators of $ E_{1,-1}^{\infty}=E_{1,-1}^{1}$ and $E_{2,-2}^{\infty}=E_{2,-2}^{1}$. 
				(e) Relation between the stacked 2D $\mZ_2$-TSCs and a boundary-gapped patchwork composed of 1D $\mZ_2$-TSCs.
			}
		\end{center}
	\end{figure}

	Our cell decomposition is in Fig.~\ref{app:fig_p4} (a), which is composed of inequivalent a $2$-cell, two $1$-cells, and three $0$-cells. 
	
	Next, we identify EAZ classes for each $p$-cell. 
	For each of $0$-cells $\bx = (a/2, a/2)$ with $a=0\text{ or }1$, the site-symmetry group $G_{\bx} = \calG_{\bx} + \calA_{\bx} + \calP_{\bx} + \calJ_{\bx}$ is
	\begin{align}
		\calG_{\bx} &= \{e, C_{4}^{z}T_{x}^{a}, C_{2}^{z}T_{x}^{a}T_{y}^{a}, (C_{4}^{z})^3T_{y}^{a}\};\\
		\calA_{\bx} &= \calG_{\bx}\calT;\\
		\calP_{\bx} &=  \calG_{\bx}\calC;\\
		\calJ_{\bx} &=  \calG_{\bx}\Gamma,
	\end{align}
	where $T_{x}$ and $T_{y}$ are translations along $x$- and $y$-directions. 
	The projective factors we used are shown in Tables~\ref{tab:z_C4_G}-\ref{tab:z_C4_gamma2}.
	\begin{table}[t]
		\begin{center}
			\caption{\label{tab:z_C4_G}Projective factors $z_{g,g'}$ for $g, g' \in \calG$}
			\begin{tabular}{c|cccc}
				\hline
				$z_{g,g'}$ & $e$ & $C_{4}^{z}T_{x}^{a}$ & $C_{2}^{z}T_{x}^{a}T_{y}^{a}$ & $(C_{4}^{z})^3T_{y}^{a}$ \\
				\hline
				$e$ & $1$ & $1$ & $1$ & $1$ \\
				$C_{4}^{z}T_{x}^{a}$& $1$ & $1$ & $-1$ & $1$ \\
				$C_{2}^{z}T_{x}^{a}T_{y}^{a}$ & $1$ & $-1$ & $-1$ & $-1$ \\
				$(C_{4}^{z})^3T_{y}^{a}$ & $1$ & $1$ & $-1$ & $-1$ \\
				\hline
			\end{tabular}
		\end{center}
	\end{table}

\begin{table}[t]
	\begin{center}
		\caption{\label{tab:z_C4_A}Projective factors $z_{g,g'}$ for $g, g' \in \calA$}
		\begin{tabular}{c|cccc}
			\hline
			$z_{g,g'}$ & $\calT$ & $C_{4}^{z}T_{x}^{a}\calT$ & $C_{2}^{z}T_{x}^{a}T_{y}^{a}\calT$ & $(C_{4}^{z})^3T_{y}^{a}\calT$ \\
			\hline
			$\calT$ & $-1$ & $-1$ & $-1$ & $-1$ \\
			$C_{4}^{z}T_{x}^{a}\calT$& $-1$ & $-1$ & $1$ & $-1$ \\
			$C_{2}^{z}T_{x}^{a}T_{y}^{a}\calT$ & $-1$ & $1$ & $1$ & $1$ \\
			$(C_{4}^{z})^3T_{y}^{a}\calT$ & $-1$ & $-1$ & $1$ & $1$ \\
			\hline
		\end{tabular}
	\end{center}
\end{table}

\begin{table}[t]
	\begin{center}
		\caption{\label{tab:z_C4_C}Projective factors $z_{g,g'}$ for $g, g' \in \calP$}
		\begin{tabular}{c|cccc}
			\hline
			$z_{g,g'}$ & $\calC$ & $C_{4}^{z}T_{x}^{a}\calC$ & $C_{2}^{z}T_{x}^{a}T_{y}^{a}\calC$ & $(C_{4}^{z})^3T_{y}^{a}\calC$ \\
			\hline
			$\calC$ & $1$ & $1$ & $1$ & $1$ \\
			$C_{4}^{z}T_{x}^{a}\calC$& $1$ & $1$ & $-1$ & $1$ \\
			$C_{2}^{z}T_{x}^{a}T_{y}^{a}\calC$ & $1$ & $-1$ & $-1$ & $-1$ \\
			$(C_{4}^{z})^3T_{y}^{a}\calC$ & $1$ & $1$ & $-1$ & $-1$ \\
			\hline
		\end{tabular}
	\end{center}
\end{table}

\begin{table}[t]
	\begin{center}
		\caption{\label{tab:z_C4_gamma1}Projective factors $z_{g,g'}$ for $g \in \calG$ and$g' \in \calJ$}
		\begin{tabular}{c|cccc}
			\hline
			$z_{g,g'}$ & $\Gamma$ & $C_{4}^{z}T_{x}^{a}\Gamma$ & $C_{2}^{z}T_{x}^{a}T_{y}^{a}\Gamma$ & $(C_{4}^{z})^3T_{y}^{a}\Gamma$ \\
			\hline
			$e$ & $1$ & $1$ & $1$ & $1$ \\
			$C_{4}^{z}T_{x}^{a}$& $1$ & $1$ & $-1$ & $1$ \\
			$C_{2}^{z}T_{x}^{a}T_{y}^{a}$ & $1$ & $-1$ & $-1$ & $-1$ \\
			$(C_{4}^{z})^3T_{y}^{a}$ & $1$ & $1$ & $-1$ & $-1$ \\
			\hline
		\end{tabular}
	\end{center}
\end{table}

\begin{table}[t]
	\begin{center}
		\caption{\label{tab:z_C4_gamma2}Projective factors $z_{g,g'}$ for $g \in \calJ$ and$g' \in \calG$}
		\begin{tabular}{c|cccc}
			\hline
			$z_{g,g'}$ & $e$ & $C_{4}^{z}T_{x}^{a}$ & $C_{2}^{z}T_{x}^{a}T_{y}^{a}$ & $(C_{4}^{z})^3T_{y}^{a}$ \\
			\hline
			$\Gamma$ & $1$ & $1$ & $1$ & $1$ \\
			$C_{4}^{z}T_{x}^{a}\Gamma$& $1$ & $1$ & $-1$ & $1$ \\
			$C_{2}^{z}T_{x}^{a}T_{y}^{a}\Gamma$ & $1$ & $-1$ & $-1$ & $-1$ \\
			$(C_{4}^{z})^3T_{y}^{a}\Gamma$ & $1$ & $1$ & $-1$ & $-1$ \\
			\hline
		\end{tabular}
	\end{center}
\end{table}

	For $\chi_{\bx}^{\alpha}(e) = +1$ and $\chi_{\bx}^{\alpha}(C_{4}^{z}T_{x}^{a}T_{y}^{a}) = e^{i \frac{\alpha}{4}\pi}$ in Eq.~\eqref{eq:char_irrep},
	\begin{align}
		W^{\alpha}_{\bx}(\mathcal{P}) &= \frac{1}{4}\left( 1 + e^{i \tfrac{\alpha}{2}\pi}  -1  - e^{i \tfrac{\alpha}{2}\pi}\right) = 0,\\
		W^{\alpha}_{\bx}(\mathcal{A}) &= \frac{1}{4}\left(-1 - e^{i \tfrac{\alpha}{2}\pi}  +1  + e^{i \tfrac{\alpha}{2}\pi}\right) = 0,\\
		W^{\alpha}_{\bx}(\mathcal{J}) &= \frac{1}{4}\left( 1 + 1 + 1 + 1 \right) = 1.
	\end{align}
	This indicates that EAZ classes for each rotation eigenvalue sector are class AIII.
	Similarly, we find that EAZ classes at each of $0$-cells $\bx = (1/2, 0)$ and  $(0, 1/2)$ are class AIII.
	
	Since there are no unitary symmetries other than the identity on $1$- and $2$-cells, the EAZ classes for them are class DIII.
	Thus, we can place (i) 2D $\mZ_2$-TSC on $\alpha$; (ii) 1D $\mZ_2$-TSC on $a$; (iii) 1D $\mZ_2$-TSC on $b$, which results in $E_{1,-1}^{1} = (\mZ_2)^2$ and $E_{2,-2}^{1} = \mZ_2$ [see Table~\ref{app:E1_p4}].
	We also show building block TSCs in Fig.~\ref{app:fig_p4} (b)-(d).
	
	\begin{table}[H]
		\begin{center}
			\caption{\label{app:E1_p4}$E^1$-pages for layer group $pb$. Here, $\star$ denotes that $E^1$-pages are not involved in the  classification.}
			\begin{tabular}{c|c|c|c}
				$n=0$ & $0$ & $\star$ & $\star$\\
				$n=1$ & $\mZ^5$ & $(\mZ_2)^2$& $\mZ_2$\\
				$n=2$ & $\star$ & $(\mZ_2)^2$& $\mZ_2$\\
				\hline
				$E_{p,-n}^{1}$ & $p=0$ & $p=1$ & $p=2$
			\end{tabular}
		\end{center}
	\end{table}
	
	\subsubsection{first differential}
	Again, discussions about $d_{2,-1}^{1}$ and $d_{2,-2}^{1}$ in $p4$ are the completely same as those for $p\bar{1}$ and $p4$, i.e.,
	$d_{2,-1}^{1} = (0, 0)^T$ and $d_{2,-2}^{1} = (0, 0)^T$.
	
	Let us discuss $d_{1,-1}^{1}: E_{1,-1}^{1} = (\mZ_2)^2 \rightarrow E_{0,-1}^{1} = \mZ^5$. 
	It should be noted that the map from the abelian group that contains only torsion parts to a free abelian group must be trivial.
	As a result, we obtain $E_{1,-1}^{2} = \text{Ker }d_{1,-1}^{1}/\text{Im }d_{2,-1}^{1} = E_{1,-1}^{1}$ and $E_{2,-2}^{2} = \text{Ker }d_{2,-2}^{1} = E_{2,-2}^{1}$ [see Table~\ref{app:E2_p4}].
	\begin{table}[H]
		\begin{center}
			\caption{\label{app:E2_p4}$E^2$-pages for layer group $pb$. Here, $\star$ denotes that $E^2$-pages are not involved in the  classification.}
			\begin{tabular}{c|c|c|c}
				$n=0$ & $0$ & $\star$ & $\star$\\
				$n=1$ & $\mZ^5$ & $(\mZ_2)^2$& $\mZ_2$\\
				$n=2$ & $\star$ & $\star$& $\mZ_2$\\
				\hline
				$E_{p,-n}^{2}$ & $p=0$ & $p=1$ & $p=2$
			\end{tabular}
		\end{center}
	\end{table}

	\subsubsection{second differential}
	As is the case with $pb$, $d_{2,-2}^{2}:E_{2,-2}^{2}\rightarrow E_{0,-1}^{2}$ is trivial.
	To understand this fact, we discuss the Hamiltonian $H_{\text{DIII}}(\bm{r}; \Delta)$~\eqref{eq:DIII-SC_app} with the fourfold rotation symmetry
	\begin{align}
		U(C_{4}^{z}) &= e^{i \frac{\pi}{4}\tau_z s_z},
	\end{align}
	where $U(C_{4}^{z})$ satisfies the relation $U(C_{4}^{z}) H_{\text{DIII}}(\bm{r}; \Delta) = H_{\text{DIII}}(C_{4}^{z}\bm{r}; \Delta)U(C_{4}^{z})$.
	Then, the phase of $\Delta(\bm{r})$ can be uniformly chosen as real, which results in the absence of vortices.
	As a result, $E_{2,-2}^{\infty}=\text{Ker }d_{2,-2}^{2} = E_{2,-2}^{2}$ [see Table~\ref{app:E3_p4}].
	
	\begin{table}[H]
		\begin{center}
			\caption{\label{app:E3_p4}$E^\infty$-pages for layer group $pb$. Here, $\star$ denotes that $E^\infty$-pages are not involved in the  classification.}
			\begin{tabular}{c|c|c|c}
				$n=0$ & $0$ & $\star$ & $\star$\\
				$n=1$ & $\mZ^5$ & $(\mZ_2)^2$& $\mZ_2$\\
				$n=2$ & $\star$ & $\star$& $\mZ_2$\\
				\hline
				$E_{p,-n}^{\infty}$ & $p=0$ & $p=1$ & $p=2$
			\end{tabular}
		\end{center}
	\end{table}
	
	\subsubsection{Group extension}
	In a similar way to $pb$, we again discuss the DTSC with fourfold rotation symmetry $U(C_{4}^{z}) = \mu_0 \otimes e^{i (\pi/4)\tau_zs_z}$.
	From $U(C_{4}^{z})M({\bm{r}}) = M(C_{4}^{z}\bm{r})U(C_{4}^{z})$, we find that $m(C_{4}^{z}\br) = m(\br)$.
	This transformation implies that we can add a uniform mass term, i.e., any domain wall does not exists, and thus the DTSC is trivial.
	
	Again, let us consider the edge theory.
	We introduce two vector 
	\begin{align}
		\bm{n}_{\perp}(\bm{r}) = (\cos \theta, \sin \theta)^T,\\
		\bm{n}_{\parallel}(\bm{r}) = (-\sin \theta, \cos \theta)^T,
	\end{align} 
	where we adopt the polar coordinate.
	
	The edge Hamiltonian on the disk geometry is described by
	\begin{align}
		H_{\bm{r},\bm{k}} &= \bm{k}\cdot\bm{n}_{\parallel}(\bm{r}) \sigma_z\gamma_0,\\
		U_{\bm{r}}(\calC) &= -i e^{i\theta \sigma_z}\gamma_0, \\
		U_{\bm{r}}(\calT) &=  i \sigma_y\gamma_0,\\
		U_{\bm{r}}(C_{4}^{z}) &= e^{i (\pi/4) \sigma_z}\gamma_0,
	\end{align}
	where $\gamma$ represents the flavor space of two copies of $\mZ_2$-TSCs.
	The projected mass term is 
	\begin{align}
		M_{\bm{r}} &= m_{\bm{r}} \gamma_y (\bm{n}_{\perp}(\bm{r})\cdot (\sigma_x, -\sigma_y)).
	\end{align}
	From $U_{\bm{r}}(C_{4}^{z}) \mu_y(\bm{n}_{\perp}(\bm{r})\cdot (\sigma_x, -\sigma_y)) = \mu_y (\bm{n}_{\perp}(C_{4}^{z}\bm{r})\cdot (\sigma_x, -\sigma_y))U_{\bm{r}}(C_{4}^{z})$, $m_{\bm{r}}$ has to satisfy $m_{C_{4}^{z}\bm{r}} = m_{\bm{r}}$. This transformation implies that the mass term can fully gap out the edge modes. 
	Thus, we conclude that DTSC is the trivial phase and that the group extension is trivial and that the final classification is $(\mZ_2)^{3}$.
	
	The above discussions are easily generalized to twofold and sixfold rotation symmetries.

	\clearpage
	\section{Atiyah-Hirzebruch spectral sequence in momentum space}
	\label{app:K-AHSS}
	In this appendix, we briefly review on the momentum-space classification of topological superconducting phases, which is classified by twisted equivalent $K$-group $^{\phi}K_{G}^{(z,c)-n}(T^3)$~\cite{Freed2013, Shiozaki-Sato-Gomi2017}. The twisted equivalent $K$-group contains four data, $\phi, c, z$, and $n$. Here, $\phi, c: G \rightarrow \{\pm 1\}$ are maps. The $\phi_g = \pm 1$ is defined by $g\bk = \phi_g p_g \bk$, and the $c_g=\pm1$ is defined by
	\begin{align}
		U_{\bk}^{\text{BdG}}(g)H^{\text{BdG}}_{\bk} = c_gH^{\text{BdG}}_{g\bk}U_{\bk}(g) \ \ \text{for $\phi_g = +1$},\\
		U^{\text{BdG}}_{\bk}(g)[H_{\bk}^{\text{BdG}}]^{*} = c_gH^{\text{BdG}}_{g\bk}U^{\text{BdG}}_{\bk}(g) \ \ \text{for $\phi_g = -1$},
	\end{align}
	where $U_{\bk}(g)$ is a unitary representation of $g \in G$ and $H_{\bk}$ is the Hamiltonian. 
	Also, $z$ denotes the set of projective factors $\{z_{g,g'}\in \text{U}(1)\}_{g,g' \in G}$ such that 
	\begin{align}
		U^{\text{BdG}}_{g'\bk}(g)U^{\text{BdG}}_{\bk}(g') &= z_{g,g'}U^{\text{BdG}}_{\bk}(gg')\ \text{for $\phi_g = +1$},\\
		U^{\text{BdG}}_{g'\bk}(g)[U_{\bk}^{\text{BdG}}]^{*}(g') &= z_{g,g'}U^{\text{BdG}}_{\bk}(gg')\ \text{for $\phi_g = -1$},
	\end{align}
    The $n$ denotes the number of additional chiral symmetries, which is called {\it grading} in K-theory. It is difficult to directly compute the twisted equivalent $K$-group for a given symmetry setting. Then, solving Atiyah-Hirzebruch spectral sequences (AHSS) in momentum space provides us with ``approximated $^{\phi}K_{G}^{(z,c)-n}(T^3)$.'' 
    As mentioned in the main text, for various symmetry settings in which we are interested, AHSS can completely determine $^{\phi}K_{G}^{(z,c)-n}(T^3)$. In the following, we explain this finding.
   
	\subsection{Basics of Atiyah-Hirzebruch spectral sequence in momentum space}
	Before moving on to details of our findings, let us briefly explain AHSS in momentum space [see Ref.~\onlinecite{K-AHSS} for more detailed information].
	In a similar way to real-space classification, we first decompose the Brillouin zone into points (0-cells), line segments (1-cells), polygons (2-cells), and polyhedrons (3-cells). 
	After obtaining a cell decomposition, we define abelian group $E_{1}^{p,-n}$ for a set of $p$-cell. 
	A map from $E_{1}^{p,-n}$ to $E_{1}^{p+1,-n}$
	\begin{align}
		d_{1}^{p,-n}: E_{1}^{p,-n} \rightarrow E_{1}^{p+1,-n}
	\end{align}
	is defined, which is called \textit{first differential}~\cite{K-AHSS}. Then, we define $E_2$-pages by
	\begin{align}
		E_{2}^{0,-n} &\equiv \text{Ker}\ d_{1}^{0,-n},\\
		E_{2}^{p,-n} &\equiv \text{Ker}\ d_{1}^{p,-n}/ \text{Im}\ d_{1}^{p-1,-n}\ (\text{for }p=1,.., d-1),\\
		E_{2}^{d,-n} &\equiv E_{1}^{d,-n}/ \text{Im}\ d_{1}^{d-1,-n}.
	\end{align}
	
	Similarly, we can define higher differential and $E_r$-pages for $r \geq2$ by 
	\begin{align}
		d_{r}^{p,-n}: E_{r}^{p,-n} \rightarrow E_{r}^{p+r,-n-r+1}
	\end{align}
	and
	\begin{align}
		E_{r+1}^{p,-n} &\equiv \text{Ker}\ d_{r}^{0,-n},\\
		E_{r+1}^{p,-n} &\equiv \text{Ker}\ d_{r}^{p,-n}/ \text{Im}\ d_{r}^{p-r,-n+r-1}\ \text{(for $p \leq d-r$)},\\
		E_{r+1}^{p,-n} &\equiv E_{r}^{p,-n}\ \text{(for $p =d-r+1,.., d$)}.
	\end{align}
	For three-dimensional systems, $E_{4}^{p,-n}$ is the limiting page $E_{\infty}^{p,-n}$.
	One might think that $\bigoplus_{p}E_{\infty}^{p,-p}$ is the same as $^{\phi}K_{G}^{(z,c)-n}(T^3)$. However, this is untrue. Although $\bigoplus_{p}E_{\infty}^{p,-p}$ is equivalent to $^{\phi}K_{G}^{(z,c)-n}(T^3)$ as sets, these are generally not equivalent as abelian groups. To obtain $^{\phi}K_{G}^{(z,c)-n}(T^3)$ from $\{E_{\infty}^{p,-p}\}_{p=0}^{3}$, we must solve the group extension problems.
	
	\subsection{Findings on $E_2$-pages}
	It is easy to obtain $E_2$-pages.
	After computing $E_{2}^{p,-n}$ for all space groups and grading $n$, we observe the following facts for the conventional pairing symmetries in 159 space groups:
	\begin{enumerate}
		\setlength{\itemsep}{-2pt}
		\item[(I)] $E_{2}^{0,0} = E_{2}^{2,-2} = E_{2}^{3,-3}= 0$ ;
		\item[(II)] $E_{2}^{2,-3} =0$.
	\end{enumerate}
	These findings ensure that 
	\begin{enumerate}
		\setlength{\itemsep}{-2pt}
		\item[(i)] $E_{\infty}^{0,0} = E_{\infty}^{2,-2} =E_{\infty}^{3,-3} =0$;
		\item[(ii)] $E_{\infty}^{1,-1} \simeq E_{2}^{1,-1}$ since the fact (II) implies $d_{2}^{1,-1}= 0$.
	\end{enumerate}
	As discussed above, to determine the twisted equivalent $K$-group $^{\phi}K_{G}^{(z,c)-n}(T^3)$ from $E_{\infty}$-pages, we must solve the group extension problem.
	However, the findings (i) and (ii) immediately imply that $^{\phi}K_{G}^{(z,c)-n}(T^3)\simeq E_{\infty}^{1,-1} = E_{2}^{1,-1}$. Indeed, the same things happen for conventional pairing symmetries in layer groups. More precisely, $E_{2}^{2,-2}$ is nontrivial only for layer groups $p2\ (\text{No.}~2), p4\ (\text{No.}~49), p\bar{4}\ (\text{No.}~50), p6\ (\text{No.}~73)$.
	
	In Tables \ref{tab:SG2} -- \ref{tab:RG}, we list $^{\phi}K_{G}^{(z,c)-n}(T^3)$ for all rod, all layer, and the 159 space groups.
	%In Tables~\ref{tab:classification_SG} and \ref{tab:classification_LG}, we list $^{\phi}K_{G}^{(z,c)-n}(T^3)$ for all centrosymmetric space groups and layer groups.

	\clearpage
	
	\onecolumngrid
	
	\section{Classification tables for centrosymmetric space groups with trivial pairing symmetries}
	\label{app:tabSG}

	\begin{longtable*}{c|c|c|c|c|c}
		\caption{\label{tab:SG2}Classification table of topological phases in space groups.}\\
		\hline
		Space group & pairing symmetry & $E_{3,-3}^{\infty}$ & $E_{2,-2}^{\infty}$ & $E_{1,-1}^{\infty}$&$^{\phi}K_{G}^{(z,c)-n}(T^3)$\\
		\hline\hline
		$1$ & $A$ & $\mathbb{Z}$ & $\mathbb{Z}_2^3$ & $\mathbb{Z}_2^3$ & $\mathbb{Z}_2^6\times\mathbb{Z}$\\
		$2$ & $A_g$ & $0$ & $0$ & $0$ & $0$\\
		$3$ & $A$ & $\mathbb{Z}$ & $\mathbb{Z}_2^4$ & $\mathbb{Z}_2^3\times \mathbb{Z}^4$ & ??\\
		$4$ & $A$ & $\mathbb{Z}$ & $\mathbb{Z}_2^3$ & $\mathbb{Z}_2^3$ & ??\\
		$5$ & $A$ & $\mathbb{Z}$ & $\mathbb{Z}_2^3$ & $\mathbb{Z}_2^2\times \mathbb{Z}^2$ & ??\\
		$6$ & $A'$ & $0$ & $\mathbb{Z}_2^2$ & $\mathbb{Z}_2\times \mathbb{Z}^4$ & $\mathbb{Z}_2\times \mathbb{Z}^4$\\
		$7$ & $A'$ & $0$ & $\mathbb{Z}_2^3$ & $\mathbb{Z}_2^3$ & $\mathbb{Z}_2^4\times \mathbb{Z}_4$\\
		$8$ & $A'$ & $0$ & $\mathbb{Z}_2^2$ & $\mathbb{Z}_2\times \mathbb{Z}^2$ & $\mathbb{Z}_2^2\times \mathbb{Z}^2$\\
		$9$ & $A'$ & $0$ & $\mathbb{Z}_2^2$ & $\mathbb{Z}_2^2$ & $\mathbb{Z}_2^4$\\
		$10$ & $A_g$ & $0$ & $0$ & $0$ & $0$\\
		$11$ & $A_g$ & $0$ & $0$ & $\mathbb{Z}^2$ & $\mathbb{Z}^2$\\
		$12$ & $A_g$ & $0$ & $0$ & $0$ & $0$\\
		$13$ & $A_g$ & $0$ & $\mathbb{Z}_2$ & $\mathbb{Z}_2\times \mathbb{Z}^2$ & $\mathbb{Z}_2\times \mathbb{Z}^2$\\
		$14$ & $A_g$ & $0$ & $\mathbb{Z}_2$ & $\mathbb{Z}_2$ & $\mathbb{Z}_2^2$\\
		$15$ & $A_g$ & $0$ & $\mathbb{Z}_2$ & $\mathbb{Z}$ & $\mathbb{Z}$\\
		$16$ & $A_1$ & $\mathbb{Z}$ & $\mathbb{Z}_2^5$ & $\mathbb{Z}^{12}$ & ??\\
		$17$ & $A_1$ & $\mathbb{Z}$ & $\mathbb{Z}_2^4$ & $\mathbb{Z}_2^3\times \mathbb{Z}^4$ & ??\\
		$18$ & $A_1$ & $\mathbb{Z}$ & $\mathbb{Z}_2^3$ & $\mathbb{Z}_2^2\times \mathbb{Z}^2$ & ??\\
		$19$ & $A_1$ & $\mathbb{Z}$ & $\mathbb{Z}_2^2$ & $\mathbb{Z}_2^2$ & ??\\
		$20$ & $A_1$ & $\mathbb{Z}$ & $\mathbb{Z}_2^3$ & $\mathbb{Z}_2^2\times \mathbb{Z}^2$ & ??\\
		$21$ & $A_1$ & $\mathbb{Z}$ & $\mathbb{Z}_2^4$ & $\mathbb{Z}_2\times \mathbb{Z}^7$ & ??\\
		$22$ & $A_1$ & $\mathbb{Z}$ & $\mathbb{Z}_2^4$ & $\mathbb{Z}_2\times \mathbb{Z}^6$ & ??\\
		$23$ & $A_1$ & $\mathbb{Z}$ & $\mathbb{Z}_2^3$ & $\mathbb{Z}^6$ & ??\\
		$24$ & $A_1$ & $\mathbb{Z}$ & $\mathbb{Z}_2^3$ & $\mathbb{Z}_2^2\times \mathbb{Z}^3$ & ??\\
		$25$ & $A_1$ & $0$ & $\mathbb{Z}_2$ & $\mathbb{Z}^4$ & $\mathbb{Z}^4$\\
		$26$ & $A_1$ & $0$ & $\mathbb{Z}_2$ & $\mathbb{Z}_2\times \mathbb{Z}^2$ & $\mathbb{Z}_2\times \mathbb{Z}^2$\\
		$27$ & $A_1$ & $0$ & $\mathbb{Z}_2$ & $\mathbb{Z}_2^3$ & ??\\
		$28$ & $A_1$ & $0$ & $\mathbb{Z}_2^3$ & $\mathbb{Z}_2^2\times \mathbb{Z}^4$ & $\mathbb{Z}_2^2\times \mathbb{Z}^4$\\
		$29$ & $A_1$ & $0$ & $\mathbb{Z}_2^3$ & $\mathbb{Z}_2^3$ & $\mathbb{Z}_2^2\times \mathbb{Z}_4^2$\\
		$30$ & $A_1$ & $0$ & $\mathbb{Z}_2^3$ & $\mathbb{Z}_2^2\times \mathbb{Z}^2$ & $\mathbb{Z}_2^3\times \mathbb{Z}^2$\\
		$31$ & $A_1$ & $0$ & $\mathbb{Z}_2^2$ & $\mathbb{Z}_2\times \mathbb{Z}^2$ & $\mathbb{Z}_2^2\times \mathbb{Z}^2$\\
		$32$ & $A_1$ & $0$ & $\mathbb{Z}_2^3$ & $\mathbb{Z}_2^2\times \mathbb{Z}^2$ & $\mathbb{Z}_2^2\times \mathbb{Z}_4\times \mathbb{Z}^2$\\
		$33$ & $A_1$ & $0$ & $\mathbb{Z}_2^2$ & $\mathbb{Z}_2^2$ & $\mathbb{Z}_2^4$\\
		$34$ & $A_1$ & $0$ & $\mathbb{Z}_2^3$ & $\mathbb{Z}_2^2\times \mathbb{Z}^2$ & $\mathbb{Z}_2^3\times \mathbb{Z}^2$\\
		$35$ & $A_1$ & $0$ & $\mathbb{Z}_2^2$ & $\mathbb{Z}_2\times \mathbb{Z}^3$ & $\mathbb{Z}_2\times \mathbb{Z}^3$\\
		$36$ & $A_1$ & $0$ & $\mathbb{Z}_2^2$ & $\mathbb{Z}_2\times \mathbb{Z}$ & $\mathbb{Z}_4\times \mathbb{Z}$\\
		$37$ & $A_1$ & $0$ & $\mathbb{Z}_2^2$ & $\mathbb{Z}_2^2\times \mathbb{Z}$ & ??\\
		$38$ & $A_1$ & $0$ & $\mathbb{Z}_2$ & $\mathbb{Z}^3$ & $\mathbb{Z}^3$\\
		$39$ & $A_1$ & $0$ & $\mathbb{Z}_2$ & $\mathbb{Z}_2^2\times \mathbb{Z}$ & $\mathbb{Z}_2^2\times \mathbb{Z}$\\
		$40$ & $A_1$ & $0$ & $\mathbb{Z}_2^2$ & $\mathbb{Z}_2\times \mathbb{Z}^3$ & $\mathbb{Z}_2\times \mathbb{Z}^3$\\
		$41$ & $A_1$ & $0$ & $\mathbb{Z}_2^3$ & $\mathbb{Z}_2^2\times \mathbb{Z}$ & $\mathbb{Z}_2^2\times \mathbb{Z}_4\times \mathbb{Z}$\\
		$42$ & $A_1$ & $0$ & $\mathbb{Z}_2$ & $\mathbb{Z}_2\times \mathbb{Z}^2$ & $\mathbb{Z}_2\times \mathbb{Z}^2$\\
		$43$ & $A_1$ & $0$ & $\mathbb{Z}_2^2$ & $\mathbb{Z}_2\times \mathbb{Z}$ & $\mathbb{Z}_2^2\times \mathbb{Z}$\\
		$44$ & $A_1$ & $0$ & $\mathbb{Z}_2$ & $\mathbb{Z}^2$ & $\mathbb{Z}_2\times \mathbb{Z}^2$\\
		$45$ & $A_1$ & $0$ & $\mathbb{Z}_2^2$ & $\mathbb{Z}_2^2$ & $\mathbb{Z}_2^2\times \mathbb{Z}_4$\\
		$46$ & $A_1$ & $0$ & $\mathbb{Z}_2^2$ & $\mathbb{Z}_2\times \mathbb{Z}^2$ & $\mathbb{Z}_2\times \mathbb{Z}^2$\\
		$47$ & $A_g$ & $0$ & $0$ & $0$ & $0$\\
		$48$ & $A_g$ & $0$ & $\mathbb{Z}_2^2$ & $\mathbb{Z}^6$ & $\mathbb{Z}^6$\\
		$49$ & $A_g$ & $0$ & $\mathbb{Z}_2$ & $\mathbb{Z}^4$ & $\mathbb{Z}^4$\\
		$50$ & $A_g$ & $0$ & $\mathbb{Z}_2^2$ & $\mathbb{Z}^6$ & $\mathbb{Z}^6$\\
		$51$ & $A_g$ & $0$ & $0$ & $\mathbb{Z}$ & $\mathbb{Z}$\\
		$52$ & $A_g$ & $0$ & $\mathbb{Z}_2^2$ & $\mathbb{Z}_2\times \mathbb{Z}^2$ & $\mathbb{Z}_2\times \mathbb{Z}^2$\\
		$53$ & $A_g$ & $0$ & $\mathbb{Z}_2$ & $\mathbb{Z}_2\times \mathbb{Z}$ & $\mathbb{Z}_2\times \mathbb{Z}$\\
		$54$ & $A_g$ & $0$ & $\mathbb{Z}_2$ & $\mathbb{Z}_2^2\times \mathbb{Z}$ & $\mathbb{Z}_2^2\times \mathbb{Z}$\\
		$55$ & $A_g$ & $0$ & $0$ & $0$ & $0$\\
		$56$ & $A_g$ & $0$ & $\mathbb{Z}_2$ & $\mathbb{Z}_2$ & $\mathbb{Z}_2^2$\\
		$57$ & $A_g$ & $0$ & $\mathbb{Z}_2$ & $\mathbb{Z}_2\times \mathbb{Z}^2$ & $\mathbb{Z}_2\times \mathbb{Z}^2$\\
		$58$ & $A_g$ & $0$ & $\mathbb{Z}_2$ & $0$ & $\mathbb{Z}_2$\\
		$59$ & $A_g$ & $0$ & $0$ & $\mathbb{Z}^2$ & $\mathbb{Z}^2$\\
		$60$ & $A_g$ & $0$ & $\mathbb{Z}_2^2$ & $\mathbb{Z}_2\times \mathbb{Z}$ & $\mathbb{Z}_4\times \mathbb{Z}$\\
		$61$ & $A_g$ & $0$ & $\mathbb{Z}_2^2$ & $\mathbb{Z}_2^2$ & $\mathbb{Z}_4^2$\\
		$62$ & $A_g$ & $0$ & $\mathbb{Z}_2$ & $\mathbb{Z}$ & $\mathbb{Z}$\\
		$63$ & $A_g$ & $0$ & $0$ & $\mathbb{Z}$ & $\mathbb{Z}$\\
		$64$ & $A_g$ & $0$ & $0$ & $\mathbb{Z}_2$ & $\mathbb{Z}_2$\\
		$65$ & $A_g$ & $0$ & $0$ & $0$ & $0$\\
		$66$ & $A_g$ & $0$ & $\mathbb{Z}_2$ & $\mathbb{Z}^2$ & $\mathbb{Z}^2$\\
		$67$ & $A_g$ & $0$ & $0$ & $\mathbb{Z}$ & $\mathbb{Z}$\\
		$68$ & $A_g$ & $0$ & $\mathbb{Z}_2$ & $\mathbb{Z}_2\times \mathbb{Z}^3$ & $\mathbb{Z}_2\times \mathbb{Z}^3$\\
		$69$ & $A_g$ & $0$ & $0$ & $0$ & $0$\\
		$70$ & $A_g$ & $0$ & $\mathbb{Z}_2^2$ & $\mathbb{Z}^3$ & $\mathbb{Z}^3$\\
		$71$ & $A_g$ & $0$ & $0$ & $0$ & $0$\\
		$72$ & $A_g$ & $0$ & $0$ & $\mathbb{Z}^2$ & $\mathbb{Z}^2$\\
		$73$ & $A_g$ & $0$ & $0$ & $\mathbb{Z}_2^2$ & $\mathbb{Z}_2^2$\\
		$74$ & $A_g$ & $0$ & $0$ & $0$ & $0$\\
		$75$ & $A$ & $\mathbb{Z}$ & $\mathbb{Z}_2^3$ & $\mathbb{Z}_2^2\times \mathbb{Z}^5$ & ??\\
		$76$ & $A$ & $\mathbb{Z}$ & $\mathbb{Z}_2^2$ & $\mathbb{Z}_2^2$ & ??\\
		$77$ & $A$ & $\mathbb{Z}$ & $\mathbb{Z}_2^3$ & $\mathbb{Z}_2^2\times \mathbb{Z}^3$ & ??\\
		$78$ & $A$ & $\mathbb{Z}$ & $\mathbb{Z}_2^2$ & $\mathbb{Z}_2^2$ & ??\\
		$79$ & $A$ & $\mathbb{Z}$ & $\mathbb{Z}_2^2$ & $\mathbb{Z}_2\times \mathbb{Z}^3$ & ??\\
		$80$ & $A$ & $\mathbb{Z}$ & $\mathbb{Z}_2^2$ & $\mathbb{Z}_2\times \mathbb{Z}$ & ??\\
		$81$ & $A$ & $0$ & $\mathbb{Z}_2^3$ & $\mathbb{Z}_2^2\times \mathbb{Z}$ & ??\\
		$82$ & $A$ & $0$ & $\mathbb{Z}_2^2$ & $\mathbb{Z}_2$ & ??\\
		$83$ & $A_g$ & $0$ & $0$ & $0$ & $0$\\
		$84$ & $A_g$ & $0$ & $\mathbb{Z}_2$ & $0$ & $\mathbb{Z}_2$\\
		$85$ & $A_g$ & $0$ & $\mathbb{Z}_2$ & $\mathbb{Z}_2\times \mathbb{Z}^2$ & $\mathbb{Z}_2\times \mathbb{Z}^2$\\
		$86$ & $A_g$ & $0$ & $\mathbb{Z}_2$ & $\mathbb{Z}_2\times \mathbb{Z}$ & $\mathbb{Z}_2^2\times \mathbb{Z}$\\
		$87$ & $A_g$ & $0$ & $0$ & $0$ & $0$\\
		$88$ & $A_g$ & $0$ & $\mathbb{Z}_2$ & $0$ & $\mathbb{Z}_2$\\
		$89$ & $A_1$ & $\mathbb{Z}$ & $\mathbb{Z}_2^4$ & $\mathbb{Z}^{11}$ & ??\\
		$90$ & $A_1$ & $\mathbb{Z}$ & $\mathbb{Z}_2^3$ & $\mathbb{Z}_2\times \mathbb{Z}^5$ & ??\\
		$91$ & $A_1$ & $\mathbb{Z}$ & $\mathbb{Z}_2^3$ & $\mathbb{Z}_2^2\times \mathbb{Z}^3$ & ??\\
		$92$ & $A_1$ & $\mathbb{Z}$ & $\mathbb{Z}_2^2$ & $\mathbb{Z}_2\times \mathbb{Z}$ & ??\\
		$93$ & $A_1$ & $\mathbb{Z}$ & $\mathbb{Z}_2^4$ & $\mathbb{Z}^9$ & ??\\
		$94$ & $A_1$ & $\mathbb{Z}$ & $\mathbb{Z}_2^3$ & $\mathbb{Z}_2\times \mathbb{Z}^4$ & ??\\
		$95$ & $A_1$ & $\mathbb{Z}$ & $\mathbb{Z}_2^3$ & $\mathbb{Z}_2^2\times \mathbb{Z}^3$ & ??\\
		$96$ & $A_1$ & $\mathbb{Z}$ & $\mathbb{Z}_2^2$ & $\mathbb{Z}_2\times \mathbb{Z}$ & ??\\
		$97$ & $A_1$ & $\mathbb{Z}$ & $\mathbb{Z}_2^3$ & $\mathbb{Z}^7$ & ??\\
		$98$ & $A_1$ & $\mathbb{Z}$ & $\mathbb{Z}_2^3$ & $\mathbb{Z}_2\times \mathbb{Z}^4$ & ??\\
		$99$ & $A_1$ & $0$ & $\mathbb{Z}_2$ & $\mathbb{Z}^3$ & $\mathbb{Z}^3$\\
		$100$ & $A_1$ & $0$ & $\mathbb{Z}_2^2$ & $\mathbb{Z}_2\times \mathbb{Z}^3$ & $\mathbb{Z}_2\times \mathbb{Z}^3$\\
		$101$ & $A_1$ & $0$ & $\mathbb{Z}_2$ & $\mathbb{Z}_2\times \mathbb{Z}$ & $\mathbb{Z}_2^2\times \mathbb{Z}$\\
		$102$ & $A_1$ & $0$ & $\mathbb{Z}_2^2$ & $\mathbb{Z}_2\times \mathbb{Z}^2$ & $\mathbb{Z}_2^2\times \mathbb{Z}^2$\\
		$103$ & $A_1$ & $0$ & $\mathbb{Z}_2$ & $\mathbb{Z}_2^2$ & ??\\
		$104$ & $A_1$ & $0$ & $\mathbb{Z}_2^2$ & $\mathbb{Z}_2\times \mathbb{Z}^2$ & $\mathbb{Z}_2^2\times \mathbb{Z}^2$\\
		$105$ & $A_1$ & $0$ & $\mathbb{Z}_2$ & $\mathbb{Z}^2$ & $\mathbb{Z}_2\times \mathbb{Z}^2$\\
		$106$ & $A_1$ & $0$ & $\mathbb{Z}_2^2$ & $\mathbb{Z}_2\times \mathbb{Z}$ & $\mathbb{Z}_2^3\times \mathbb{Z}$\\
		$107$ & $A_1$ & $0$ & $\mathbb{Z}_2$ & $\mathbb{Z}^2$ & $\mathbb{Z}^2$\\
		$108$ & $A_1$ & $0$ & $\mathbb{Z}_2$ & $\mathbb{Z}_2\times \mathbb{Z}$ & $\mathbb{Z}_2\times \mathbb{Z}$\\
		$109$ & $A_1$ & $0$ & $\mathbb{Z}_2$ & $\mathbb{Z}$ & $\mathbb{Z}_2\times \mathbb{Z}$\\
		$110$ & $A_1$ & $0$ & $\mathbb{Z}_2^2$ & $\mathbb{Z}_2$ & $\mathbb{Z}_2^3$\\
		$111$ & $A_1$ & $0$ & $\mathbb{Z}_2^3$ & $\mathbb{Z}^6$ & $\mathbb{Z}^6$\\
		$112$ & $A_1$ & $0$ & $\mathbb{Z}_2^3$ & $\mathbb{Z}^5$ & $\mathbb{Z}_2\times \mathbb{Z}^5$\\
		$113$ & $A_1$ & $0$ & $\mathbb{Z}_2^2$ & $\mathbb{Z}_2\times \mathbb{Z}$ & $\mathbb{Z}_2^2\times \mathbb{Z}$\\
		$114$ & $A_1$ & $0$ & $\mathbb{Z}_2^2$ & $\mathbb{Z}_2$ & $\mathbb{Z}_2^3$\\
		$115$ & $A_1$ & $0$ & $\mathbb{Z}_2^2$ & $\mathbb{Z}^4$ & $\mathbb{Z}^4$\\
		$116$ & $A_1$ & $0$ & $\mathbb{Z}_2^2$ & $\mathbb{Z}_2\times \mathbb{Z}^2$ & $\mathbb{Z}_2^2\times \mathbb{Z}^2$\\
		$117$ & $A_1$ & $0$ & $\mathbb{Z}_2^3$ & $\mathbb{Z}_2\times \mathbb{Z}^3$ & $\mathbb{Z}_2^2\times \mathbb{Z}^3$\\
		$118$ & $A_1$ & $0$ & $\mathbb{Z}_2^3$ & $\mathbb{Z}_2\times \mathbb{Z}^3$ & $\mathbb{Z}_2^2\times \mathbb{Z}^3$\\
		$119$ & $A_1$ & $0$ & $\mathbb{Z}_2^2$ & $\mathbb{Z}^3$ & $\mathbb{Z}^3$\\
		$120$ & $A_1$ & $0$ & $\mathbb{Z}_2^3$ & $\mathbb{Z}_2\times \mathbb{Z}^2$ & $\mathbb{Z}_2^2\times \mathbb{Z}^2$\\
		$121$ & $A_1$ & $0$ & $\mathbb{Z}_2^2$ & $\mathbb{Z}^3$ & $\mathbb{Z}^3$\\
		$122$ & $A_1$ & $0$ & $\mathbb{Z}_2^2$ & $\mathbb{Z}_2\times \mathbb{Z}$ & $\mathbb{Z}_2^2\times \mathbb{Z}$\\
		$123$ & $A_{1g}$ & $0$ & $0$ & $0$ & $0$\\
		$124$ & $A_{1g}$ & $0$ & $\mathbb{Z}_2$ & $\mathbb{Z}^3$ & $\mathbb{Z}^3$\\
		$125$ & $A_{1g}$ & $0$ & $\mathbb{Z}_2$ & $\mathbb{Z}^4$ & $\mathbb{Z}^4$\\
		$126$ & $A_{1g}$ & $0$ & $\mathbb{Z}_2^2$ & $\mathbb{Z}^5$ & $\mathbb{Z}^5$\\
		$127$ & $A_{1g}$ & $0$ & $0$ & $0$ & $0$\\
		$128$ & $A_{1g}$ & $0$ & $\mathbb{Z}_2$ & $\mathbb{Z}$ & $\mathbb{Z}$\\
		$129$ & $A_{1g}$ & $0$ & $0$ & $\mathbb{Z}$ & $\mathbb{Z}$\\
		$130$ & $A_{1g}$ & $0$ & $\mathbb{Z}_2$ & $\mathbb{Z}_2\times \mathbb{Z}$ & $\mathbb{Z}_2\times \mathbb{Z}$\\
		$131$ & $A_{1g}$ & $0$ & $\mathbb{Z}_2$ & $\mathbb{Z}$ & $\mathbb{Z}$\\
		$132$ & $A_{1g}$ & $0$ & $\mathbb{Z}_2$ & $\mathbb{Z}^2$ & $\mathbb{Z}^2$\\
		$133$ & $A_{1g}$ & $0$ & $\mathbb{Z}_2^2$ & $\mathbb{Z}^4$ & $\mathbb{Z}^4$\\
		$134$ & $A_{1g}$ & $0$ & $\mathbb{Z}_2$ & $\mathbb{Z}^3$ & $\mathbb{Z}^3$\\
		$135$ & $A_{1g}$ & $0$ & $\mathbb{Z}_2$ & $\mathbb{Z}$ & $\mathbb{Z}$\\
		$136$ & $A_{1g}$ & $0$ & $\mathbb{Z}_2$ & $0$ & $\mathbb{Z}_2$\\
		$137$ & $A_{1g}$ & $0$ & $\mathbb{Z}_2$ & $\mathbb{Z}^2$ & $\mathbb{Z}^2$\\
		$138$ & $A_{1g}$ & $0$ & $\mathbb{Z}_2$ & $0$ & $\mathbb{Z}_2$\\
		$139$ & $A_{1g}$ & $0$ & $0$ & $0$ & $0$\\
		$140$ & $A_{1g}$ & $0$ & $0$ & $\mathbb{Z}$ & $\mathbb{Z}$\\
		$141$ & $A_{1g}$ & $0$ & $\mathbb{Z}_2$ & $\mathbb{Z}$ & $\mathbb{Z}$\\
		$142$ & $A_{1g}$ & $0$ & $\mathbb{Z}_2$ & $\mathbb{Z}_2\times \mathbb{Z}$ & $\mathbb{Z}_2\times \mathbb{Z}$\\
		$143$ & $A$ & $\mathbb{Z}$ & $\mathbb{Z}_2$ & $\mathbb{Z}_2\times \mathbb{Z}^3$ & ??\\
		$144$ & $A$ & $\mathbb{Z}$ & $\mathbb{Z}_2$ & $\mathbb{Z}_2$ & ??\\
		$145$ & $A$ & $\mathbb{Z}$ & $\mathbb{Z}_2$ & $\mathbb{Z}_2$ & ??\\
		$146$ & $A$ & $\mathbb{Z}$ & $\mathbb{Z}_2$ & $\mathbb{Z}_2\times \mathbb{Z}$ & ??\\
		$147$ & $A_g$ & $0$ & $0$ & $\mathbb{Z}$ & $\mathbb{Z}$\\
		$148$ & $A_g$ & $0$ & $0$ & $0$ & $0$\\
		$149$ & $A_1$ & $\mathbb{Z}$ & $\mathbb{Z}_2^2$ & $\mathbb{Z}_2\times \mathbb{Z}^5$ & ??\\
		$150$ & $A_1$ & $\mathbb{Z}$ & $\mathbb{Z}_2^2$ & $\mathbb{Z}_2\times \mathbb{Z}^4$ & ??\\
		$151$ & $A_1$ & $\mathbb{Z}$ & $\mathbb{Z}_2^2$ & $\mathbb{Z}_2\times \mathbb{Z}^2$ & ??\\
		$152$ & $A_1$ & $\mathbb{Z}$ & $\mathbb{Z}_2^2$ & $\mathbb{Z}_2\times \mathbb{Z}^2$ & ??\\
		$153$ & $A_1$ & $\mathbb{Z}$ & $\mathbb{Z}_2^2$ & $\mathbb{Z}_2\times \mathbb{Z}^2$ & ??\\
		$154$ & $A_1$ & $\mathbb{Z}$ & $\mathbb{Z}_2^2$ & $\mathbb{Z}_2\times \mathbb{Z}^2$ & ??\\
		$155$ & $A_1$ & $\mathbb{Z}$ & $\mathbb{Z}_2^2$ & $\mathbb{Z}_2\times \mathbb{Z}^3$ & ??\\
		$156$ & $A_1$ & $0$ & $\mathbb{Z}_2$ & $\mathbb{Z}^2$ & $\mathbb{Z}^2$\\
		$157$ & $A_1$ & $0$ & $\mathbb{Z}_2$ & $\mathbb{Z}^3$ & $\mathbb{Z}^3$\\
		$158$ & $A_1$ & $0$ & $\mathbb{Z}_2$ & $\mathbb{Z}_2$ & $\mathbb{Z}_2^2$\\
		$159$ & $A_1$ & $0$ & $\mathbb{Z}_2$ & $\mathbb{Z}_2\times \mathbb{Z}$ & $\mathbb{Z}_2^2\times \mathbb{Z}$\\
		$160$ & $A_1$ & $0$ & $\mathbb{Z}_2$ & $\mathbb{Z}^2$ & $\mathbb{Z}^2$\\
		$161$ & $A_1$ & $0$ & $\mathbb{Z}_2$ & $\mathbb{Z}_2$ & $\mathbb{Z}_2^2$\\
		$162$ & $A_{1g}$ & $0$ & $0$ & $\mathbb{Z}$ & $\mathbb{Z}$\\
		$163$ & $A_{1g}$ & $0$ & $\mathbb{Z}_2$ & $\mathbb{Z}^2$ & $\mathbb{Z}^2$\\
		$164$ & $A_{1g}$ & $0$ & $0$ & $0$ & $0$\\
		$165$ & $A_{1g}$ & $0$ & $\mathbb{Z}_2$ & $\mathbb{Z}$ & $\mathbb{Z}$\\
		$166$ & $A_{1g}$ & $0$ & $0$ & $0$ & $0$\\
		$167$ & $A_{1g}$ & $0$ & $\mathbb{Z}_2$ & $\mathbb{Z}$ & $\mathbb{Z}$\\
		$168$ & $A$ & $\mathbb{Z}$ & $\mathbb{Z}_2^2$ & $\mathbb{Z}_2\times \mathbb{Z}^5$ & ??\\
		$169$ & $A$ & $\mathbb{Z}$ & $\mathbb{Z}_2$ & $\mathbb{Z}_2$ & ??\\
		$170$ & $A$ & $\mathbb{Z}$ & $\mathbb{Z}_2$ & $\mathbb{Z}_2$ & ??\\
		$171$ & $A$ & $\mathbb{Z}$ & $\mathbb{Z}_2^2$ & $\mathbb{Z}_2\times \mathbb{Z}^2$ & ??\\
		$172$ & $A$ & $\mathbb{Z}$ & $\mathbb{Z}_2^2$ & $\mathbb{Z}_2\times \mathbb{Z}^2$ & ??\\
		$173$ & $A$ & $\mathbb{Z}$ & $\mathbb{Z}_2$ & $\mathbb{Z}_2\times \mathbb{Z}^2$ & ??\\
		$174$ & $A'$ & $0$ & $0$ & $\mathbb{Z}_2$ & $\mathbb{Z}_2$\\
		$175$ & $A_g$ & $0$ & $0$ & $0$ & $0$\\
		$176$ & $A_g$ & $0$ & $0$ & $0$ & $0$\\
		$177$ & $A_1$ & $\mathbb{Z}$ & $\mathbb{Z}_2^3$ & $\mathbb{Z}^9$ & ??\\
		$178$ & $A_1$ & $\mathbb{Z}$ & $\mathbb{Z}_2^2$ & $\mathbb{Z}_2\times \mathbb{Z}^2$ & ??\\
		$179$ & $A_1$ & $\mathbb{Z}$ & $\mathbb{Z}_2^2$ & $\mathbb{Z}_2\times \mathbb{Z}^2$ & ??\\
		$180$ & $A_1$ & $\mathbb{Z}$ & $\mathbb{Z}_2^3$ & $\mathbb{Z}^6$ & ??\\
		$181$ & $A_1$ & $\mathbb{Z}$ & $\mathbb{Z}_2^3$ & $\mathbb{Z}^6$ & ??\\
		$182$ & $A_1$ & $\mathbb{Z}$ & $\mathbb{Z}_2^2$ & $\mathbb{Z}_2\times \mathbb{Z}^4$ & ??\\
		$183$ & $A_1$ & $0$ & $\mathbb{Z}_2$ & $\mathbb{Z}^2$ & $\mathbb{Z}^2$\\
		$184$ & $A_1$ & $0$ & $\mathbb{Z}_2$ & $\mathbb{Z}_2$ & ??\\
		$185$ & $A_1$ & $0$ & $\mathbb{Z}_2$ & $\mathbb{Z}$ & $\mathbb{Z}$\\
		$186$ & $A_1$ & $0$ & $0$ & $\mathbb{Z}$ & $\mathbb{Z}$\\
		$187$ & $A'_1$ & $0$ & $0$ & $\mathbb{Z}$ & $\mathbb{Z}$\\
		$188$ & $A'_1$ & $0$ & $\mathbb{Z}_2$ & $\mathbb{Z}_2\times \mathbb{Z}$ & $\mathbb{Z}_2\times \mathbb{Z}$\\
		$189$ & $A'_1$ & $0$ & $0$ & $\mathbb{Z}$ & $\mathbb{Z}$\\
		$190$ & $A'_1$ & $0$ & $\mathbb{Z}_2$ & $\mathbb{Z}_2\times \mathbb{Z}$ & $\mathbb{Z}_2\times \mathbb{Z}$\\
		$191$ & $A_{1g}$ & $0$ & $0$ & $0$ & $0$\\
		$192$ & $A_{1g}$ & $0$ & $\mathbb{Z}_2$ & $\mathbb{Z}^2$ & $\mathbb{Z}^2$\\
		$193$ & $A_{1g}$ & $0$ & $0$ & $0$ & $0$\\
		$194$ & $A_{1g}$ & $0$ & $0$ & $0$ & $0$\\
		$195$ & $A$ & $\mathbb{Z}$ & $\mathbb{Z}_2$ & $\mathbb{Z}^5$ & ??\\
		$196$ & $A$ & $\mathbb{Z}$ & $0$ & $\mathbb{Z}_2\times \mathbb{Z}^3$ & ??\\
		$197$ & $A$ & $\mathbb{Z}$ & $\mathbb{Z}_2$ & $\mathbb{Z}^3$ & ??\\
		$198$ & $A$ & $\mathbb{Z}$ & $0$ & $\mathbb{Z}$ & ??\\
		$199$ & $A$ & $\mathbb{Z}$ & $\mathbb{Z}_2$ & $\mathbb{Z}^2$ & ??\\
		$200$ & $A_g$ & $0$ & $0$ & $0$ & $0$\\
		$201$ & $A_g$ & $0$ & $0$ & $\mathbb{Z}^2$ & $\mathbb{Z}^2$\\
		$202$ & $A_g$ & $0$ & $0$ & $0$ & $0$\\
		$203$ & $A_g$ & $0$ & $0$ & $\mathbb{Z}$ & $\mathbb{Z}$\\
		$204$ & $A_g$ & $0$ & $0$ & $0$ & $0$\\
		$205$ & $A_g$ & $0$ & $0$ & $0$ & $0$\\
		$206$ & $A_g$ & $0$ & $0$ & $0$ & $0$\\
		$207$ & $A_1$ & $\mathbb{Z}$ & $\mathbb{Z}_2^2$ & $\mathbb{Z}^8$ & ??\\
		$208$ & $A_1$ & $\mathbb{Z}$ & $\mathbb{Z}_2^2$ & $\mathbb{Z}^6$ & ??\\
		$209$ & $A_1$ & $\mathbb{Z}$ & $\mathbb{Z}_2$ & $\mathbb{Z}^6$ & ??\\
		$210$ & $A_1$ & $\mathbb{Z}$ & $\mathbb{Z}_2$ & $\mathbb{Z}_2\times \mathbb{Z}^3$ & ??\\
		$211$ & $A_1$ & $\mathbb{Z}$ & $\mathbb{Z}_2^2$ & $\mathbb{Z}^6$ & ??\\
		$212$ & $A_1$ & $\mathbb{Z}$ & $\mathbb{Z}_2$ & $\mathbb{Z}^2$ & ??\\
		$213$ & $A_1$ & $\mathbb{Z}$ & $\mathbb{Z}_2$ & $\mathbb{Z}^2$ & ??\\
		$214$ & $A_1$ & $\mathbb{Z}$ & $\mathbb{Z}_2^2$ & $\mathbb{Z}^4$ & ??\\
		$215$ & $A_1$ & $0$ & $\mathbb{Z}_2$ & $\mathbb{Z}^2$ & $\mathbb{Z}^2$\\
		$216$ & $A_1$ & $0$ & $0$ & $\mathbb{Z}$ & $\mathbb{Z}$\\
		$217$ & $A_1$ & $0$ & $\mathbb{Z}_2$ & $\mathbb{Z}$ & $\mathbb{Z}$\\
		$218$ & $A_1$ & $0$ & $\mathbb{Z}_2$ & $\mathbb{Z}$ & $\mathbb{Z}_2\times \mathbb{Z}$\\
		$219$ & $A_1$ & $0$ & $\mathbb{Z}_2$ & $\mathbb{Z}_2$ & $\mathbb{Z}_2^2$\\
		$220$ & $A_1$ & $0$ & $\mathbb{Z}_2$ & $0$ & $\mathbb{Z}_2$\\
		$221$ & $A_{1g}$ & $0$ & $0$ & $0$ & $0$\\
		$222$ & $A_{1g}$ & $0$ & $\mathbb{Z}_2$ & $\mathbb{Z}^3$ & $\mathbb{Z}^3$\\
		$223$ & $A_{1g}$ & $0$ & $\mathbb{Z}_2$ & $\mathbb{Z}$ & $\mathbb{Z}$\\
		$224$ & $A_{1g}$ & $0$ & $0$ & $\mathbb{Z}$ & $\mathbb{Z}$\\
		$225$ & $A_{1g}$ & $0$ & $0$ & $0$ & $0$\\
		$226$ & $A_{1g}$ & $0$ & $0$ & $\mathbb{Z}$ & $\mathbb{Z}$\\
		$227$ & $A_{1g}$ & $0$ & $0$ & $0$ & $0$\\
		$228$ & $A_{1g}$ & $0$ & $0$ & $\mathbb{Z}_2$ & $\mathbb{Z}_2$\\
		$229$ & $A_{1g}$ & $0$ & $0$ & $0$ & $0$\\
		$230$ & $A_{1g}$ & $0$ & $\mathbb{Z}_2$ & $\mathbb{Z}$ & $\mathbb{Z}$\\
		
		\hline
	\end{longtable*}
	\clearpage
	
	\section{Classification tables for layer groups with trivial pairing symmetries}
	\label{app:tabLG}
	\begin{longtable}{c|c|c|c|c|c}
		\caption{\label{tab:LG2} Classification table of topological phases in layer groups.}\\
		\hline
		Layer group & Corresponding space group & pairing symmetry & $E_{2,-2}^{\infty}$ & $E_{1,-1}^{\infty}$ & $^{\phi}K_{G}^{(z,c)-n}(T^2)$\\
		\hline\hline
		$1$ & $1$ & $A$ & $\mathbb{Z}_2$ & $\mathbb{Z}_2^2$ & \
		$\mathbb{Z}_2^3$\\
		$2$ & $2$ & $A_g$ & $0$ & $0$ & $0$\\
		$3$ & $3$ & $A$ & $\mathbb{Z}_2$ & $\mathbb{Z}_2^3$ & $\mathbb{Z}_2^4$\\
		$4$ & $6$ & $A'$ & $0$ & $\mathbb{Z}^2$ & $\mathbb{Z}^2$\\
		$5$ & $7$ & $A'$ & $\mathbb{Z}_2$ & $\mathbb{Z}_2^2$ & \
		$\mathbb{Z}_2^3$\\
		$6$ & $10$ & $A_g$ & $0$ & $0$ & $0$\\
		$7$ & $13$ & $A_g$ & $0$ & $\mathbb{Z}_2$ & $\mathbb{Z}_2$\\
		$8$ & $3$ & $A$ & $\mathbb{Z}_2$ & $\mathbb{Z}_2\times \
		\mathbb{Z}^2$ & $\mathbb{Z}_2\times \mathbb{Z}^2$\\
		$9$ & $4$ & $A$ & $\mathbb{Z}_2$ & $\mathbb{Z}_2^2$ & \
		$\mathbb{Z}_2\times \mathbb{Z}_4$\\
		$10$ & $5$ & $A$ & $\mathbb{Z}_2$ & $\mathbb{Z}_2\times \
		\mathbb{Z}$ & $\mathbb{Z}_2\times \mathbb{Z}$\\
		$11$ & $6$ & $A'$ & $\mathbb{Z}_2$ & $\mathbb{Z}_2\times \
		\mathbb{Z}^2$ & $\mathbb{Z}_2\times \mathbb{Z}^2$\\
		$12$ & $7$ & $A'$ & $\mathbb{Z}_2$ & $\mathbb{Z}_2^2$ & \
		$\mathbb{Z}_2\times \mathbb{Z}_4$\\
		$13$ & $8$ & $A'$ & $\mathbb{Z}_2$ & $\mathbb{Z}_2\times \
		\mathbb{Z}$ & $\mathbb{Z}_2\times \mathbb{Z}$\\
		$14$ & $10$ & $A_g$ & $0$ & $0$ & $0$\\
		$15$ & $11$ & $A_g$ & $0$ & $\mathbb{Z}$ & $\mathbb{Z}$\\
		$16$ & $13$ & $A_g$ & $0$ & $\mathbb{Z}$ & $\mathbb{Z}$\\
		$17$ & $14$ & $A_g$ & $0$ & $\mathbb{Z}_2$ & $\mathbb{Z}_2$\\
		$18$ & $12$ & $A_g$ & $0$ & $0$ & $0$\\
		$19$ & $16$ & $A_1$ & $\mathbb{Z}_2$ & $\mathbb{Z}^4$ & \
		$\mathbb{Z}^4$\\
		$20$ & $17$ & $A_1$ & $\mathbb{Z}_2$ & $\mathbb{Z}_2^2\times \
		\mathbb{Z}$ & $\mathbb{Z}_2^2\times \mathbb{Z}$\\
		$21$ & $18$ & $A_1$ & $\mathbb{Z}_2$ & $\mathbb{Z}_2^2$ & \
		$\mathbb{Z}_2\times \mathbb{Z}_4$\\
		$22$ & $21$ & $A_1$ & $\mathbb{Z}_2$ & $\mathbb{Z}_2\times \
		\mathbb{Z}^2$ & $\mathbb{Z}_2\times \mathbb{Z}^2$\\
		$23$ & $25$ & $A_1$ & $\mathbb{Z}_2$ & $\mathbb{Z}^4$ & \
		$\mathbb{Z}^4$\\
		$24$ & $28$ & $A_1$ & $\mathbb{Z}_2$ & $\mathbb{Z}_2^2\times \
		\mathbb{Z}$ & $\mathbb{Z}_2^2\times \mathbb{Z}$\\
		$25$ & $32$ & $A_1$ & $\mathbb{Z}_2$ & $\mathbb{Z}_2^2$ & \
		$\mathbb{Z}_2\times \mathbb{Z}_4$\\
		$26$ & $35$ & $A_1$ & $\mathbb{Z}_2$ & $\mathbb{Z}_2\times \
		\mathbb{Z}^2$ & $\mathbb{Z}_2\times \mathbb{Z}^2$\\
		$27$ & $25$ & $A_1$ & $0$ & $\mathbb{Z}$ & $\mathbb{Z}$\\
		$28$ & $26$ & $A_1$ & $0$ & $\mathbb{Z}_2$ & $\mathbb{Z}_2$\\
		$29$ & $26$ & $A_1$ & $0$ & $\mathbb{Z}$ & $\mathbb{Z}$\\
		$30$ & $27$ & $A_1$ & $0$ & $\mathbb{Z}_2$ & $\mathbb{Z}_2$\\
		$31$ & $28$ & $A_1$ & $\mathbb{Z}_2$ & $\mathbb{Z}_2\times \
		\mathbb{Z}^2$ & $\mathbb{Z}_2\times \mathbb{Z}^2$\\
		$32$ & $31$ & $A_1$ & $\mathbb{Z}_2$ & $\mathbb{Z}_2\times \
		\mathbb{Z}$ & $\mathbb{Z}_2\times \mathbb{Z}$\\
		$33$ & $29$ & $A_1$ & $\mathbb{Z}_2$ & $\mathbb{Z}_2^2$ & \
		$\mathbb{Z}_2\times \mathbb{Z}_4$\\
		$34$ & $30$ & $A_1$ & $\mathbb{Z}_2$ & $\mathbb{Z}_2\times \
		\mathbb{Z}$ & $\mathbb{Z}_2\times \mathbb{Z}$\\
		$35$ & $38$ & $A_1$ & $0$ & $\mathbb{Z}$ & $\mathbb{Z}$\\
		$36$ & $39$ & $A_1$ & $0$ & $\mathbb{Z}_2$ & $\mathbb{Z}_2$\\
		$37$ & $47$ & $A_g$ & $0$ & $0$ & $0$\\
		$38$ & $49$ & $A_g$ & $0$ & $\mathbb{Z}$ & $\mathbb{Z}$\\
		$39$ & $50$ & $A_g$ & $0$ & $\mathbb{Z}^2$ & $\mathbb{Z}^2$\\
		$40$ & $51$ & $A_g$ & $0$ & $0$ & $0$\\
		$41$ & $51$ & $A_g$ & $0$ & $\mathbb{Z}$ & $\mathbb{Z}$\\
		$42$ & $53$ & $A_g$ & $0$ & $\mathbb{Z}_2$ & $\mathbb{Z}_2$\\
		$43$ & $54$ & $A_g$ & $0$ & $\mathbb{Z}_2$ & $\mathbb{Z}_2$\\
		$44$ & $55$ & $A_g$ & $0$ & $0$ & $0$\\
		$45$ & $57$ & $A_g$ & $0$ & $\mathbb{Z}_2$ & $\mathbb{Z}_2$\\
		$46$ & $59$ & $A_g$ & $0$ & $\mathbb{Z}^2$ & $\mathbb{Z}^2$\\
		$47$ & $65$ & $A_g$ & $0$ & $0$ & $0$\\
		$48$ & $67$ & $A_g$ & $0$ & $0$ & $0$\\
		$49$ & $75$ & $A$ & $\mathbb{Z}_2$ & $\mathbb{Z}_2^2$ & $\mathbb{Z}_2^3$\\
		$50$ & $81$ & $A$ & $\mathbb{Z}_2$ & $\mathbb{Z}_2^2$ & $\mathbb{Z}_2^3$\\
		$51$ & $83$ & $A_g$ & $0$ & $0$ & $0$\\
		$52$ & $85$ & $A_g$ & $0$ & $\mathbb{Z}_2$ & $\mathbb{Z}_2$\\
		$53$ & $89$ & $A_1$ & $\mathbb{Z}_2$ & $\mathbb{Z}^3$ & \
		$\mathbb{Z}^3$\\
		$54$ & $90$ & $A_1$ & $\mathbb{Z}_2$ & $\mathbb{Z}_2\times \
		\mathbb{Z}$ & $\mathbb{Z}_2\times \mathbb{Z}$\\
		$55$ & $99$ & $A_1$ & $\mathbb{Z}_2$ & $\mathbb{Z}^3$ & \
		$\mathbb{Z}^3$\\
		$56$ & $100$ & $A_1$ & $\mathbb{Z}_2$ & $\mathbb{Z}_2\times \
		\mathbb{Z}$ & $\mathbb{Z}_2\times \mathbb{Z}$\\
		$57$ & $111$ & $A_1$ & $\mathbb{Z}_2$ & $\mathbb{Z}^3$ & \
		$\mathbb{Z}^3$\\
		$58$ & $113$ & $A_1$ & $\mathbb{Z}_2$ & $\mathbb{Z}_2\times \
		\mathbb{Z}$ & $\mathbb{Z}_2\times \mathbb{Z}$\\
		$59$ & $115$ & $A_1$ & $\mathbb{Z}_2$ & $\mathbb{Z}^3$ & \
		$\mathbb{Z}^3$\\
		$60$ & $117$ & $A_1$ & $\mathbb{Z}_2$ & $\mathbb{Z}_2\times \
		\mathbb{Z}$ & $\mathbb{Z}_2\times \mathbb{Z}$\\
		$61$ & $123$ & $A_{1g}$ & $0$ & $0$ & $0$\\
		$62$ & $125$ & $A_{1g}$ & $0$ & $\mathbb{Z}$ & $\mathbb{Z}$\\
		$63$ & $127$ & $A_{1g}$ & $0$ & $0$ & $0$\\
		$64$ & $129$ & $A_{1g}$ & $0$ & $\mathbb{Z}$ & $\mathbb{Z}$\\
		$65$ & $143$ & $A$ & $\mathbb{Z}_2$ & $0$ & $\mathbb{Z}_2$\\
		$66$ & $147$ & $A_g$ & $0$ & $0$ & $0$\\
		$67$ & $149$ & $A_1$ & $\mathbb{Z}_2$ & $\mathbb{Z}$ & $\mathbb{Z}$\\
		$68$ & $150$ & $A_1$ & $\mathbb{Z}_2$ & $\mathbb{Z}$ & $\mathbb{Z}$\\
		$69$ & $156$ & $A_1$ & $\mathbb{Z}_2$ & $\mathbb{Z}$ & $\mathbb{Z}$\\
		$70$ & $157$ & $A_1$ & $\mathbb{Z}_2$ & $\mathbb{Z}$ & $\mathbb{Z}$\\
		$71$ & $162$ & $A_{1g}$ & $0$ & $0$ & $0$\\
		$72$ & $164$ & $A_{1g}$ & $0$ & $0$ & $0$\\
		$73$ & $168$ & $A$ & $\mathbb{Z}_2$ & $\mathbb{Z}_2$ & $\mathbb{Z}_2^2$\\
		$74$ & $174$ & $A'$ & $0$ & $0$ & $0$\\
		$75$ & $175$ & $A_g$ & $0$ & $0$ & $0$\\
		$76$ & $177$ & $A_1$ & $\mathbb{Z}_2$ & $\mathbb{Z}^2$ & \
		$\mathbb{Z}^2$\\
		$77$ & $183$ & $A_1$ & $\mathbb{Z}_2$ & $\mathbb{Z}^2$ & \
		$\mathbb{Z}^2$\\
		$78$ & $187$ & $A'_1$ & $0$ & $0$ & $0$\\
		$79$ & $189$ & $A'_1$ & $0$ & $0$ & $0$\\
		$80$ & $191$ & $A_{1g}$ & $0$ & $0$ & $0$\\
		\hline
	\end{longtable}
	
	\section{Classification tables for rod groups with trivial pairing symmetries}
	\label{app:tabRG}
	\begin{longtable*}{c|c|c|c}
		\caption{\label{tab:RG}Classification table of topological phases in rod groups.}\\
		\hline
		Rod group & Corresponding space group & pairing symmetry & $^{\phi}K_{G}^{(z,c)-n}(T^1)$\\
		\hline\hline
		$1$ & $1$ & $A$ & $\mathbb{Z}_2$\\
		$2$ & $2$ & $A_g$ & $0$\\
		$3$ & $3$ & $A$ & $\mathbb{Z}_2$\\
		$4$ & $6$ & $A'$ & $\mathbb{Z}$\\
		$5$ & $7$ & $A'$ & $\mathbb{Z}_2$\\
		$6$ & $10$ & $A_g$ & $0$\\
		$7$ & $13$ & $A_g$ & $0$\\
		$8$ & $3$ & $A$ & $\mathbb{Z}$\\
		$9$ & $4$ & $A$ & $\mathbb{Z}_2$\\
		$10$ & $6$ & $A'$ & $\mathbb{Z}_2$\\
		$11$ & $10$ & $A_g$ & $0$\\
		$12$ & $11$ & $A_g$ & $0$\\
		$13$ & $16$ & $A_1$ & $\mathbb{Z}$\\
		$14$ & $17$ & $A_1$ & $\mathbb{Z}_2$\\
		$15$ & $25$ & $A_1$ & $0$\\
		$16$ & $27$ & $A_1$ & $0$\\
		$17$ & $26$ & $A_1$ & $0$\\
		$18$ & $25$ & $A_1$ & $\mathbb{Z}$\\
		$19$ & $28$ & $A_1$ & $\mathbb{Z}_2$\\
		$20$ & $47$ & $A_g$ & $0$\\
		$21$ & $49$ & $A_g$ & $0$\\
		$22$ & $51$ & $A_g$ & $0$\\
		$23$ & $75$ & $A$ & $\mathbb{Z}^2$\\
		$24$ & $76$ & $A$ & $\mathbb{Z}_2$\\
		$25$ & $77$ & $A$ & $\mathbb{Z}$\\
		$26$ & $78$ & $A$ & $\mathbb{Z}_2$\\
		$27$ & $81$ & $A$ & $0$\\
		$28$ & $83$ & $A_g$ & $0$\\
		$29$ & $84$ & $A_g$ & $0$\\
		$30$ & $89$ & $A_1$ & $\mathbb{Z}^2$\\
		$31$ & $91$ & $A_1$ & $\mathbb{Z}_2$\\
		$32$ & $93$ & $A_1$ & $\mathbb{Z}$\\
		$33$ & $95$ & $A_1$ & $\mathbb{Z}_2$\\
		$34$ & $99$ & $A_1$ & $0$\\
		$35$ & $101$ & $A_1$ & $0$\\
		$36$ & $103$ & $A_1$ & $0$\\
		$37$ & $111$ & $A_1$ & $0$\\
		$38$ & $112$ & $A_1$ & $0$\\
		$39$ & $123$ & $A_{1g}$ & $0$\\
		$40$ & $124$ & $A_{1g}$ & $0$\\
		$41$ & $131$ & $A_{1g}$ & $0$\\
		$42$ & $143$ & $A$ & $\mathbb{Z}_2\times\mathbb{Z}$\\
		$43$ & $144$ & $A$ & $\mathbb{Z}_2$\\
		$44$ & $145$ & $A$ & $\mathbb{Z}_2$\\
		$45$ & $147$ & $A_g$ & $0$\\
		$46$ & $149$ & $A_1$ & $\mathbb{Z}_2\times\mathbb{Z}$\\
		$47$ & $151$ & $A_1$ & $\mathbb{Z}_2$\\
		$48$ & $153$ & $A_1$ & $\mathbb{Z}_2$\\
		$49$ & $156$ & $A_1$ & $\mathbb{Z}$\\
		$50$ & $158$ & $A_1$ & $\mathbb{Z}_2$\\
		$51$ & $162$ & $A_{1g}$ & $0$\\
		$52$ & $163$ & $A_{1g}$ & $0$\\
		$53$ & $168$ & $A$ & $\mathbb{Z}^3$\\
		$54$ & $169$ & $A$ & $\mathbb{Z}_2$\\
		$55$ & $171$ & $A$ & $\mathbb{Z}$\\
		$56$ & $173$ & $A$ & $\mathbb{Z}_2\times\mathbb{Z}$\\
		$57$ & $172$ & $A$ & $\mathbb{Z}$\\
		$58$ & $170$ & $A$ & $\mathbb{Z}_2$\\
		$59$ & $174$ & $A'$ & $\mathbb{Z}_2$\\
		$60$ & $175$ & $A_g$ & $0$\\
		$61$ & $176$ & $A_g$ & $0$\\
		$62$ & $177$ & $A_1$ & $\mathbb{Z}^3$\\
		$63$ & $178$ & $A_1$ & $\mathbb{Z}_2$\\
		$64$ & $180$ & $A_1$ & $\mathbb{Z}$\\
		$65$ & $182$ & $A_1$ & $\mathbb{Z}_2\times\mathbb{Z}$\\
		$66$ & $181$ & $A_1$ & $\mathbb{Z}$\\
		$67$ & $179$ & $A_1$ & $\mathbb{Z}_2$\\
		$68$ & $183$ & $A_1$ & $0$\\
		$69$ & $184$ & $A_1$ & $0$\\
		$70$ & $186$ & $A_1$ & $0$\\
		$71$ & $187$ & $A'_1$ & $\mathbb{Z}$\\
		$72$ & $188$ & $A'_1$ & $\mathbb{Z}_2$\\
		$73$ & $191$ & $A_{1g}$ & $0$\\
		$74$ & $192$ & $A_{1g}$ & $0$\\
		$75$ & $194$ & $A_{1g}$ & $0$\\
		\hline
	\end{longtable*}

\end{document}